\documentclass[journal,comsoc]{IEEEtran}

\usepackage[T1]{fontenc}
\usepackage{graphicx}
\usepackage{cite}
\usepackage[dvipsnames,svgnames]{xcolor}
\usepackage{subfig}
\usepackage{dblfloatfix}

\usepackage{amsmath}

\usepackage{amssymb}
\usepackage{amsmath}

\usepackage[cmintegrals]{newtxmath}

\renewcommand{\IEEEQED}{\IEEEQEDopen}
\newtheorem{lemma}{Lemma}
\newtheorem{assumption}{Assumption}
\newtheorem{theorem}{Theorem}
\newtheorem{definition}{Definition}

\begin{document}

\title{Massive MIMO with Spatially Correlated Rician Fading Channels}

\author{\"{O}zgecan~\"{O}zdogan,~\IEEEmembership{Student Member,~IEEE,}
        Emil Bj\"{o}rnson,~\IEEEmembership{Senior Member,~IEEE,}
        and~Erik~G.~Larsson,~\IEEEmembership{Fellow,~IEEE}\\Email: \{ozgecan.ozdogan, emil.bjornson, erik.g.larsson\}@liu.se
        \thanks{Manuscript received May 17, 2018; revised September 3, 2018 and November 14, 2018; accepted January 5, 2019. Date of publication January 21, 2019; date of current version May 15, 2019. This work was supported in part by ELLIIT and in part by the Swedish Research Council. This paper was presented at the IEEE SPAWC 2018. The associate editor coordinating the review of this paper and approving it for publication was X. Chen. \emph{(Corresponding author: \"{O}zgecan \"{O}zdogan)}}
\thanks{The authors are with the Department of Electrical Engineering (ISY), Link\"oping University, 581 83 Link\"oping, Sweden (e-mail: ozgecan.ozdogan@liu.se; emil.bjornson@liu.se, erik.g.larsson@liu.se).}
\thanks{Color version of one or more of the figures in this paper are available online at http://ieeexplore.ieee.org.}}

\markboth{IEEE Transactions on Communications, Vol. 67, No. 5, May 2019}%
{Shell \MakeLowercase{\textit{et al.}}: Bare Demo of IEEEtran.cls for IEEE Communications Society Journals}
%



\maketitle

\begin{abstract}
This paper considers multi-cell Massive MIMO (multiple-input multiple-output) systems where the channels are spatially correlated Rician fading. The channel model is composed of a deterministic line-of-sight (LoS) path and a stochastic non-line-of-sight (NLoS) component describing a practical spatially correlated multipath environment. We derive the statistical properties of the minimum mean squared error (MMSE), element-wise MMSE (EW-MMSE), and least-square (LS) channel estimates for this model. Using these estimates for maximum ratio (MR) combining and precoding, rigorous closed-form uplink (UL) and downlink (DL)   achievable spectral efficiency (SE) expressions are derived and analyzed. The asymptotic SE behavior when using the different channel estimators are also analyzed. Numerical results show that the SE is higher when using the MMSE estimator than the other estimators, and the performance gap increases with the number of antennas.
\end{abstract}

\begin{IEEEkeywords}
Massive MIMO, spatially correlated Rician fading, channel estimation, spectral efficiency.
\end{IEEEkeywords}

\IEEEpeerreviewmaketitle

\section{Introduction}

Massive MIMO (multiple-input multiple-output) is the key technology for increasing the spectral efficiency (SE) in future cellular networks, by virtue of beamforming and spatial multiplexing \cite{Larsson2014}. A Massive MIMO base station (BS) is equipped with a massive number (e.g., a hundred)  of individually steerable antennas, which can be effectively used to serve tens of user equipments (UEs) simultaneously on the same time-frequency resource. The canonical form of Massive MIMO operates in time-division duplex (TDD) mode and acquires channel state information (CSI) by using uplink (UL) pilot signaling and channel reciprocity \cite{EmilsBook}. The CSI is used for coherent UL receive combining and downlink (DL) transmit precoding. 

The achievable SEs of Massive MIMO systems with imperfect CSI have been rigorously characterized and optimized for fading channels modeled by either spatially uncorrelated~\cite{Marzetta2016a} or spatially correlated~\cite{EmilsBook,hoydis2013massive} Rayleigh fading. Communication with fading-free (line-of-sight) LoS propagation has also be treated \cite{EmilsBook,Yang2017}. However, practical channels can consist of a combination of a deterministic LoS path and small-scale fading caused by multipath propagation, which can be modeled by the Rician fading model \cite{Tse2005a}.

The performance of Massive MIMO with Rician fading channels is much less analyzed than with Rayleigh fading. The single-cell case was studied in \cite{Zhang2014,Kong2015,Hu2017} under the assumption of spatially uncorrelated Rician fading channels and zero-forcing (ZF) processing. Approximate SE expressions for the UL and DL were provided in \cite{Zhang2014} and \cite{Kong2015,Hu2017}, respectively. The multi-cell case was studied in \cite{Zhao2016,Wu2017,Sanguinetti2017}, assuming spatially uncorrelated Rician fading within each cell and spatially uncorrelated Rayleigh fading across cells. Approximate SE expressions were derived in the UL with ZF combining \cite{Wu2017} and in the DL with ZF \cite{Zhao2016} or regularized ZF precoding \cite{Sanguinetti2017}. Note that these are the prior works that consider imperfect CSI, which is the practically relevant scenario, while prior works assuming perfect CSI can be found in the reference lists of \cite{Zhang2014,Kong2015,Hu2017,Zhao2016,Wu2017,Sanguinetti2017}.

\subsection{Main Contributions}

There are three major limitations of the prior works. First, the fading was modeled as spatially uncorrelated, although practical channels are correlated, due the finite number of scattering clusters \cite{EmilsBook}. Second, the inter-cell channels were modeled by Rayleigh fading, although it may happen that a UE has LoS paths to multiple BSs (e.g., in parks, dense small-cell deployments, or when serving unmanned aerial vehicles (UAVs)). The existence of an LoS path depends on environmental factors. In the simulation part, we use a probabilistic approach based on the 3GPP model \cite{3gpp} to achieve a realistic scenario. Third, only approximate SE expressions were derived in closed form in prior works, which only provide insights into special operational regimes, such as having asymptotically many antennas. In this paper, we address these shortcomings:

\begin{itemize}
	\item We consider a multi-cell scenario with spatially correlated Rician fading channels between the pairs of BSs and UEs that are determined statistically where other pairs have spatially correlated Rayleigh fading channels. Previously, this channel model has only been used for single-cell scenarios with perfect CSI \cite{Zhang2013,Tataria2017}.
	
	\item We derive the minimum mean squared error (MMSE), element-wise MMSE (EW-MMSE) and least-square (LS) channel estimators and characterize their statistics. Using these estimates for MR combining and precoding, we compute rigorous closed-form UL and DL SEs and discuss their structure.
	
	\item We analyze asymptotic behavior of UL and DL SEs under spatially correlated Rician fading when using the different estimators.
	
	\item We compare  the UL and DL SEs with MMSE, EW-MMSE and LS estimation numerically, considering both correlated and uncorrelated Rician and Rayleigh fading.
\end{itemize}

The conference version of this paper \cite{Ozdogan2018a} only considered the UL and only used the MMSE and LS estimators.

\emph{\textbf{Reproducible research:}} All the simulation results can be reproduced using the Matlab code and data files available at: https://github.com/emilbjornson/rician-massive-mimo

\section{Channel and System Model}

We consider a Massive MIMO system with $L$ cells where each cell consists of  one base station (BS) with $M_j$ antennas that serves $K$ single-antenna user equipments (UEs). The system operates in TDD mode where the channel responses remain constant over a coherence block of $\tau_c$ samples. Also, we assume that the channel realizations are independent between any pair of coherence blocks. The size of $\tau_c$ is determined by the carrier frequency and external factors such as the propagation environment and UE mobility \cite{Marzetta2016a}. The samples are used for three different tasks: $\tau_p$ samples for uplink (UL) pilot signals, $\tau_u$  samples for UL data transmission and $\tau_d$  samples for downlink (DL) data transmission where $\tau_c = \tau_p +\tau_u + \tau_d$. Both UL and DL channels are estimated by uplink pilot signals by exploiting channel reciprocity in the TDD protocol.\footnote{We assume that the hardware is fully synchronized for reciprocity; see \cite[Sec.~6]{EmilsBook} for a review of calibration techniques.}

The channel response between UE $k$ in cell $l$ and the BS in cell $j$ is denoted by $\mathbf{h}^j_{lk} \in \mathbb{C}^{M_j}$. Each element of $\mathbf{h}^j_{lk}$ corresponds to the propagation channel from the UE to one of the BS's $M_j$ antennas. The superscript of $\mathbf{h}^j_{lk}$ indicates the BS index and the subscript identifies the index of the cell and the UE. While the channel responses are the same in the UL and DL of a coherence block, for notational convenience, we use $\mathbf{h}^j_{lk}$ for the UL channel and $(\mathbf{h}^j_{lk})^H$ for DL channel.

In this paper, we consider spatially correlated Rician fading channels. Each channel vector $\mathbf{h}^j_{lk}$, $\forall j,l \in 1,\dots,L$ and $\forall k \in 1,\dots,K$, is modeled as a realization of the circularly symmetric complex Gaussian distribution\footnote{Strictly speaking, circularly symmetric random variable must have zero mean, but we consider the common extension of this terminology to the case when it is sufficient that the non-zero-mean part is circularly symmetric.}
\begin{equation}
\mathbf{h}^j_{lk} \sim \mathcal{N}_\mathbb{C}\left(  \bar{\mathbf{h}}^j_{lk}, \mathbf{R}^j_{lk} \right),
\end{equation}
where the mean $ \bar{\mathbf{h}}^j_{lk} \in \mathbb{C}^{M_j}$ corresponds to the LoS component and $\mathbf{R}^j_{lk} \in \mathbb{C}^{M_j \times M_j}$ is the positive semi-definite covariance matrix describing the spatial correlation of the non-line-of-sight (NLoS) components. The small-scale fading is described by the Gaussian distribution whereas  $\mathbf{R}^j_{lk}$ and $\bar{\mathbf{h}}^j_{lk}$ model the macroscopic propagation effects, including pathloss, shadow-fading, and the antenna gains and radiation patterns at the transmitter and receiver. 
The average channel gain from one of the antennas at BS $j$ to UE $k$ in cell $l$ is determined by the normalized trace as
\begin{equation}
	\beta^j_{lk}= \frac{1}{M_j} \mathrm{tr}(\mathbf{R}^j_{lk}).
\end{equation}
where $\beta^j_{lk}$ is called the large-scale fading coefficient.

\section{Channel Estimation} \label{sec3}

Each BS requires CSI for receive processing. Therefore, $\tau_p$ samples are reserved for performing UL pilot-based channel estimation in each coherence block, giving room for $\tau_p$ mutually orthogonal pilot sequences. 
These pilot sequences are allocated to different UEs and the same sequences are reused by UEs in multiple cells. The deterministic pilot sequence of UE $k$ in cell $j$ is denoted by $\boldsymbol{\phi}_{jk} \in \mathbb{C}^{\tau_p}$ and  $\|\boldsymbol{\phi}_{jk} \|^2= \tau_p$.
We define the set
\begin{equation}
\mathcal{P}_{jk} = \left\lbrace (l,i): \boldsymbol{\phi}_{li} = \boldsymbol{\phi}_{jk}, l=1,\dots,L, i=1,\dots,K_l \right\rbrace ,
\end{equation}
with indices of all UEs in the system that utilize the same pilot sequence as UE $k$ in cell $j$ (including the UE itself). The received pilot signal $\mathbf{Y}^p_{j} \in \mathbb{C}^{M_j \times \tau_p }$   at BS $j$ is
\begin{equation}
\mathbf{Y}^p_{j} = \sum_{k=1}^{K_j} \sqrt{p_{jk}} \mathbf{h}^j_{jk} \boldsymbol{\phi}^T_{jk} + \mathop{\sum_{l=1 }}^{L}_{l \neq j} \sum_{i=1}^{K_l} \sqrt{p_{li}} \mathbf{h}^j_{li} \boldsymbol{\phi}^T_{li} + \mathbf{N}^p_j ,
\end{equation}
where $\mathbf{N}^p_j \in \mathbb{C}^{M_j\times \tau_p}$ has independent and identically distributed $\mathcal{N}_\mathbb{C}(0, \sigma^2_{\mathrm{ul}} )$-elements, with variance $\sigma^2_{\mathrm{ul}}$. 
To estimate the channel $\mathbf{h}^j_{li}$, BS~$j$ multiplies $\mathbf{Y}^p_{j}$ with the UE's pilot sequence $\boldsymbol{\phi}^*_{li}$ to obtain 
\begin{equation}\label{sec3eq1}
\mathbf{y}^p_{jli}= \mathbf{Y}^p_{j}  \boldsymbol{\phi}^*_{li}=\sqrt{p_{li}} \tau_p \mathbf{h}^j_{li} + \!\sum_{(l',i') \in \mathcal{P}_{li} \backslash (l,i)} \!\!\!\! \sqrt{p_{l'i'}} \tau_p \mathbf{h}^j_{l'i'} +\mathbf{N}^p_j \boldsymbol{\phi}^*_{li}.
\end{equation}
The processed received pilot signal $\mathbf{y}^p_{jli} \in \mathbb{C}^{M_j}$ is a sufficient statistics for estimating $\mathbf{h}^j_{li}$ \cite{EmilsBook}. We will now consider three different channel estimators, which rely on different amounts of statistical channel knowledge. The statistical distributions (the mean vector and covariance matrices) can be estimated using the sample mean and sample covariance matrices in practice \cite{EmilsBook,Bjornson2016c,Neumann-2017-A,Vorobyov2018,Haghighatshoar2017a}.  Note that a small change in the UE location may result in a significant phase-shift of the LoS component. More specifically, if the UE moves half a wavelength away from the BS, the phase of the channel response changes by $180^\circ $. This phase shift, however, will be identical for all BS antennas, and may therefore be accurately tracked in practice.

\subsection{MMSE Channel Estimator}
Based on the processed received pilot signal in $\eqref{sec3eq1}$, the BS can apply MMSE estimation to obtain an estimate of ${\mathbf{h}}^j_{li}$ as shown in the following lemma. Notice that the Bayesian MMSE estimator requires that the statistical distributions are fully known. 
\begin{lemma} \label{lemma:MMSE-estimate}
	The MMSE estimate of channel from BS $j$  to UE $i$ in cell $l$ is
	\begin{equation}\label{mmse1}
	\hat{\mathbf{h}}^j_{li} = \bar{\mathbf{h}}^j_{li} + \sqrt{p_{li}}  \mathbf{R}^j_{li} \boldsymbol{\Psi}^j_{li} \left( \mathbf{y}^p_{jli} - \bar{\mathbf{y}}^p_{jli}  \right),
	\end{equation}
	where $\bar{\mathbf{y}}^p_{jli}=\sum_{(l',i') \in \mathcal{P}_{li} } \sqrt{p_{l'i'}} \tau_p \bar{\mathbf{h}}^j_{l'i'} $	and 
	\begin{align}
	\boldsymbol{\Psi}^j_{li}={\tau_p} \mathrm{Cov}\left\lbrace {\mathbf{y}^p_{jli}}\right\rbrace^{-1} \!=\!\Bigg(  \sum_{(l',i') \in \mathcal{P}_{li} } \!\!{p_{l'i'}} \tau_p \mathbf{R}^j_{l'i'} +  \sigma^2 \mathbf{I}_{M_j} \Bigg)^{\!-1}\!\!.
	\end{align}
	The estimation error $\tilde{\mathbf{h}}^j_{li} = \mathbf{h}^j_{li} - \hat{\mathbf{h}}^j_{li}$ has the covariance matrix
	\begin{equation}
	\mathbf{C}^j_{li}=  \mathbf{R}^j_{li} - p_{li} \tau_p \mathbf{R}^j_{li} \boldsymbol{\Psi}^j_{li} \mathbf{R}^j_{li}
	\end{equation}
	and the mean-squared error is $\mathrm{MSE}= \mathbb{E}\lbrace \|  \mathbf{h}^j_{li} - \hat{\mathbf{h}}^j_{li}\|^2 \rbrace  =\mathrm{tr}( \mathbf{C}^j_{li} ) $. The MMSE estimate  $\hat{\mathbf{h}}^j_{li}$ and the estimation error $\tilde{\mathbf{h}}^j_{li} $ are independent random variables and distributed as
	\begin{align}  \label{eq:statistics-MMSE-estimate}
	\hat{\mathbf{h}}^j_{li} &\sim \mathcal{N}_\mathbb{C}\left( \bar{\mathbf{h}}^j_{li},  \mathbf{R}^j_{li} - \mathbf{C}^j_{li} \right), \\
	\tilde{\mathbf{h}}^j_{li} &\sim \mathcal{N}_\mathbb{C}\left( \mathbf{0}_M, \mathbf{C}^j_{li} \right).
	\end{align}
\end{lemma}
\begin{IEEEproof}
	The proof follows from the standard MMSE estimation of Gaussian random variables that are observed in Gaussian noise \cite{KayBookESt,EmilsBook}.
\end{IEEEproof}

Note that the estimation error covariance matrix $\mathbf{C}^j_{li}$ does not depend on the mean values. In other words, the estimation error is not affected by the LoS components since these are known and can be subtracted from the received signals. Moreover, the channel estimates of UEs in the set $\mathcal{P}_{li}$ are not independent, despite the assumption that the channels are independent.
This is known as pilot contamination and happens since the UEs use the same pilot sequence. UE $(j,k) \in \mathcal{P}_{li}$ has the channel estimate
\begin{equation}
\hat{\mathbf{h}}^j_{jk} = \bar{\mathbf{h}}^j_{jk} + \sqrt{p_{jk}}  \mathbf{R}^j_{jk} \boldsymbol{\Psi}^j_{li} \left( \mathbf{y}^p_{jli} - \bar{\mathbf{y}}^p_{jli}  \right)
\end{equation}
and it is correlated with $\hat{\mathbf{h}}^j_{li}$ in \eqref{mmse1} since $\mathbf{y}^p_{jli} $ appears in both expressions and $\boldsymbol{\Psi}^j_{li}=\boldsymbol{\Psi}^j_{jk}$. We will utilize the distributions of the channel estimates and estimation errors in Sections~\ref{section4} and \ref{section5} when analyzing the UL and DL SE.

\subsection{Element-wise MMSE Channel Estimator}
If the BS does not have knowledge of the entire covariance matrices, the EW-MMSE estimator can be implemented as an alternative \cite{EmilsBook,Shariati2014}. In this method, only the diagonals of the covariance matrices are needed and the correlation between the elements are ignored by the estimator. As a result, there are no matrix inversions and thus the computational complexity  is greatly reduced as compared to the MMSE estimator in Lemma~\ref{lemma:MMSE-estimate}.
\begin{lemma}
	The EW-MMSE estimate of channel from BS $j$  to UE $i$ in cell $l$ is
\begin{equation}\label{estew1}
\hat{\mathbf{h}}^j_{li} = \bar{\mathbf{h}}^j_{li} + \sqrt{p_{li}} \mathbf{D}^j_{li} \boldsymbol{\Lambda}^{j}_{li} \left( \mathbf{y}^p_{jli} - \bar{\mathbf{y}}^p_{jli}\right), 
\end{equation}
where $\mathbf{D}^j_{li} \in \mathbb{C}^{M_j \times M_j}$ and $\boldsymbol{\Lambda}^j_{li} \in \mathbb{C}^{M_j \times M_j}$ are diagonal matrices with $\mathbf{D}^j_{li}=\mathrm{diag}\left( \left[ \mathbf{R}^j_{li}\right]_{mm} : m=1,\dots,M_j\right)  $ and $\boldsymbol{\Lambda}^j_{li}=\mathrm{diag}\left(\left[ \sum_{(l',i') \in \mathcal{P}_{li} } {p_{l'i'}} \tau_p \mathbf{R}^j_{l'i'} +  \sigma^2 \mathbf{I}_{M_j}\right]_{mm}  : m=1,\dots,M_j\right)^{-1} $. The distributions of EW-MMSE estimate $\hat{\mathbf{h}}^j_{li}$ and the estimation error $\tilde{\mathbf{h}}^j_{li}$ are correlated and distributed as
\begin{equation}\label{ewmmse1}
\hat{\mathbf{h}}^j_{li} \sim \mathcal{N}_\mathbb{C}\left( \bar{\mathbf{h}}^j_{li},   \boldsymbol{\Sigma}^j_{li}\right),
\end{equation}
\begin{equation}
\tilde{\mathbf{h}}^j_{li} \sim \mathcal{N}_\mathbb{C}\left( \mathbf{0}_M,    \tilde{ \boldsymbol{\Sigma}}^j_{li} \right),
\end{equation}
where $\boldsymbol{\Sigma}^j_{li}= p_{li}\tau_p  \mathbf{D}^j_{li} \boldsymbol{\Lambda}^{j}_{li} \left( \boldsymbol{\Psi}^j_{li}\right)^{-1}   \boldsymbol{\Lambda}^{j}_{li} \mathbf{D}^j_{li}$ and $\tilde{ \boldsymbol{\Sigma}}^j_{li}= \mathbf{R}^j_{li} - p_{li}\tau_p\mathbf{R}^j_{li} \boldsymbol{\Lambda}^{j}_{li} \mathbf{D}^j_{li} -p_{li}\tau_p\mathbf{D}^j_{li}\boldsymbol{\Lambda}^{j}_{li}\mathbf{R}^j_{li} + \boldsymbol{\Sigma}^j_{li}$.
\end{lemma}

\begin{IEEEproof}
We can easily get the desired result using the same approach as was employed to derive the MMSE estimator, but estimating each element separately using only the signal obtained at that antenna and then computing the resulting statistics.
\end{IEEEproof}

 In contrast to MMSE estimation, $	\hat{\mathbf{h}}^j_{li} $ and $\tilde{\mathbf{h}}^j_{li}$ are correlated  with $\mathbb{E}\left\lbrace 	\hat{\mathbf{h}}^j_{li}   ( \tilde{\mathbf{h}}^j_{li})^H \right\rbrace= p_{li}\tau_p\mathbf{D}^j_{li}\boldsymbol{\Lambda}^{j}_{li}\mathbf{R}^j_{li} - \boldsymbol{\Sigma}^j_{li} $ except in the special case when all the covariance matrices are diagonal.

\subsection{LS Channel Estimator}
If the BS has no prior information regarding $ \mathbf{R}^j_{li}$ and $\bar{\mathbf{h}}^j_{li}$, the non-Bayesian LS estimator can be utilized to get an estimate of the propagation channel $\mathbf{h}^j_{li}$. The LS estimate is defined as the value of $\hat{\mathbf{h}}^j_{li}$ that minimizes $\| \mathbf{y}^p_{jli} - \sqrt{p_{li}} \tau_p\hat{\mathbf{h}}^j_{li} \|^2 $, which in this case is 
\begin{equation}\label{ls1}
\hat{\mathbf{h}}^j_{li} = \frac{1}{\sqrt{p_{li}} \tau_p} \mathbf{y}^p_{jli}.
\end{equation}

\begin{lemma}\label{lemmaLS}
	The LS estimator and estimation error are correlated random variables and distributed as
	\begin{align} 
	\hat{\mathbf{h}}^j_{li}& \sim \mathcal{N}_\mathbb{C}\left( \frac{1}{\sqrt{p_{li}} \tau_p} \bar{\mathbf{y}}^p_{jli},  \ \ \frac{1}{p_{li} \tau_p} (\boldsymbol{\Psi}^j_{li})^{-1} \right), \\
	\tilde{\mathbf{h}}^j_{li} &\sim \mathcal{N}_\mathbb{C}\left( \bar{\mathbf{h}}^j_{li}\! -\! \frac{1}{\sqrt{p_{li}} \tau_p} \bar{\mathbf{y}}^p_{jli} ,  \frac{1}{p_{li} \tau_p} (\boldsymbol{\Psi}^j_{li})^{-1} \!-\mathbf{R}^j_{li}     \!\right).
	\end{align}
\end{lemma}	

\begin{IEEEproof}
	The proof is given in Appendix \ref{ProoflemmaLS}.
\end{IEEEproof}

This lemma shows that the statistics are more complicated than when using the MMSE and EW-MMSE estimators. For example, the estimation error has non-zero mean, which needs to be accounted for when analyzing the communication performance.
	\begin{figure*}[bp]
	\normalsize
	\hrulefill
	\setcounter{equation}{22}
	\begin{align}\label{sec4eq4}
	&\mathbb{E}\left\lbrace \left| {\mathbf{v}}_{jk}^H \mathbf{h}^j_{li}  \right| ^2 \right\rbrace = p_{jk} \tau_p \mathrm{tr}\left( \mathbf{R}^j_{li} \mathbf{R}^j_{jk} \boldsymbol{\Psi}^j_{jk} \mathbf{R}^j_{jk}\right) + p_{jk} \tau_p ( \bar{\mathbf{h}}^j_{li})^H    \mathbf{R}^j_{jk} \boldsymbol{\Psi}^j_{jk} \mathbf{R}^j_{jk} \bar{\mathbf{h}}^j_{li} +  ( \bar{\mathbf{h}}^j_{jk})^H    \mathbf{R}^j_{li} \bar{\mathbf{h}}^j_{jk} +  \left|  (\bar{\mathbf{h}}^j_{jk})^H    \bar{\mathbf{h}}^j_{li}   \right| ^2 \nonumber \\  
	&+\begin{cases}
	p_{jk} p_{li}  \tau^2_p \left|  \mathrm{tr}\left( \mathbf{R}^j_{li} \boldsymbol{\Psi}^j_{jk} \mathbf{R}^j_{jk}\right) \right| ^2 +  2 \sqrt{ p_{jk} p_{li}}\tau_p \mathrm{Re}\left\lbrace\mathrm{tr}\left(    \mathbf{R}^j_{li} \boldsymbol{\Psi}^j_{jk} \mathbf{R}^j_{jk}  \right)  ( \bar{\mathbf{h}}^j_{li})^H \bar{\mathbf{h}}^j_{jk}  \right\rbrace   &  (l,i) \in \mathcal{P}_{jk}   \\
	0 &  (l,i) \notin \mathcal{P}_{jk}.
	\end{cases}
	\end{align}
	
	\vspace*{-10pt}
\end{figure*}
 	\begin{figure*}[bp]
	\normalsize
	\hrulefill
	\setcounter{equation}{24}
	\begin{equation}\label{xi1}
	{\xi}^{\mathrm{ul}}_{li}= \frac{p_{jk} \tau_p \mathrm{tr}\left( \mathbf{R}^j_{li} \mathbf{R}^j_{jk} \boldsymbol{\Psi}^j_{jk} \mathbf{R}^j_{jk}\right) + p_{jk} \tau_p \left( \bar{\mathbf{h}}^j_{li}\right)^H    \mathbf{R}^j_{jk} \boldsymbol{\Psi}^j_{jk} \mathbf{R}^j_{jk} \bar{\mathbf{h}}^j_{li} +  \left( \bar{\mathbf{h}}^j_{jk}\right)^H    \mathbf{R}^j_{li} \bar{\mathbf{h}}^j_{jk} + \left|  (\bar{\mathbf{h}}^j_{jk})^H    \bar{\mathbf{h}}^j_{li}   \right| ^2}{ p_{jk} \tau_p \mathrm{tr}\left( \mathbf{R}^j_{jk} \boldsymbol{\Psi}^j_{jk} \mathbf{R}^j_{jk}\right) +\| \bar{\mathbf{h}}^j_{jk}   \|^2} ,
	\end{equation}
		\begin{equation}\label{coherentInterference}
	{\Gamma}^{\mathrm{ul}}_{li}=    \frac{p_{jk} p_{li}  \tau^2_p\left|  \mathrm{tr}\left( \mathbf{R}^j_{li} \boldsymbol{\Psi}^j_{jk} \mathbf{R}^j_{jk}\right) \right| ^2 +  2 \sqrt{ p_{jk} p_{li}}\tau_p \mathrm{Re}\left\lbrace \mathrm{tr}\left(    \mathbf{R}^j_{li} \boldsymbol{\Psi}^j_{jk} \mathbf{R}^j_{jk}  \right) \left( \bar{\mathbf{h}}^j_{li}\right)^H \bar{\mathbf{h}}^j_{jk}  \right\rbrace }{p_{jk} \tau_p \mathrm{tr}\left( \mathbf{R}^j_{jk} \boldsymbol{\Psi}^j_{jk} \mathbf{R}^j_{jk}\right) +\| \bar{\mathbf{h}}^j_{jk}   \|^2} .
	\end{equation}
	
\end{figure*}

	\setcounter{equation}{17}
\section{Uplink Spectral efficiency with MR combining}\label{section4}
\label{section:SE}

During data transmission, the received signal $\mathbf{y}_j \in \mathbb{C}^{M_j}$ at BS $j$ is
\begin{align}\label{s1eq1}
\mathbf{y}_j = \displaystyle\sum_{k=1}^{K_j} \mathbf{h}^j_{jk} s_{jk} + \mathop{\sum_{l=1 }}^{L}_{l \neq j}\sum_{i=1}^{K_l} \mathbf{h}^j_{li} s_{li} + \mathbf{n}_j ,
\end{align}
where $\mathbf{n}_j \sim \mathcal{N}_\mathbb{C}\left(  \mathbf{0}_{M_j}, \sigma^2_{\mathrm{ul}} \mathbf{I}_{M_j}    \right) $ is additive noise. The UL signal from UE $k$ in cell $l$ is denoted by $s_{lk} \in \mathbb{C}$ and has power $p_{lk} = \mathbb{E}\left\lbrace |s_{lk}|^2\right\rbrace $. The first term in \eqref{s1eq1} is the desired signal and the latter terms denote interference and noise, respectively. 

BS~$j$ selects the receive combining vector $\mathbf{v}_{jk} \in \mathbb{C}^{M_j}$ based on its CSI and multiplies it with $\mathbf{y}_j$ to separate the desired signal from its UE $k$ from interference. 
As in \cite[Th.~4.4]{EmilsBook}, the ergodic UL capacity of UE $k$ in cell $j$ is lower bounded by 
\begin{equation} \label{eq:SEexpression}
\mathrm{SE}^{\mathrm{ul}}_{jk}=\frac{\tau_u}{\tau_c}\log_2\left(1 + \gamma^{\mathrm{ul}}_{jk} \right) \  \ \mathrm{[bit/ s/ Hz]},
\end{equation}
with the effective SINR
\begin{equation}\label{sec4eq1}
\gamma^{\mathrm{ul}}_{jk} =\frac{p_{jk} |  \mathbb{E}\lbrace \mathbf{v}^H_{jk} \mathbf{h}^j_{jk}  \rbrace   | ^2}{\displaystyle\sum_{l=1}^{L}\sum_{i=1}^{K_l}   p_{li}  \mathbb{E}\lbrace | \mathbf{v}^H_{jk} \mathbf{h}^j_{li}  | ^2 \rbrace  - p_{jk} |  \mathbb{E}\lbrace \mathbf{v}^H_{jk} \mathbf{h}^j_{jk}  \rbrace   | ^2  +\sigma^2_{\mathrm{ul}}  \mathbb{E}\lbrace  \| \mathbf{v}_{jk} \|^2 \rbrace } ,
\end{equation}
where the expectations are with respect to all sources of randomness. Since $\mathrm{SE}^{\mathrm{ul}}_{jk}$ is below the capacity, it is an ergodic achievable SE.
The effective SINR $\gamma^{\mathrm{ul}}_{jk}$ can be computed numerically for any combining scheme and channel estimator. We will show that it can be computed in closed form when using MR combining, based on each of the three channel estimators derived in Section~\ref{sec3}.

\subsection{Uplink Spectral Efficiency with MMSE estimator}
If the MMSE estimator in \eqref{mmse1} is used, we obtain a closed-form expression for the SE in \eqref{eq:SEexpression} as in the next theorem.

\begin{theorem}\label{thULMMSE}
	If MR  combining with $\mathbf{v}_{jk} = \hat{\mathbf{h}}^j_{jk}$ is used based on the MMSE estimator, then 
	\begin{equation}\label{sec4eq2}
	\mathbb{E}\left\lbrace \mathbf{v}^H_{jk} \mathbf{h}^j_{jk}  \right\rbrace = p_{jk} \tau_p \mathrm{tr}\left( \mathbf{R}^j_{jk} \boldsymbol{\Psi}^j_{jk} \mathbf{R}^j_{jk}\right) +\| \bar{\mathbf{h}}^j_{jk}   \|^2 ,
	\end{equation}
	\begin{equation}\label{sec4eq3}
	\mathbb{E}\left\lbrace  \| \mathbf{v}_{jk} \|^2 \right\rbrace  = p_{jk} \tau_p \mathrm{tr}\left( \mathbf{R}^j_{jk} \boldsymbol{\Psi}^j_{jk} \mathbf{R}^j_{jk}\right) +\| \bar{\mathbf{h}}^j_{jk}   \|^2 ,
	\end{equation}
	and $\mathbb{E}\lbrace | {\mathbf{v}}_{jk}^H \mathbf{h}^j_{li}  | ^2 \rbrace$ is given in \eqref{sec4eq4}, at the bottom of this page. 	Plugging these expressions into the SINR in $\eqref{sec4eq1}$ yields \setcounter{equation}{23}
	 \begin{equation} \label{MMSEeq1}
	 \gamma^{\mathrm{ul,mmse}}_{jk} =\frac{ p_{jk}^2 \tau_p \mathrm{tr}\left( \mathbf{R}^j_{jk} \boldsymbol{\Psi}^j_{jk} \mathbf{R}^j_{jk}\right) + p_{jk}\| \bar{\mathbf{h}}^j_{jk}   \|^2}{\displaystyle\sum_{l=1}^{L}\sum_{i=1}^{K_l}   p_{li} {\xi}^{\mathrm{ul}}_{li}  +  \displaystyle\sum_{(l,i) \in \mathcal{P}_{jk} \backslash (j,k)} p_{li} {\Gamma}^{\mathrm{ul}}_{li}  - p_{jk}\nu_{jk}^{\mathrm{ul}} + \sigma^2_{\mathrm{ul}} },
	 \end{equation}
	 where  $\nu_{jk}^{\mathrm{ul}}=\frac{\| \bar{\mathbf{h}}^j_{jk}   \|^4}{p_{jk} \tau_p \mathrm{tr}\left( \mathbf{R}^j_{jk} \boldsymbol{\Psi}^j_{jk} \mathbf{R}^j_{jk}\right) +\| \bar{\mathbf{h}}^j_{jk}   \|^2}$, ${\xi}^{\mathrm{ul}}_{li}$ and ${\Gamma}^{\mathrm{ul}}_{li}$  correspond to LoS-related interference, non-coherent interference, and coherent interference, respectively. The latter two are given at the bottom of this page.

\end{theorem}

\begin{IEEEproof}
	The proof is given in Appendix \ref{ProofthULMMSE}.
\end{IEEEproof}

The rigorous closed-from SINR expression in \eqref{MMSEeq1}  provides important and exact insights into the behaviors of Rician fading Massive MIMO systems. The signal terms in the numerator depend on the estimation quality and the LoS component. The former is reduced by pilot contamination, since $ p_{jk}^2 \tau_p \mathrm{tr}\left( \mathbf{R}^j_{jk} \boldsymbol{\Psi}^j_{jk} \mathbf{R}^j_{jk}\right) = p_{jk} \mathrm{tr}\left( \mathbf{R}^j_{jk} - \mathbf{C}^j_{jk}\right) $, which is the transmit power multiplied with the trace of the covariance matrix of the channel estimate in \eqref{eq:statistics-MMSE-estimate}.

In the denominator, the relation between the covariance matrices $\mathbf{R}^j_{jk}$ and $\mathbf{R}^j_{li}$, and  the inner product of LoS components $\bar{\mathbf{h}}^j_{jk} $ and $\bar{\mathbf{h}}^j_{li} $ determine how large the interference terms are. If the covariance matrices span different subspaces, or one has very small eigenvalues (e.g., due to weak large-scale fading), there will be little interference from the NLoS propagation. Similarly, there is little interference from the LoS propagation when $\bar{\mathbf{h}}^j_{jk} $ and $\bar{\mathbf{h}}^j_{li}$ are nearly orthogonal.

The non-coherent interference term ${\xi}^{\mathrm{ul}}_{li}$ does not increase with $M_j$, unless $\bar{\mathbf{h}}^j_{jk} $ and $ \bar{\mathbf{h}}^j_{li} $ are nearly parallel vectors. The coherent interference term ${\Gamma}^{\mathrm{ul}}_{li}$ involves the pilot-contaminating UEs, which are $(l,i) \in \mathcal{P}_{jk} \backslash (j,k)$, and it grows linearly with $M_j$. The term $\nu_{jk}^{\mathrm{ul}}$ grows with $M_j$ and  depends on the norm of desired UE's LoS component.

\setcounter{equation}{26}

\subsection{Uplink Spectral Efficiency with EW-MMSE Estimator}
If the EW-MMSE estimator in \eqref{estew1} is used, we obtain a closed-form expression for the SE in \eqref{eq:SEexpression} as in the next theorem.

\begin{theorem}\label{thEWMMSE}
	If MR  combining with $\mathbf{v}_{jk} = \hat{\mathbf{h}}^j_{jk}$ is used based on the EW-MMSE estimator, then 
	\begin{equation}\label{ewmmse2}
	\mathbb{E}\left\lbrace \mathbf{v}^H_{jk} \mathbf{h}^j_{jk}  \right\rbrace = p_{jk} \tau_p \mathrm{tr}\left(  \mathbf{D}^j_{jk}  \boldsymbol{\Lambda}^j_{jk} \mathbf{D}^j_{jk}\right) + \| \bar{\mathbf{h}}^j_{jk}   \|^2,
	\end{equation}
	\begin{equation}\label{ewmmse3}
	\mathbb{E}\left\lbrace  \| \mathbf{v}_{jk} \|^2 \right\rbrace  =  \mathrm{tr}\left(  \boldsymbol{\Sigma}^j_{jk} \right) +\| \bar{\mathbf{h}}^j_{jk}   \|^2,
	\end{equation}
	and $\mathbb{E}\lbrace | {\mathbf{v}}_{jk}^H \mathbf{h}^j_{li}  | ^2 \rbrace$ is given in \eqref{eq2EWMMSE}, at the top of next page. Plugging these expressions into the SINR in $\eqref{sec4eq1}$ gives $\gamma^{\mathrm{ul,ew}}_{jk}$ in \eqref{gammaew}.	
	 
	\begin{figure*}[!h]
		\normalsize
		\setcounter{equation}{28}
		\begin{align}\label{eq2EWMMSE}
	&\mathbb{E}\left\lbrace \left| {\mathbf{v}}_{jk}^H \mathbf{h}^j_{li}  \right| ^2 \right\rbrace = {\chi}^{\mathrm{ul}}_{li}= \mathrm{tr}\left(\mathbf{R}^j_{li} \boldsymbol{\Sigma}^j_{jk}\right) + (\bar{\mathbf{h}}^j_{jk} )^H\mathbf{R}^j_{li} \bar{\mathbf{h}}^j_{jk} 
	+ (\bar{\mathbf{h}}^j_{li} )^H \boldsymbol{\Sigma}^j_{jk} \bar{\mathbf{h}}^j_{li} +|(\bar{\mathbf{h}}^j_{jk} )^H \bar{\mathbf{h}}^j_{li}|^2 \nonumber \\  
	&+\begin{cases}
	{p_{jk}} {p_{li}}\tau^2_p \left( \mathrm{tr}\left(\mathbf{D}^j_{li}  \boldsymbol{\Lambda}^{j}_{jk} \mathbf{D}^j_{jk}   \right)\right) ^2  +   2 \sqrt{p_{jk}p_{li}} \tau_p\mathrm{tr}\left( \mathbf{D}^j_{li}  \boldsymbol{\Lambda}^{j}_{jk} \mathbf{D}^j_{jk}  \right)\mathrm{Re}\left\lbrace (\bar{\mathbf{h}}^j_{jk})^H \bar{\mathbf{h}}^j_{li}  \right\rbrace  &  (l,i) \in \mathcal{P}_{jk}   \\
	0 &  (l,i) \notin \mathcal{P}_{jk}.
	\end{cases}
	\end{align}
		\hrulefill
		\vspace*{2pt}
	\end{figure*}

 \begin{figure*}[!h]
 	\normalsize
		\begin{equation} \label{gammaew}
		\gamma^{\mathrm{ul,ew}}_{jk} =\frac{ p_{jk} \left(  p_{jk} \tau_p \mathrm{tr}\left(  \mathbf{D}^j_{jk}  \boldsymbol{\Lambda}^j_{jk} \mathbf{D}^j_{jk}\right) + \| \bar{\mathbf{h}}^j_{jk}   \|^2\right) ^2}{\displaystyle\sum_{l=1}^{L}\sum_{i=1}^{K_l}   p_{li} {\chi}^{\mathrm{ul}}_{li}  - p_{jk}\left(  p_{jk} \tau_p \mathrm{tr}\left(  \mathbf{D}^j_{jk}  \boldsymbol{\Lambda}^j_{jk} \mathbf{D}^j_{jk}\right) + \| \bar{\mathbf{h}}^j_{jk}   \|^2\right) ^2 + \sigma^2_{\mathrm{ul}} \left( \mathrm{tr}\left(  \boldsymbol{\Sigma}^j_{jk} \right) + \| \bar{\mathbf{h}}^j_{jk}   \|^2\right)  }.
		\end{equation}
		\hrulefill
		\vspace*{2pt}
	\end{figure*}	
\end{theorem}
\begin{IEEEproof}
	The proof is given in Appendix \ref{ProofthEWMMSE}.
\end{IEEEproof}

This SINR expression in Theorem~\ref{thEWMMSE} is more complicated than when using the MMSE estimator, but can be interpreted in an analogous way.

\subsection{Uplink Spectral Efficiency with LS Estimator}
If the LS estimator in \eqref{ls1} is used, we obtain a closed-form expression for the SE in \eqref{eq:SEexpression} as in the next theorem.

\begin{theorem}\label{thULLS}
	If MR combining with $\mathbf{v}_{jk} = \frac{1}{\sqrt{p_{jk}} \tau_p} \mathbf{y}^p_{jjk} $ is used based on the LS estimator, then 
	\begin{equation}\label{lseq1}
	\mathbb{E}\left\lbrace \mathbf{v}^H_{jk} \mathbf{h}^j_{jk}  \right\rbrace = \mathrm{tr}(\mathbf{R}^j_{jk} )+ \sum_{(l,i) \in \mathcal{P}_{jk} } \frac{\sqrt{p_{li}}}{\sqrt{p_{jk}}} (\bar{\mathbf{h}}^j_{li} )^H \bar{\mathbf{h}}^j_{jk} ,
	\end{equation}
	\begin{equation}\label{lseq2}
	\mathbb{E}\left\lbrace \|\mathbf{v}^H_{jk}\|^2 \right\rbrace =  \frac{1}{p_{jk} \tau_p} \mathrm{tr}\left( (\boldsymbol{\Psi}^j_{jk})^{-1} \right) + \frac{1}{p_{jk} \tau^2_p} \|\bar{\mathbf{y}}^p_{jjk}\|^2 ,
	\end{equation} 
and $\mathbb{E}\lbrace | {\mathbf{v}}_{jk}^H \mathbf{h}^j_{li}  | ^2 \rbrace$ is given in \eqref{lseq3}, at the top of next page where $\bar{\mathbf{x}}_{jk} = \bar{\mathbf{y}}^p_{jjk} - \sqrt{p_{li}} \tau_p\bar{\mathbf{h}}^j_{li}$ and $(\boldsymbol{\Omega}^j_{jk})^{-1} = (\boldsymbol{\Psi}^j_{jk})^{-1} -p_{li}\tau_p \mathbf{R}^j_{li}  $. Plugging these into $\eqref{sec4eq1}$ gives $\gamma^{\mathrm{ul,ls}}_{jk}$ in \eqref{LSeq1} where ${\chi}_{li}^{\mathrm{ul,ls}}$ is defined in \eqref{lseq3}.

		\begin{figure*}[!h]
		\normalsize
		\vspace*{-2pt}
		\setcounter{equation}{32}
			
		\begin{align}\label{lseq3}
		&{{p_{jk}} \tau^2_p}\mathbb{E}\left\lbrace |\mathbf{v}^H_{jk} \mathbf{h}^j_{li} |^2 \right\rbrace ={{p_{jk}} \tau^2_p}{\chi}_{li}^{\mathrm{ul,ls}}= \tau_p\mathrm{tr}\left( \mathbf{R}^j_{li}(\boldsymbol{\Psi}^j_{jk})^{-1} \right) + |(\bar{\mathbf{y}}^p_{jjk})^H \bar{\mathbf{h}}^j_{li}|^2   \nonumber  \\ 
		&+\begin{cases}
		(\bar{\mathbf{y}}^p_{jjk})^H \mathbf{R}^j_{li}\bar{\mathbf{y}}^p_{jjk} + \tau_p (\bar{\mathbf{h}}^j_{li})^H (\boldsymbol{\Psi}^j_{jk})^{-1}\bar{\mathbf{h}}^j_{li}   &  (l,i) \!\notin \mathcal{P}_{jk}\\ 
		p_{li} \tau^2_p  |  \mathrm{tr}(  \mathbf{R}^j_{li})| ^2 +  \bar{\mathbf{x}}^H_{jk}\mathbf{R}^j_{li}\bar{\mathbf{x}}_{jk}+ \tau_p (\bar{\mathbf{h}}^j_{li})^H (\boldsymbol{\Omega}^j_{jk})^{-1} \bar{\mathbf{h}}^j_{li}  \\ + 2\sqrt{p_{li}} \tau_p  \mathrm{Re}\left\lbrace (\bar{\mathbf{y}}^p_{jjk})^H \bar{\mathbf{h}}^j_{li} \mathrm{tr}\left(  \mathbf{R}^j_{li} \right) +  (\bar{\mathbf{y}}^p_{jjk})^H \mathbf{R}^j_{li} \bar{\mathbf{h}}^j_{li} +  \bar{\mathbf{x}}^H_{jk}   \bar{\mathbf{h}}^j_{li} \| \bar{\mathbf{h}}^j_{li} \|^2 \right\rbrace \!\!  &  (l,i) \!\in \mathcal{P}_{jk}.
		\end{cases}
		\end{align}
	\hrulefill
		
	\end{figure*}

\begin{figure*}[!h]
	\normalsize
	\vspace*{-2pt}
	\begin{equation}\label{LSeq1} \gamma^{\mathrm{ul,ls}}_{jk} \!=\!
\frac{p_{jk} \left| \mathrm{tr}(\mathbf{R}^j_{jk} )+ \sum_{(l,i) \in \mathcal{P}_{jk} } \frac{\sqrt{p_{li}}}{\sqrt{p_{jk}}} (\bar{\mathbf{h}}^j_{li} )^H \bar{\mathbf{h}}^j_{jk}\right|^2 }{\displaystyle\sum_{l=1}^{L}\sum_{i=1}^{K_l} {p_{li}{\chi}_{li}^{\mathrm{ul,ls}}}  -p_{jk}{\left| \mathrm{tr}(\mathbf{R}^j_{jk} )+ \sum_{(l,i) \in \mathcal{P}_{jk} } \frac{\sqrt{p_{li}}}{\sqrt{p_{jk}}} (\bar{\mathbf{h}}^j_{li} )^H \bar{\mathbf{h}}^j_{jk}\right|^2} +  \sigma^2_{\mathrm{ul}}\left( \frac{\mathrm{tr}\left( (\boldsymbol{\Psi}^j_{jk})^{-1} \right)}{p_{jk} \tau_p}  + \frac{\|\bar{\mathbf{y}}^p_{jjk}\|^2}{p_{jk} \tau^2_p}  \right) 	 },
\end{equation}

	\hrulefill
	
\end{figure*}	

\begin{figure*}[!h]
	\normalsize
	\vspace*{-4pt}
	\setcounter{equation}{42}
	\begin{align}\label{dlmmse1}
	&\!\!{{\mathbb{E}\left\lbrace  \| \hat{\mathbf{h}}^l_{li} \|^2 \right\rbrace}}\mathbb{E}\left\lbrace \left| \mathbf{w}_{li}^H \mathbf{h}^l_{jk}  \right|^2 \right\rbrace =p_{li} \tau_p \mathrm{tr}\left( \mathbf{R}^l_{jk} \mathbf{R}^l_{li} \boldsymbol{\Psi}^l_{li} \mathbf{R}^j_{li}\right) + p_{li} \tau_p \left( \bar{\mathbf{h}}^l_{jk}\right)^H   \mathbf{R}^l_{li} \boldsymbol{\Psi}^l_{li} \mathbf{R}^l_{li} \bar{\mathbf{h}}^l_{jk}    +  \left|  (\bar{\mathbf{h}}^l_{jk})^H    \bar{\mathbf{h}}^l_{li}   \right| ^2\\ 
	& \!\!+  \left( \bar{\mathbf{h}}^l_{li}\right)^H    \mathbf{R}^l_{jk} \bar{\mathbf{h}}^l_{li} +\begin{cases}
	p_{jk} p_{li}  \tau^2_p \left|  \mathrm{tr}\left( \mathbf{R}^l_{jk} \boldsymbol{\Psi}^l_{li} \mathbf{R}^l_{li}\right) \right| ^2 +  2 \sqrt{ p_{jk} p_{li}}\tau_p \mathrm{Re}\left\lbrace \mathrm{tr}\left(   \mathbf{R}^l_{jk} \boldsymbol{\Psi}^l_{li} \mathbf{R}^l_{li}  \right) \left( \bar{\mathbf{h}}^l_{li}\right)^H \bar{\mathbf{h}}^l_{jk}  \right\rbrace   &(l,i) \in \mathcal{P}_{jk}   \\
	0 & (l,i) \notin \mathcal{P}_{jk} \nonumber ,
	\end{cases}
	\end{align} 
	\hrulefill
	\vspace*{-2pt}
\end{figure*}

\end{theorem}
\begin{IEEEproof}
	The proof is given in Appendix \ref{ProofthULLS}.
\end{IEEEproof}
	\setcounter{equation}{34}
Note that the Rayleigh fading counterpart of \eqref{LSeq1} can be easily obtained by setting all the mean vectors to zero. In this case, the difference in SE between the MMSE and LS/EW-MMSE estimators can be rather small \cite{EmilsBook}. However, the loss in SE incurred by using the LS estimator under Rician fading can be quite large depending on the dominance of LoS paths. 

Since the mean values are not utilized as prior information, the interference terms are larger than when using the MMSE estimator. The LS estimates of the pilot-contaminating UEs are equal up to a scaling factor. Compared to the SE with MMSE estimator, the inner product of  $\bar{\mathbf{y}}^p_{jjk}$ and $\bar{\mathbf{h}}^j_{li}$ determines how large  the corresponding interference terms are instead of the inner product of $\bar{\mathbf{h}}^j_{jk}$ and $\bar{\mathbf{h}}^j_{li}$.

\subsection{Uplink Spectral Efficiency with Mean Only Estimator}

 If the BS only knows the mean values of the UEs' channels but not the covariance matrices, we can use this information as a channel estimate, without the need for sending pilots. The mean values vary much slower compared to the small-scale fading coefficients. Thus, using the MO estimator may greatly reduce the computational complexity since the channels are not estimated for each coherence block. However, it is  only meaningful when the UEs have dominant LoS paths since it ignores the NLoS paths.  If MR combining with $\mathbf{v}_{jk} = \bar{\mathbf{h}}^j_{jk} $ is used based on such a Mean Only (MO) estimator, then 
	\begin{equation}
	\mathbb{E}\left\lbrace \mathbf{v}^H_{jk} \mathbf{h}^j_{jk}  \right\rbrace = \mathbb{E}\left\lbrace \|\mathbf{v}^H_{jk}\|^2 \right\rbrace =  \|\bar{\mathbf{h}}^j_{jk}\|^2,
	\end{equation}	
	\begin{equation}
	\mathbb{E}\left\lbrace |\mathbf{v}^H_{jk} \mathbf{h}^j_{li} |^2 \right\rbrace = ( \bar{\mathbf{h}}^j_{jk})^H \mathbf{R}^j_{li} \bar{\mathbf{h}}^j_{jk} + |( \bar{\mathbf{h}}^j_{jk})^H\bar{\mathbf{h}}^j_{li}  |^2.
	\end{equation}
	Plugging these expressions into the SINR in $\eqref{sec4eq1}$ gives $\gamma^{\mathrm{ul,mo}}_{jk}$ as
	\begin{equation}\label{gammaulmo}\!\!	
\frac{p_{jk} \|\bar{\mathbf{h}}^j_{jk}\|^2}{\displaystyle\sum_{l=1}^{L}\sum_{i=1}^{K_l}  \frac{p_{li}}{\|\bar{\mathbf{h}}^j_{jk}\|^2} \left( ( \bar{\mathbf{h}}^j_{jk})^H \mathbf{R}^j_{li} \bar{\mathbf{h}}^j_{jk} + |( \bar{\mathbf{h}}^j_{jk})^H\bar{\mathbf{h}}^j_{li} |^2 \right)  - p_{jk} \|\bar{\mathbf{h}}^j_{jk}\|^2 + \sigma^2_{\mathrm{ul}}}.
	\end{equation}

Note that if UE $(j,k)$ does not have an LoS component then we have no information regarding its channel and the SINR in \eqref{gammaulmo} becomes zero. Thus, the MO estimator is only useful for UEs that have an LoS path.


\section{Downlink Spectral Efficiency with MR Precoding}\label{section5}

Each coherence block contains $\tau_d$ DL data transmissions, where the transmitted signal from BS $l$ is  
\begin{equation}
\mathbf{x}_l= \sum_{k=1}^{K_l} \mathbf{w}_{lk} \varsigma_{lk},
\end{equation}
where $\varsigma_{lk} \sim \mathcal{N}_\mathbb{C}\left( 0, \rho_{lk} \right)$ is the DL data signal intended for UE $k$ in the cell and $\rho_{lk}$ is the signal power. The transmit precoding vector $\mathbf{w}_{lk}$ determines the spatial directivity of the transmission. The precoding vector satisfies $\mathbb{E}\left\lbrace \|\mathbf{w}_{lk}\|^2 \right\rbrace =1 $, such that $\mathbb{E}\left\lbrace \|\mathbf{w}_{lk} \varsigma_{lk}\|^2 \right\rbrace = \rho_{lk} $ is the transmit power allocated to this UE. The received signal $y_{jk} \in \mathbb{C}$ at UE $k$ in cell $j$ is  
\begin{align}\label{sec1eq2}
&y_{jk}=( \mathbf{h}^l_{jk})^H \mathbf{w}_{jk} \varsigma_{jk} + \ \mathop{\sum_{i=1 }}^{K_{j}}_{i \neq k} ( \mathbf{h}^j_{jk})^H \mathbf{w}_{ji} \varsigma_{ji} \nonumber \\
&+ \ \mathop{\sum_{l=1 }}^{L}_{l \neq j} \sum_{i=1}^{K_l}( \mathbf{h}^l_{jk})^H \mathbf{w}_{li} \varsigma_{li} + n_{jk} ,
\end{align}
where $n_{jk} \sim \mathcal{N}_\mathbb{C}\left( 0, \sigma^2_{\mathrm{dl}}  \right)$ is i.i.d. additive receiver noise with variance $\sigma^2_{\mathrm{dl}}$. The first term in the above equation denotes the desired signal, the second term is the intra-cell interference, and the third term is the inter-cell interference.
The ergodic DL capacity of UE $k$ in cell $j$ is lower bounded by \cite[Th.~4.6]{EmilsBook}
\begin{equation}\label{dleq1}
\mathrm{SE}^{\mathrm{dl}}_{jk}=\frac{\tau_d}{\tau_c}\log_2\left(1 + \gamma^{\mathrm{dl}}_{jk} \right) \  \ \mathrm{[bit/ s/ Hz]}
\end{equation}
with 
\begin{equation}\label{dlSINR}
\gamma^{\mathrm{dl}}_{jk} =\frac{\rho_{jk} \left|  \mathbb{E}\left\lbrace \mathbf{w}^H_{jk} \mathbf{h}^j_{jk}  \right\rbrace   \right| ^2}{\displaystyle\sum_{l=1}^{L}\sum_{i=1}^{K_l}   \rho_{li}  \mathbb{E}\left\lbrace \left| \mathbf{w}^H_{li} \mathbf{h}^l_{jk}  \right| ^2 \right\rbrace  - \rho_{jk} \left|  \mathbb{E}\left\lbrace \mathbf{w}^H_{jk} \mathbf{h}^j_{jk}  \right\rbrace   \right| ^2  +\sigma^2_{\mathrm{dl}} } ,
\end{equation}
where the expectations are with respect to all sources of randomness. In the following subsections, the effective SINR $\gamma^{\mathrm{dl}}_{jk}$ is computed for MR precoding when using the different channel estimators.

\subsection{Downlink Spectral Efficiency with MMSE Estimator }
If the MMSE estimator in \eqref{mmse1} is used,  we obtain a closed-form expression for the DL SE in \eqref{dleq1} as in the next theorem.
\begin{theorem}
If MR precoding with $\mathbf{w}_{jk} = \frac{\hat{\mathbf{h}}^j_{jk}}{\sqrt{\mathbb{E}\left\lbrace  \| \hat{\mathbf{h}}^j_{jk} \|^2 \right\rbrace}}$ is used based on the MMSE estimator, then 
\begin{equation}\label{dlmmse2}
\mathbb{E}\left\lbrace \mathbf{w}^H_{jk} \mathbf{h}^j_{jk}  \right\rbrace = \sqrt{ p_{jk} \tau_p \mathrm{tr}\left( \mathbf{R}^j_{jk} \boldsymbol{\Psi}^j_{jk} \mathbf{R}^j_{jk}\right) + \| \bar{\mathbf{h}}^j_{jk}   \|^2 },
\end{equation}
and $\mathbb{E}\left\lbrace \left| \mathbf{w}_{li}^H \mathbf{h}^l_{jk}  \right|^2 \right\rbrace$ is given in \eqref{dlmmse1}, at the top of this page where ${{\mathbb{E}\left\lbrace  \| \hat{\mathbf{h}}^l_{li} \|^2 \right\rbrace}} =p_{li} \tau_p \mathrm{tr}\left( \mathbf{R}^l_{li} \boldsymbol{\Psi}^l_{li} \mathbf{R}^l_{li}\right) + \| \bar{\mathbf{h}}^l_{li}   \|^2$.  Inserting these into the DL SINR in \eqref{dlSINR} gives\setcounter{equation}{43}
\begin{equation}
\gamma^{\mathrm{dl,mmse}}_{jk}=\frac{ \rho_{jk} p_{jk} \tau_p \mathrm{tr}\left( \mathbf{R}^j_{jk} \boldsymbol{\Psi}^j_{jk} \mathbf{R}^j_{jk}\right) + \rho_{jk}\| \bar{\mathbf{h}}^j_{jk}   \|^2}{\displaystyle\sum_{l=1}^{L}\sum_{i=1}^{K_l}   \rho_{li} {\xi}^{\mathrm{dl}}_{li}  +  \displaystyle\sum_{(l,i) \in \mathcal{P}_{jk} \backslash (j,k)} \rho_{li} {\Gamma}^{\mathrm{dl}}_{li}  - \rho_{jk}\nu_{jk}^{\mathrm{dl}} + \sigma^2_{\mathrm{dl}} },
\end{equation}
where $\nu_{jk}^{\mathrm{dl}}={\| \bar{\mathbf{h}}^l_{jk}   \|^4}\big/\left( {p_{jk} \tau_p \mathrm{tr}\left( \mathbf{R}^l_{jk} \boldsymbol{\Psi}^l_{jk} \mathbf{R}^l_{jk}\right) + \| \bar{\mathbf{h}}^l_{jk}   \|^2}\right) $, ${\xi}^\mathrm{dl}_{li}$ and ${\Gamma}^\mathrm{dl}_{li}$ correspond to non-coherent interference in \eqref{dleq3}, coherent interference in \eqref{dleq4} respectively.
\begin{figure*}[!h]
	\normalsize
\vspace*{-2pt}
	\begin{equation}\label{dleq3}
	{\xi}^{\mathrm{dl}}_{li}= \frac{p_{li} \tau_p \mathrm{tr}\left( \mathbf{R}^l_{jk} \mathbf{R}^l_{li} \boldsymbol{\Psi}^l_{li} \mathbf{R}^l_{li}\right) + p_{li} \tau_p \left( \bar{\mathbf{h}}^l_{jk}\right)^H   \mathbf{R}^l_{li} \boldsymbol{\Psi}^l_{li} \mathbf{R}^l_{li} \bar{\mathbf{h}}^l_{jk} \\
		+  \left( \bar{\mathbf{h}}^l_{li}\right)^H    \mathbf{R}^l_{jk} \bar{\mathbf{h}}^l_{li} +  \left|  (\bar{\mathbf{h}}^l_{jk})^H    \bar{\mathbf{h}}^l_{li}   \right| ^2 }{ p_{li} \tau_p \mathrm{tr}\left( \mathbf{R}^l_{li} \boldsymbol{\Psi}^l_{li} \mathbf{R}^l_{li}\right) +|| \bar{\mathbf{h}}^l_{li}   ||^2} ,
	\end{equation}
	\begin{equation}\label{dleq4}
	{\Gamma}^{\mathrm{dl}}_{li}=    \frac{p_{jk} p_{li}  \tau^2_p \left| \mathrm{tr}\left( \mathbf{R}^l_{jk} \boldsymbol{\Psi}^l_{li} \mathbf{R}^l_{li}\right) \right| ^2 +  2 \sqrt{ p_{jk} p_{li}}\tau_p\mathrm{Re}\left\lbrace \mathrm{tr}\left(   \mathbf{R}^l_{jk} \boldsymbol{\Psi}^l_{li} \mathbf{R}^l_{li}  \right)  \left( \bar{\mathbf{h}}^l_{li}\right)^H \bar{\mathbf{h}}^l_{jk}  \right\rbrace }{ p_{li} \tau_p \mathrm{tr}\left( \mathbf{R}^l_{li} \boldsymbol{\Psi}^l_{li} \mathbf{R}^l_{li}\right) +|| \bar{\mathbf{h}}^l_{li}   ||^2} .
	\end{equation}
	\hrulefill	
	\vspace*{-2pt}
\end{figure*}

\end{theorem}
\begin{IEEEproof}
	The proof is similar to the uplink case in Appendix \ref{ProofthULMMSE} and is omitted.
\end{IEEEproof}	

This SINR expression resembles the UL counterpart in Theorem~\ref{thULMMSE} to a large extent. The transmit power and noise variance are denoted differently and some of the indices are switched in the interference terms, as expected from the UL-DL duality \cite[Sec.~4.3]{EmilsBook}. Hence, the interpretation of the SINR is qualitatively the same as in the uplink.

\begin{figure*}
	\normalsize
	\vspace*{-4pt}
\setcounter{equation}{47}
	\begin{align}\label{dlewmmse2}
	&{{\mathbb{E}\left\lbrace  \| \hat{\mathbf{h}}^l_{li} \|^2 \right\rbrace}}\mathbb{E}\left\lbrace \left| \mathbf{w}_{li}^H \mathbf{h}^l_{jk}  \right| ^2 \right\rbrace = {\chi}^{\mathrm{dl}}_{li}= \mathrm{tr}\left(\mathbf{R}^l_{jk} \boldsymbol{\Sigma}^l_{li}\right) + (\bar{\mathbf{h}}^l_{li} )^H\mathbf{R}^l_{jk} \bar{\mathbf{h}}^l_{li} 
	+ (\bar{\mathbf{h}}^l_{jk} )^H \boldsymbol{\Sigma}^l_{li} \bar{\mathbf{h}}^l_{jk} +|(\bar{\mathbf{h}}^l_{li} )^H \bar{\mathbf{h}}^l_{jk}|^2 \nonumber \\  
	&+\begin{cases}
	{p_{jk}} {p_{li}}\tau^2_p\left(  \mathrm{tr}\left(\mathbf{D}^l_{jk}  \boldsymbol{\Lambda}^{l}_{li} \mathbf{D}^l_{li}   \right) \right)^2  +   2 \sqrt{p_{jk}p_{li}} \tau_p\mathrm{tr}\left( \mathbf{D}^l_{jk}  \boldsymbol{\Lambda}^{l}_{li} \mathbf{D}^l_{li}  \right)\mathrm{Re}\left\lbrace (\bar{\mathbf{h}}^l_{li})^H \bar{\mathbf{h}}^l_{jk}  \right\rbrace  &  (l,i) \in \mathcal{P}_{jk}   \\
	0 &  (l,i) \notin \mathcal{P}_{jk}.
	\end{cases}
	\end{align}
	\hrulefill
\vspace*{-2pt}
\end{figure*}

\begin{figure*}[!h]
	\normalsize
	\vspace*{-2pt}
	\setcounter{equation}{48}
	\begin{equation} \label{dlgammaew}
	\gamma^{\mathrm{dl,ew}}_{jk} =\frac{ \rho_{jk} \left(   p_{jk} \tau_p \mathrm{tr}\left(  \mathbf{D}^j_{jk}  \boldsymbol{\Lambda}^j_{jk} \mathbf{D}^j_{jk}\right) + \| \bar{\mathbf{h}}^j_{jk}   \right) ^2\Big/ \left( \mathrm{tr}\left(  \boldsymbol{\Sigma}^j_{jk} \right) + \| \bar{\mathbf{h}}^j_{jk}   \|^2\right) }{\displaystyle\sum_{l=1}^{L}\sum_{i=1}^{K_l}   \rho_{li} \frac{{\chi}^{\mathrm{dl}}_{li}}{\mathrm{tr}\left(  \boldsymbol{\Sigma}^l_{li} \right) + \| \bar{\mathbf{h}}^l_{li}   \|^2}  - \rho_{jk}\frac{\left(   p_{jk} \tau_p \mathrm{tr}\left(  \mathbf{D}^j_{jk}  \boldsymbol{\Lambda}^j_{jk} \mathbf{D}^j_{jk}\right) + \| \bar{\mathbf{h}}^j_{jk}   \|^2\right) ^2 }{\mathrm{tr}\left(  \boldsymbol{\Sigma}^j_{jk} \right) + \| \bar{\mathbf{h}}^j_{jk}   \|^2}+ \sigma^2_{\mathrm{dl}}  }.
	\end{equation}	
	\hrulefill
\end{figure*}	


\setcounter{equation}{46}
\subsection{Downlink Spectral Efficiency with EW-MMSE Estimator}
If the EW-MMSE estimator in \eqref{estew1} is used,  we obtain a closed-form expression for the DL SE in \eqref{dleq1} as in the next theorem.

\begin{theorem}
If MR precoding with $\mathbf{w}_{jk} = \frac{\hat{\mathbf{h}}^j_{jk}}{\sqrt{\mathbb{E}\left\lbrace  \| \hat{\mathbf{h}}^j_{jk} \|^2 \right\rbrace}}$ is used based on the EW-MMSE estimator, then 
		\begin{equation}\label{dlewmmse1}
			\mathbb{E}\left\lbrace \mathbf{w}^H_{jk} \mathbf{h}^j_{jk}  \right\rbrace=\frac{ p_{jk} \tau_p \mathrm{tr}\left(  \mathbf{D}^j_{jk}  \boldsymbol{\Lambda}^j_{jk} \mathbf{D}^j_{jk}\right) + \| \bar{\mathbf{h}}^j_{jk}   \|^2 }{\sqrt{ \mathrm{tr}\left(  \boldsymbol{\Sigma}^j_{jk} \right) + \| \bar{\mathbf{h}}^j_{jk}   \|^2}} ,
		\end{equation}
		and $\mathbb{E}\left\lbrace \left| \mathbf{w}_{li}^H \mathbf{h}^l_{jk}  \right|^2 \right\rbrace$ is given in \eqref{dlewmmse2}, at the top of next page where ${{\mathbb{E}\left\lbrace  \| \hat{\mathbf{h}}^l_{li} \|^2 \right\rbrace}}=\mathrm{tr}\left(  \boldsymbol{\Sigma}^l_{li} \right) + \| \bar{\mathbf{h}}^l_{li}   \|^2 $. Inserting these expressions into the DL SINR in $\eqref{dlSINR}$ gives $	\gamma^{\mathrm{dl,ew}}_{jk}$ in \eqref{dlgammaew}.

\end{theorem}
\begin{IEEEproof}
	The proof is similar to the uplink case in Appendix \ref{ProofthEWMMSE} and is omitted.
\end{IEEEproof}	

\subsection{Downlink Spectral Efficiency with LS Estimator}
\setcounter{equation}{49}
If the LS estimator in \eqref{ls1} is used,  we obtain a closed-form expression for the DL SE in \eqref{dleq1} as in the next theorem.
\begin{theorem}
	If MR precoding with $\mathbf{w}_{jk} = \frac{\hat{\mathbf{h}}^j_{jk}}{\sqrt{\mathbb{E}\left\lbrace  \| \hat{\mathbf{h}}^j_{jk} \|^2 \right\rbrace}}$ is used based on the LS estimator, then
	\begin{equation}\label{lsdleq1}
\!\!\!\!\!\!\!\!\mathbb{E}\left\lbrace \mathbf{w}^H_{jk} \mathbf{h}^j_{jk}  \right\rbrace=	\frac{\mathbb{E}\left\lbrace \mathbf{v}^H_{jk} \mathbf{h}^j_{jk}  \right\rbrace}{\sqrt{\mathbb{E}\left\lbrace \|\mathbf{v}_{jk}\|^2 \right\rbrace}} =\frac{\mathrm{tr}(\mathbf{R}^j_{jk} )+ \sum_{(l,i) \in \mathcal{P}_{jk} } \frac{\sqrt{p_{li}}}{\sqrt{p_{jk}}} (\bar{\mathbf{h}}^j_{li} )^H \bar{\mathbf{h}}^j_{jk} }{\sqrt{\frac{1}{p_{jk} \tau^2_p}  \left( \tau_p \mathrm{tr}\left( (\boldsymbol{\Psi}^j_{jk})^{-1} \right) + \|\bar{\mathbf{y}}^p_{jjk}\|^2\right)}  },
	\end{equation}
and $\mathbb{E}\left\lbrace \left| \mathbf{w}_{li}^H \mathbf{h}^l_{jk}  \right|^2 \right\rbrace$ is given in \eqref{lsdleq2}, at the top of next page where ${{\mathbb{E}\left\lbrace  \| \hat{\mathbf{h}}^l_{li} \|^2 \right\rbrace}}=  \frac{1}{p_{li} \tau^2_p}  \left( \tau_p \mathrm{tr}\left( (\boldsymbol{\Psi}^l_{li})^{-1} \right) + \|\bar{\mathbf{y}}^p_{lli}\|^2\right)$.  Inserting these expressions into the DL SINR in \eqref{dlSINR} gives $\gamma^{\mathrm{dl,ls}}_{jk}$ in \eqref{lsdlgamma}
where the interference term ${\chi}_{li}^{\mathrm{dl,ls}}$ is defined  in \eqref{lsdleq2}.
\end{theorem}
\begin{IEEEproof}
	The proof is similar to the uplink case in Appendix \ref{ProofthULLS} and is omitted.
\end{IEEEproof}

\setcounter{equation}{52}
\subsection{Downlink Spectral Efficiency with Mean Only Estimator}
If MR precoding with $\mathbf{w}_{jk} = \frac{\bar{\mathbf{h}}^j_{jk}}{\left\| \bar{\mathbf{h}}^j_{jk}\right\|}  $ is used based on the MO estimator, then 
\begin{equation}\label{dlmo1}
\mathbb{E}\left\lbrace \mathbf{w}^H_{jk} \mathbf{h}^j_{jk}  \right\rbrace = \|\bar{\mathbf{h}}^j_{jk}\|,
\end{equation}
\begin{equation}\label{dlmo2}
\mathbb{E}\left\lbrace |\mathbf{w}^H_{li} \mathbf{h}^l_{jk} |^2 \right\rbrace = \left( {\left( \bar{\mathbf{h}}^l_{li}\right)^H \mathbf{R}^l_{jk} \bar{\mathbf{h}}^l_{li} + \left|\left( \bar{\mathbf{h}}^l_{li}\right)^H\bar{\mathbf{h}}^l_{jk}  \right|^2}\right) \bigg/ {\|\bar{\mathbf{h}}^l_{li}\|^2 }.
\end{equation}
Inserting these expressions into the DL SINR in \eqref{dlSINR} gives  $\gamma^{\mathrm{dl,mo}}_{jk}$ as
\begin{equation}
\frac{\rho_{jk} \|\bar{\mathbf{h}}^j_{jk}\|^2 }{\displaystyle\sum_{l=1}^{L}\sum_{i=1}^{K_l} \frac{\rho_{li}}{\|\bar{\mathbf{h}}^l_{li}\|^2} \left( ( \bar{\mathbf{h}}^l_{li})^H \mathbf{R}^l_{jk} \bar{\mathbf{h}}^l_{li} + |( \bar{\mathbf{h}}^l_{li})^H\bar{\mathbf{h}}^l_{jk}|^2 \right)  - p_{jk} \|\bar{\mathbf{h}}^j_{jk}\|^2 + \sigma^2_\mathrm{dl}}.
\end{equation}

\begin{figure*}[!h]
	\normalsize

	\setcounter{equation}{50}
	\begin{align}\label{lsdleq2}
	&{{\mathbb{E}\left\lbrace  \| \hat{\mathbf{h}}^l_{li} \|^2 \right\rbrace}} \mathbb{E}\left\lbrace |\mathbf{w}^H_{li} \mathbf{h}^l_{jk} |^2 \right\rbrace  = {{p_{li}} \tau^2_p}{\chi}_{li}^{\mathrm{dl,ls}} =\tau_p\mathrm{tr}\left( \mathbf{R}^l_{jk}(\boldsymbol{\Psi}^l_{li})^{-1} \right) + |(\bar{\mathbf{y}}^p_{jli})^H \bar{\mathbf{h}}^l_{jk}|^2  \nonumber   \\
	&+\begin{cases}
	(\bar{\mathbf{y}}^p_{jli})^H \mathbf{R}^l_{jk}\bar{\mathbf{y}}^p_{jli} + \tau_p (\bar{\mathbf{h}}^l_{jk})^H (\boldsymbol{\Psi}^l_{li})^{-1}\bar{\mathbf{h}}^l_{jk}   &  (l,i) \notin \mathcal{P}_{jk}  \\ 
	p_{jk} \tau^2_p  \left|  \mathrm{tr}(  \mathbf{R}^l_{jk})\right| ^2 +  \bar{\mathbf{x}}^H_{li}\mathbf{R}^l_{jk}\bar{\mathbf{x}}_{li}+ \tau_p (\bar{\mathbf{h}}^l_{jk})^H (\boldsymbol{\Omega}^l_{li})^{-1} \bar{\mathbf{h}}^l_{jk}  \\
	+ 2\sqrt{p_{jk}} \tau_p  \mathrm{Re}\left\lbrace    (\bar{\mathbf{y}}^p_{jli})^H \bar{\mathbf{h}}^l_{jk} \mathrm{tr}\left(  \mathbf{R}^l_{jk} \right)
	+  (\bar{\mathbf{y}}^p_{jli})^H \mathbf{R}^l_{jk} \bar{\mathbf{h}}^l_{jk} + \bar{\mathbf{x}}^H_{li}   \bar{\mathbf{h}}^l_{jk} \|\bar{\mathbf{h}}^l_{jk} \|^2 \right\rbrace   &  (l,i) \in \mathcal{P}_{jk}.
	\end{cases}
	\end{align}
	\hrulefill
	\vspace*{-2pt}
\end{figure*}

\begin{figure*}[!h]
	\normalsize

\begin{equation}\label{lsdlgamma}
\gamma^{\mathrm{dl,ls}}_{jk}=\frac{\rho_{jk}p_{jk} \tau^2_p \left| \mathrm{tr}(\mathbf{R}^j_{jk} )+ \sum_{(l,i) \in \mathcal{P}_{jk} } \frac{\sqrt{p_{li}}}{\sqrt{p_{jk}}} (\bar{\mathbf{h}}^j_{li} )^H \bar{\mathbf{h}}^j_{jk}\right| ^2 \bigg/   \left( \tau_p \mathrm{tr}\left( (\boldsymbol{\Psi}^j_{jk})^{-1} \right) + \|\bar{\mathbf{y}}^p_{jjk}\|^2\right)}{ \displaystyle\sum_{l=1}^{L}\sum_{i=1}^{K_l} \rho_{li}\frac{p_{li} \tau^2_p{\chi}_{li}^{\mathrm{dl,ls}}}{    \tau_p \mathrm{tr}\left( (\boldsymbol{\Psi}^l_{li})^{-1} \right) + \|\bar{\mathbf{y}}^p_{lli}\|^2}  -  \rho_{jk} \frac{p_{jk} \tau^2_p\left| \mathrm{tr}(\mathbf{R}^j_{jk} )+ \sum_{(l,i) \in \mathcal{P}_{jk} } \frac{\sqrt{p_{li}}}{\sqrt{p_{jk}}} (\bar{\mathbf{h}}^j_{li} )^H \bar{\mathbf{h}}^j_{jk}\right| ^2}{ \tau_p \mathrm{tr}\left( (\boldsymbol{\Psi}^j_{jk})^{-1} \right) + \|\bar{\mathbf{y}}^p_{jjk}\|^2} +  \sigma^2_{\mathrm{dl} }},
\end{equation}
	\hrulefill
	\vspace*{-2pt}
\end{figure*}

\setcounter{equation}{55}

\section{Asymptotic  Analysis}
In this section, we will analyze the asymptotic behavior of Rician fading channels when using MR based on the different channel estimators. We make the following technical assumptions:

\begin{assumption}
	For $l,j=1, \dots,L$ and $i=1, \dots,K_l$,  the spatial covariance matrix $\mathbf{R}^j_{li}$ satisfies $\limsup\limits_{M_j} \|\mathbf{R}^j_{li}\|_2 < \infty $ and $\liminf\limits_{M_j}  \frac{1}{M_j}\mathrm{tr}\left( \mathbf{R}^j_{li}\right) > 0 $. 
\end{assumption}

\begin{assumption}
 	For $l,j=1, \dots,L$ and $i=1, \dots,K_l$,  the LoS component $\bar{\mathbf{h}}^j_{li}$ satisfies $\limsup\limits_{M_j} \frac{1}{M_j}\|\bar{\mathbf{h}}^j_{li}\|^2 <\infty $. 
\end{assumption}

\begin{assumption}
	For $l,j=1, \dots,L$, $i=1, \dots,K_l$ and  $k=1, \dots,K_j$, the LoS components $\bar{\mathbf{h}}^j_{jk}$ and $\bar{\mathbf{h}}^j_{li}$ satisfy $\lim\limits_{M_j} \frac{1}{M_j} \left| (\bar{\mathbf{h}}^j_{jk})^H    \bar{\mathbf{h}}^j_{li}  \right| \rightarrow 0 $ if $(j,k) \neq (l,i)$.
\end{assumption}

The first assumption is standard in the asymptotic analysis for Massive MIMO \cite{hoydis2013massive} and implies that array gathers an amount of signal energy that is proportional to the number of antennas and this energy originates from many spatial directions. The other assumptions are discussed in Section~\ref{subsec:discussion-asymptotics}. We  recall the definition of spatially orthogonal matrices from \cite{EmilsBook}.

\begin{definition}
	For  $l=1, \dots,L$ and $i=1, \dots,K_l$, two spatial covariance matrices $\mathbf{R}^j_{li}$ and $\mathbf{R}^j_{jk}$ 

	are asymptotically spatially orthogonal if
	\begin{equation}
	\frac{1}{M_j} \mathrm{tr}\left(\mathbf{R}^j_{li} \mathbf{R}^j_{jk} \right) \rightarrow 0 \ \ \mathrm{as} \ \  M_j \rightarrow \infty.
	\end{equation}
\end{definition}

Definition 1 implies that both covariance matrices are strongly rank-deficient. In spatially correlated fading, it holds if the BS is equipped with a ULA and the channels from two UEs have non-overlapping supports of their multipath angular distributions \cite[Theorem 1]{Yin2013a}. However, such angular separations are unlikely to occur in practice, at least in sub-6 GHz cellular networks.

\subsection{Asymptotic Analysis of Spectral Efficiency with MMSE Estimator}

\begin{theorem}\label{thAsMMSE}
	Under Assumptions 1--3, it follows that $\gamma^{\mathrm{ul,mmse}}_{jk}$ grows without bound as $M_j \rightarrow \infty$ if $\mathbf{R}^j_{jk}$ is asymptotically spatially orthogonal to $\mathbf{R}^j_{li}$  for all $(l,i) \in \mathcal{P}_{jk} \backslash (j,k)$. If this is not the case, then under Assumption 1--3, as $M_j \rightarrow \infty$, it follows that 
		\begin{equation}\label{astheorem1}
		\gamma^{\mathrm{ul,mmse}}_{jk} -\frac{{p^2_{jk} \tau_p}\mathrm{tr}( \mathbf{R}^j_{jk} \boldsymbol{\Psi}^j_{jk} \mathbf{R}^j_{jk}) + {p_{jk} }\| \bar{\mathbf{h}}^j_{jk}   \|^2}{ \displaystyle \sum_{(l,i) \in \mathcal{P}_{jk} \backslash (j,k)}   p^2_{li} \frac{p_{jk}  \tau^2_p \left|  \mathrm{tr}\left( \mathbf{R}^j_{li} \boldsymbol{\Psi}^j_{jk} \mathbf{R}^j_{jk}\right) \right| ^2 }{p_{jk} \tau_p \mathrm{tr}\left( \mathbf{R}^j_{jk} \boldsymbol{\Psi}^j_{jk} \mathbf{R}^j_{jk}\right) + \| \bar{\mathbf{h}}^j_{jk}   \|^2}  } \rightarrow 0.
		\end{equation}
\end{theorem}
\begin{IEEEproof}
The proof is given in Appendix  \ref{ProofthAsMMSE}.
\end{IEEEproof}

\begin{theorem}\label{thAsDLMMSE}
	Under Assumptions 1--3, it follows that $\gamma^{\mathrm{dl,mmse}}_{jk}$ grows without bound as $M_1=\dots =M_L \rightarrow \infty$ if $\mathbf{R}^l_{jk}$ is asymptotically spatially orthogonal to $\mathbf{R}^l_{li}$  for all $(l,i) \in \mathcal{P}_{jk} \backslash (j,k)$. If this is not the case, then under Assumption 1--3, as $M_1=\dots =M_L \rightarrow \infty$, it follows that 	
		\begin{equation}\label{asDLtheorem1}
		 \gamma^{\mathrm{dl,mmse}}_{jk} -\frac{{\rho_{jk}p_{jk} \tau_p}\mathrm{tr}\left( \mathbf{R}^j_{jk} \boldsymbol{\Psi}^j_{jk} \mathbf{R}^j_{jk}\right) + {\rho_{jk} }\| \bar{\mathbf{h}}^j_{jk}   \|^2}{ \displaystyle \sum_{(l,i) \in \mathcal{P}_{jk} \backslash (j,k)}  \rho_{li}\frac{ p_{jk} p_{li}  \tau^2_p\left|  \mathrm{tr}\left( \mathbf{R}^l_{jk} \boldsymbol{\Psi}^l_{li} \mathbf{R}^l_{li}\right) \right| ^2 }{p_{li} \tau_p \mathrm{tr}\left( \mathbf{R}^l_{li} \boldsymbol{\Psi}^l_{li} \mathbf{R}^l_{li}\right) +\| \bar{\mathbf{h}}^l_{li}   \|^2}  } \rightarrow 0.
		\end{equation}
\end{theorem}
\begin{IEEEproof}
	The proof is given in Appendix  \ref{ProofthAsMMSE}.
\end{IEEEproof}

These theorems show that the SINRs are generally upper bounded by simplified asymptotic SINR expressions, which depend on the covariance matrices and mean values. It is only in the special case of asymptotically spatially orthogonal matrices that the SE grows without limit  with rate $\log_2(M)$, which is consistent with the results for correlated Rayleigh fading in \cite{Yin2013a,EmilsBook}.

\subsection{Asymptotic Analysis of Spectral Efficiency with EW-MMSE Estimator}

\begin{theorem}\label{thAsEWMMSE}
	Under Assumptions 1--3, it follows that $\gamma^{\mathrm{ul,ew}}_{jk}$ grows without bound as $M_j \rightarrow \infty$ if $\mathbf{D}^j_{jk}$ is asymptotically spatially orthogonal to $\mathbf{D}^j_{li}$  for all $(l,i) \in \mathcal{P}_{jk} \backslash (j,k)$. If this is not the case, then under Assumption 1--3, as $M_j \rightarrow \infty$, it follows that
		\begin{equation}\label{asewmmse1}
		 \gamma^{\mathrm{ul,ew}}_{jk} -\frac{ p^2_{jk} \tau_p \mathrm{tr}\left(  \mathbf{D}^j_{jk}  \boldsymbol{\Lambda}^j_{jk} \mathbf{D}^j_{jk}\right) + p_{jk} \| \bar{\mathbf{h}}^j_{jk}   \|^2}{  \displaystyle \sum_{(l,i) \in \mathcal{P}_{jk} \backslash (j,k)} p_{li} \frac{{p_{jk}} {p_{li}}\tau^2_p\left(  \mathrm{tr}\left(\mathbf{D}^j_{li}  \boldsymbol{\Lambda}^{j}_{jk} \mathbf{D}^j_{jk}   \right) \right) ^2 }{ p_{jk} \tau_p \mathrm{tr}\left(  \mathbf{D}^j_{jk}  \boldsymbol{\Lambda}^j_{jk} \mathbf{D}^j_{jk}\right) + \| \bar{\mathbf{h}}^j_{jk}   \|^2}} \rightarrow 0.
		\end{equation}
\end{theorem}
\begin{IEEEproof}
	The proof is given in Appendix  \ref{ProofthAsEWMMSE}.
\end{IEEEproof}

\begin{theorem}\label{thAsDLEWMMSE}
	Under Assumptions 1--3, it follows that $\gamma^{\mathrm{dl,ew}}_{jk}$ grows without bound as $M_1=\dots =M_L \rightarrow \infty$ if $\mathbf{D}^l_{li}$ is asymptotically spatially orthogonal to $\mathbf{D}^l_{jk}$  for all $(l,i) \in \mathcal{P}_{jk} \backslash (j,k)$. If this is not the case, then under Assumption 1--3, as $M_1=\dots =M_L \rightarrow \infty$, it follows	that
		\begin{equation}\label{asewmmse2}
		 \gamma^{\mathrm{dl,ew}}_{jk}- \frac{ \frac{\rho_{jk} \left(  p_{jk} \tau_p \mathrm{tr}\left(  \mathbf{D}^j_{jk}  \boldsymbol{\Lambda}^j_{jk} \mathbf{D}^j_{jk}\right) + \| \bar{\mathbf{h}}^j_{jk}   \|^2 \right) ^2 }{ \mathrm{tr}\left(  \boldsymbol{\Sigma}^j_{jk} \right) + \| \bar{\mathbf{h}}^j_{jk}   \|^2}  }{ \displaystyle \sum_{(l,i) \in \mathcal{P}_{jk} \backslash (j,k)} \rho_{li} \frac{{p_{jk}} {p_{li}}\tau^2_p\left(  \mathrm{tr}\left(\mathbf{D}^l_{jk}  \boldsymbol{\Lambda}^{l}_{li} \mathbf{D}^l_{li}   \right) \right) ^2  }{ \mathrm{tr}\left(  \boldsymbol{\Sigma}^l_{li} \right) + \| \bar{\mathbf{h}}^l_{li}   \|^2}} \rightarrow 0.
		\end{equation}
\end{theorem}
\begin{IEEEproof}
	The proof is given in Appendix  \ref{ProofthAsEWMMSE}.
\end{IEEEproof}

The implications from these theorems are similar to the MMSE estimation case, except that it is the diagonals of the covariance matrices that need to be asymptotically spatially orthogonal to achieve an asymptotically unbounded SE. This is a more restrictive condition.

\subsection{Asymptotic Analysis of Spectral Efficiency with LS Estimator}
\begin{theorem}\label{thAsLS}
	Under Assumptions 1--3, as $M_j \rightarrow \infty$, it follows that  

	\begin{equation}\label{aslseq1}
 \gamma^{\mathrm{ul,ls}}_{jk} - \frac{ p^2_{jk}\tau^2_p  \left( \mathrm{tr}(  \mathbf{R}^j_{jk}) +  \|\bar{\mathbf{h}}^j_{jk} \|^2\right) }{  \sum_{(l,i) \in \mathcal{P}_{jk} \backslash (j,k) } \frac{    {p^2_{li}  \tau^2_p \left(   \mathrm{tr}(  \mathbf{R}^j_{li})   +   \|\bar{\mathbf{h}}^j_{li} \|^2\right)^2  }
  }{\mathrm{tr}(\mathbf{R}^j_{jk} ) + \sum_{(l,i) \in \mathcal{P}_{jk} } \frac{\sqrt{p_{jk}}}{\sqrt{p_{li}} } \left| (\bar{\mathbf{h}}^j_{li} )^H \bar{\mathbf{h}}^j_{jk} \right|} } \rightarrow 0.
	\end{equation}

\end{theorem}
\begin{IEEEproof}
	The proof is given in Appendix  \ref{proofthAsLS}.
\end{IEEEproof}

\begin{theorem}\label{thAsDLLS}
	Under Assumptions 1--3, as $M_1=\dots =M_L \rightarrow \infty$, it follows that 
		\begin{equation}\label{asdleq1}
		 \gamma^{\mathrm{dl,ls}}_{jk}- \frac{ \frac{{\rho_{jk} p_{jk}  \tau^2_p\left(  \mathrm{tr}(  \mathbf{R}^j_{jk}) + \|\bar{\mathbf{h}}^j_{jk} \|^2\right)^2}}{ {  \tau_p \mathrm{tr}\left( (\boldsymbol{\Psi}^j_{jk})^{-1} \right) + \|\bar{\mathbf{y}}^p_{jjk}\|^2}}    }{   \sum_{(l,i) \in \mathcal{P}_{jk} \backslash (j,k) }\frac{ \rho_{li} p_{jk}  \tau^2_p  \left(  \mathrm{tr}(  \mathbf{R}^l_{jk}) +  \|\bar{\mathbf{h}}^l_{jk} \|^2\right)^2 }{   \tau_p \mathrm{tr}\left( (\boldsymbol{\Psi}^l_{li})^{-1} \right) + \|\bar{\mathbf{y}}^p_{lli}\|^2  } 
		} \rightarrow 0.
		\end{equation}
		
\end{theorem}
\begin{IEEEproof}
	The proof is given in Appendix  \ref{proofthAsLS}.
\end{IEEEproof}

In contrast to the case with the MMSE and EW-MMSE estimators, we notice that the SE does not grow without bound when using the LS estimator.

\subsection{Asymptotic Analysis of Spectral Efficiency with Mean Only Estimator}
Under Assumptions 1-3, if UE $(j,k)$ has an LoS path, it follows that $\gamma^{\mathrm{ul,mo}}_{jk}$ grows without bound as $M_j \rightarrow \infty$. In the DL, under Assumptions 1-3, if  UE $(j,k)$ has an LoS path, it follows that $\gamma^{\mathrm{ul,mo}}_{jk}$ grows without bound as $M_1,\dots, M_L \rightarrow \infty$. 

Proof: In the UL, the LoS related term  $ \frac{p_{li}  \left| \left( \bar{\mathbf{h}}^j_{jk}\right)^H\bar{\mathbf{h}}^j_{li} \right| ^2 }{\|\bar{\mathbf{h}}^j_{jk}\|^2}  $ in (37) goes to zero as $M_j \rightarrow \infty$ for $(l,i)  \neq (j,k)$ due to Assumption 2 and 3. Noting the $	\frac{\left( \bar{\mathbf{h}}^j_{jk}\right)^H  {\mathbf{R}^j_{li} }\bar{\mathbf{h}}^j_{jk} }{\|\bar{\mathbf{h}}^j_{jk} \|^2} \leq \|\mathbf{R}^j_{li} \|_2  $ is finite as $M_j \rightarrow \infty$ due to Assumption 1 completes the proof.  In the DL, the expressions for $\gamma^{\mathrm{dl,mo}}_{jk}$ contains the same matrix expressions as $\gamma^{\mathrm{ul,mo}}_{jk}$, except that indices $(l,i)$ and $(j,k)$ are swapped in the interference terms.

Note that we assume that the mean vectors are perfectly known, however small phase-shifts may effect the performance severely as discussed in Section III. 

\subsection{Discussion on the Asymptotic Behaviors of Rician Fading Channels}
\label{subsec:discussion-asymptotics}

To describe the intuition behind Assumptions 2-3 and Definition 1, we consider a uniform linear array (ULA) with omni-directional antennas, where the  LoS component is modeled as \cite[Sec.~1.3]{EmilsBook}
\begin{equation}\label{LoS}
\bar{\mathbf{h}}^j_{li} = \sqrt{\beta^{j,\mathrm{LoS}}_{li}} \left[ 1 \, e^{j2\pi d_H \sin(\varphi^j_{li})} \, \dots \, e^{j2\pi d_H (M-1)\sin(\varphi^j_{li})}  \right] ^T,
\end{equation}
where $\beta^{j,\mathrm{LoS}}_{li}$ is the large-scale fading coefficient,
$d_H \leq 0.5$ is the antenna spacing parameter (in fractions of the wavelength), and ${\varphi}^j_{li} $ is the angle of arrival (AoA) to the UE seen from the BS. 
Utilizing this model, we have $\frac{1}{M_j}\|\bar{\mathbf{h}}^j_{li}\|^2 = \beta^{j,\mathrm{LoS}}_{li}$, which is a finite value for any $M_j$. Hence, Assumption 2 holds.

The magnitude of the inner product of the LoS components of two different UEs is 
\begin{align}
&\left| (\bar{\mathbf{h}}^j_{jk})^H    \bar{\mathbf{h}}^j_{li} \right| = \sqrt{\beta^j_{jk} \beta^j_{li}} \left| \sum_{m=0}^{M_j -1} \left( e^{\jmath 2\pi d_H  (\sin(\varphi^j_{jk}) -\sin(\varphi^j_{li}))}\right)^m\right|  \nonumber \\ 
&=\sqrt{\beta^j_{jk} \beta^j_{li}}g_1(\varphi^j_{jk}, \varphi^j_{li}),
\end{align}
where
\begin{equation}\label{g1}
g_1(\varphi^j_{jk}, \varphi^j_{li})=\begin{cases}
	\frac{\sin \left( \pi d_H M_j (\sin(\varphi^j_{jk}) -\sin(\varphi^j_{li}))\right) }{\sin \left( \pi d_H  (\sin(\varphi^j_{jk}) -\sin(\varphi^j_{li}))\right)} &  \text{if $\sin(\varphi^j_{jk}) \neq \sin(\varphi^j_{li})$},   \\
M_j  & \text{if $\sin(\varphi^j_{jk}) = \sin(\varphi^j_{li})$} .
\end{cases}
\end{equation}

Hence, if the LoS components $\bar{\mathbf{h}}^j_{jk}$ and $\bar{\mathbf{h}}^j_{li}$ do not have same AoA (more precisely, if $\sin(\varphi^j_{jk}) \neq \sin(\varphi^j_{li})$) then, $\lim\limits_{M_j} \frac{1}{M_j} \left| (\bar{\mathbf{h}}^j_{jk})^H    \bar{\mathbf{h}}^j_{li} \right|  \rightarrow 0 $ holds \cite[Sec.~1.3]{EmilsBook}. This implies that Assumption 3 holds, except in the extreme case when AoAs are exactly the same.

\section{Numerical Results}

In this section, the closed-form SE expressions derived in the previous sections are validated and evaluated by simulating a Massive MIMO cellular network. We have a 16-cell setup where each cell covers a square of $250 \times 250$ m. The network has a wrap-around topology. This layout is selected to guarantee that all BSs receive equally much interference from all directions. There are $K=10$ UEs per cell and these are uniformly and independently distributed in each cell, at distances  larger than $35$\,m from the BS.  Each UE is assigned to the BS that provides largest channel gain considering all the combinations of the UE and BSs in the system. The location of each UE is used when computing the large-scale fading and nominal angle between the UE and BSs. 

  Each BS is equipped with a ULA with half-wavelength antenna spacing \cite[Sec.~1]{EmilsBook}. Thus, the LoS component from UE $i$ in cell $l$ to BS $j$ is given by \eqref{LoS}. For the covariance matrices, we consider $N=6$ scattering clusters and the covariance matrix of each cluster is modeled by the (approximate\footnote{This expression gives accurate results when each cluster has a small ASD, as is the case here.}) Gaussian local scattering model \cite[Sec.~2.6]{EmilsBook}, such that
\begin{equation} \label{NLoS}
	\left[ \mathbf{R}^j_{li}\right] _{s,m} = \frac{\beta^{j,\mathrm{NLoS}}_{li}}{N} \sum_{n=1}^{N} e^{\jmath \pi  (s-m)\sin({\varphi}^j_{li,n})}  e^{-\frac{\sigma^2_\varphi}{2}\left(\pi (s-m)\cos({\varphi}^j_{li,n})\right)^2 } ,
\end{equation}
where $\beta^{j,\mathrm{NLoS}}_{li}$ is the large-scale fading coefficient and ${\varphi}^j_{li,n} \sim \mathcal{U}[{\varphi}^j_{li}-40^\circ, \ {\varphi}^j_{li} + 40^\circ]$ is the nominal AoA for the $n$ cluster. The multipath components of a cluster have Gaussian distributed AoAs, distributed around the nominal AoA with the angular standard deviation (ASD) $\sigma_\varphi=5^\circ$. We consider communication over a 20\,MHz channel and the total receiver noise power is $-94$ dBm. Each coherence block consists of $\tau_c$ = 200 samples and the same $\tau_p$ = 10 pilots are allocated in each cell. The pilots are randomly assigned to the UEs in every cell in the sense that the $k$th UE in two cells, that belong to the same pilot group, uses the same pilot.

 Based on the 3GPP model in \cite{3gpp}, the existence of an LoS path depends on the distance. The probability of LoS for the channel between UE $(l,i)$ and BS $j$ is 
 \begin{equation}\label{probLOS}
	\mathrm{Pr}(\mathrm{LoS})= \begin{cases}
	\frac{300-d^j_{li}}{300}, & 0<d^j_{li}< 300\,\textrm{m}, \\
	0, & d^j_{li}> 300\,\textrm{m}.
	\end{cases}
\end{equation}
 If the LoS path exists then the corresponding large-scale fading coefficient is modeled (in dB) as 
 \begin{equation}\label{betalos}
 \beta^j_{li}= -30.18 -26  \log_{10}\left( {d^j_{li}} \right) + F^j_{li},
 \end{equation}
where $F^j_{li} \sim \mathcal{N}(  0, \sigma^2_\mathrm{sf} )$ is the shadow fading with $\sigma_\mathrm{sf} = 4$. 
The Rician factor is calculated as $\kappa^j_{li}=13 - 0.03d^j_{li}$ [dB] and is used to compute the large-scale fading parameters for the LoS and NLoS paths in \eqref{LoS} and \eqref{NLoS} as $	\beta^{j,\mathrm{LoS}}_{li}= \sqrt{\frac{\kappa^j_{li}}{\kappa^j_{li} +1}} \beta^j_{li}$  and $\beta^{j,\mathrm{NLoS}}_{li}= \sqrt{\frac{1}{\kappa^j_{li} +1}} \beta^j_{li}$ (in linear scale). If the LoS path does not exist then the large-scale fading parameter is modeled (in dB) as 
\begin{equation}\label{betanlos}
\beta^{j}_{li} = -34.53 -38  \log_{10}\left( {d^j_{li}} \right) + F^j_{li},
\end{equation}
where $F^j_{li} \sim \mathcal{N}(  0, \sigma^2_\mathrm{sf} )$ is the shadow fading with $\sigma_\mathrm{sf} = 10$. In this case, $\beta^{j,\mathrm{LoS}}_{li}=0$  and $\beta^{j,\mathrm{NLoS}}_{li}=\beta^{j}_{li}$ (in linear scale).

 We apply the heuristic UL power control policy from \cite[Sec.~7.3]{EmilsBook}, where the transmit power UE $k$ in the cell $j$ is
   \begin{equation}
  p_{jk}=\begin{cases}
    p^\mathrm{ul}_{\mathrm{max}}, &  \Delta > \frac{\beta^j_{jk}}{\beta^j_{j,\mathrm{min}}} ,   \\
    p^\mathrm{ul}_{\mathrm{max}} \Delta \frac{\beta^j_{j,\mathrm{min}}}{\beta^j_{jk}}, & \Delta \leq  \frac{\beta^j_{jk}}{\beta^j_{j,\mathrm{min}}}, 
  \end{cases}
  \end{equation}
where the maximum power is $ p^\mathrm{ul}_{\mathrm{max}}=10$\,dBm and $\beta^j_{j,\mathrm{min}}=\min(\beta^j_{j1},\dots,\beta^j_{jk},\dots,\beta^j_{jK})$. This policy allows the UE with the weakest channel to transmit at full power and require the remaining UEs to reduce their power so their uplink SNRs are at most $\Delta=10$\,dB higher.  For simplicity, the same transmit power is used in the DL and as in the UL: $\rho_{jk}=p_{jk}$ for each UE.

\begin{figure}[h!]%
	\centering
	{\includegraphics[width=\columnwidth]{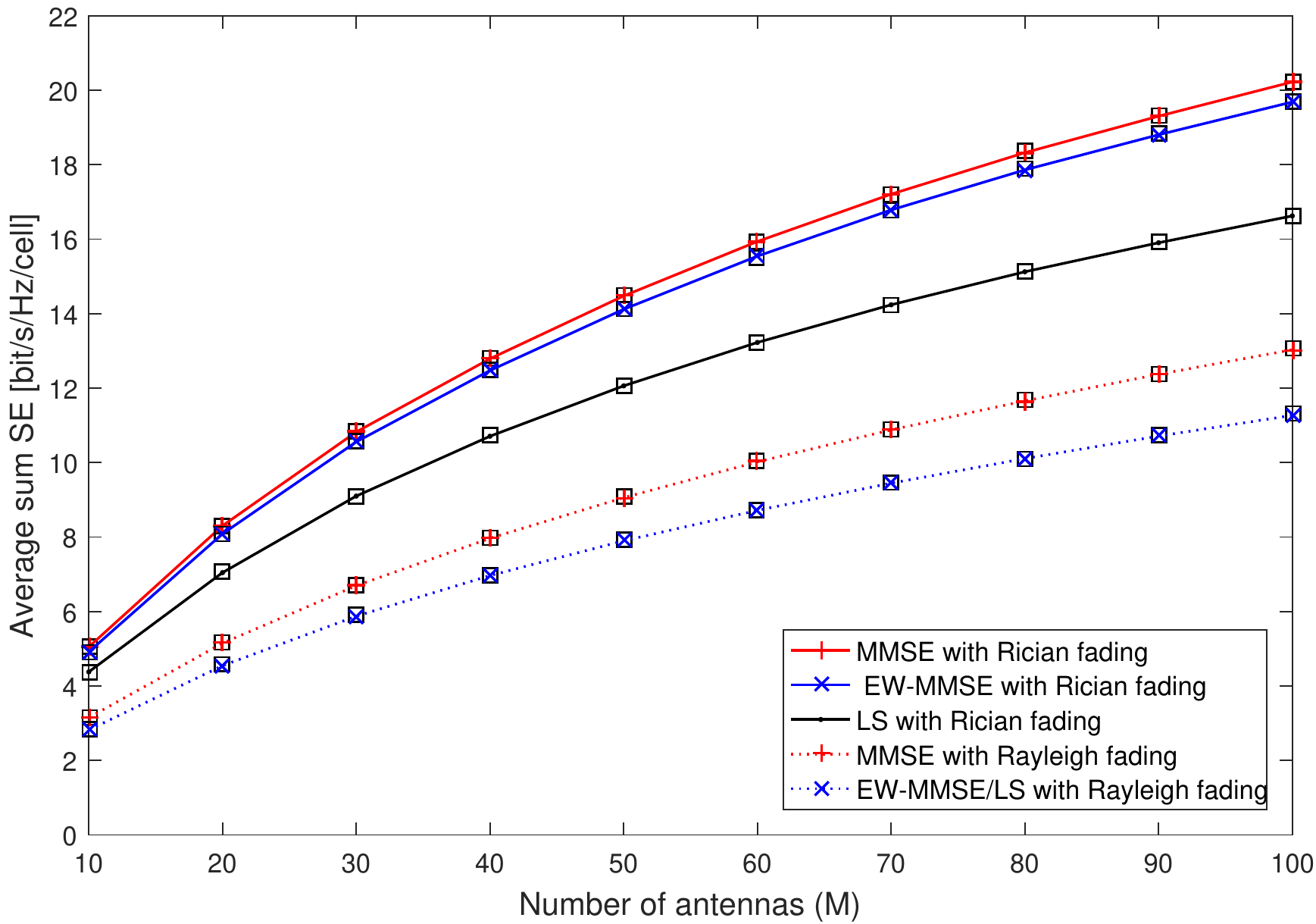} \label{fig1a}}%
		\caption{Average UL sum SE for $K=10$ as a function of the number of BS antennas for different channel estimators. }
	\label{fig1}
\end{figure}

Fig.~\ref{fig1} shows the sum UL SE averaged over different UE locations and shadow fading realizations, when using MR combining based on either the MMSE, LS or EW-MMSE estimators. As a reference, we also provide curves for Rayleigh fading with the same covariance matrices, representing the case when all the LoS components are blocked but the small-scale fading remains (i.e., the average channel gain $\mathbb{E}\{ \| \mathbf{h}^j_{li}\|^2 \}$ is smaller, as it would be the case in practice). The curves are generated using the closed-form expressions from Section~\ref{section4} and the ``$\square$'' markers are generated by Monte Carlo simulations. The fact that the markers overlap with the curves confirms the validity of our analytical results. As expected, the highest UL SE is obtained when the MMSE estimator is employed, since the LoS component and spatial correlation are known and utilized. For Rician fading, the performance of MMSE and EW-MMSE are very close and EW-MMSE performs better than LS since it utilizes knowledge of the channels' mean values.
In the case of Rayleigh fading, MMSE estimation is still the better choice, while the sum SE with LS and EW-MMSE are identical since the estimates are the same up to a scaling factor.

\begin{figure}[h!]
	\centering
	\includegraphics[width=\columnwidth]{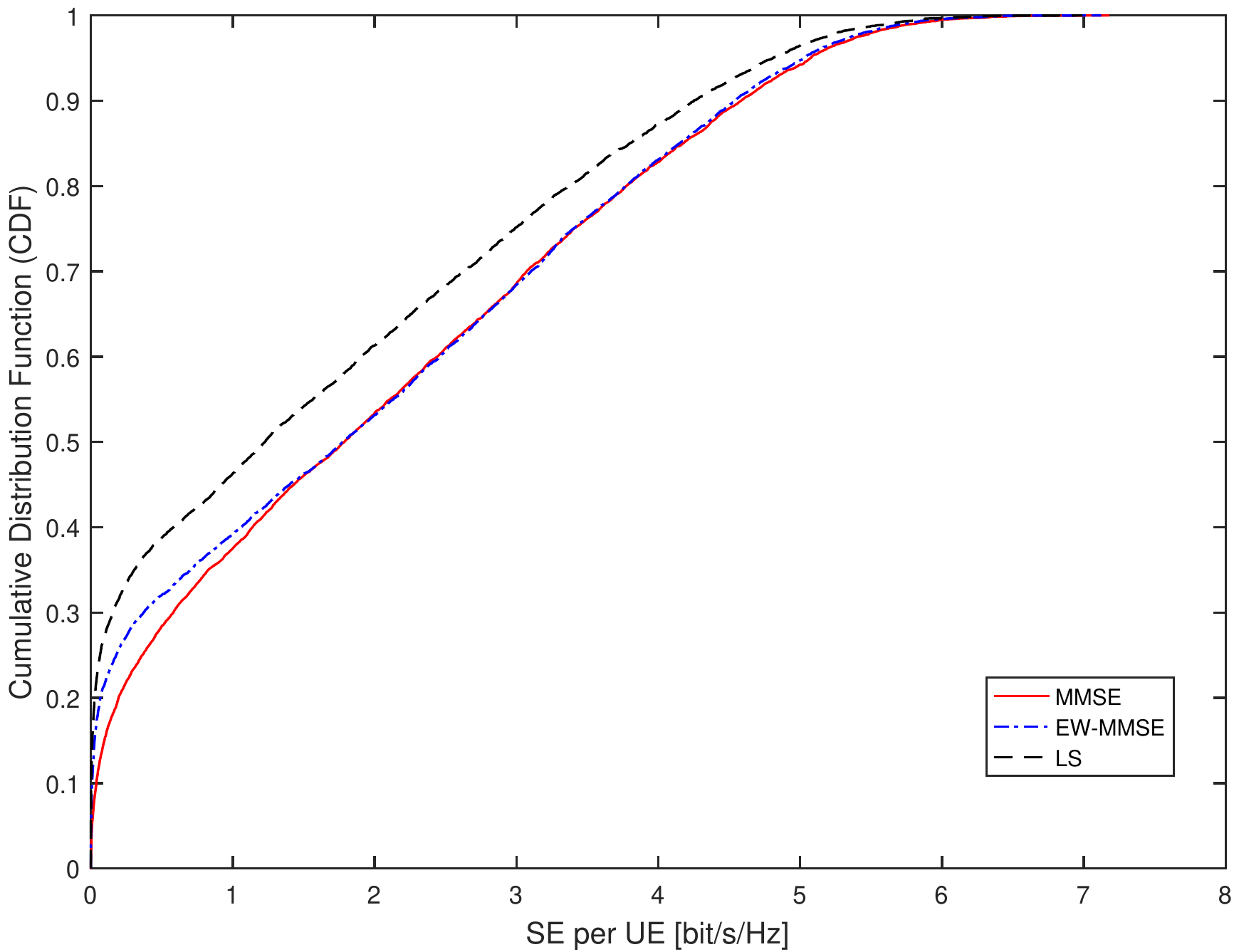}

	\caption{CDF of the UL SE per UE with $M=100$ for different channel}
	\label{fig2}
\end{figure}

\begin{figure}[h!]
	\centering
	\includegraphics[width=\columnwidth]{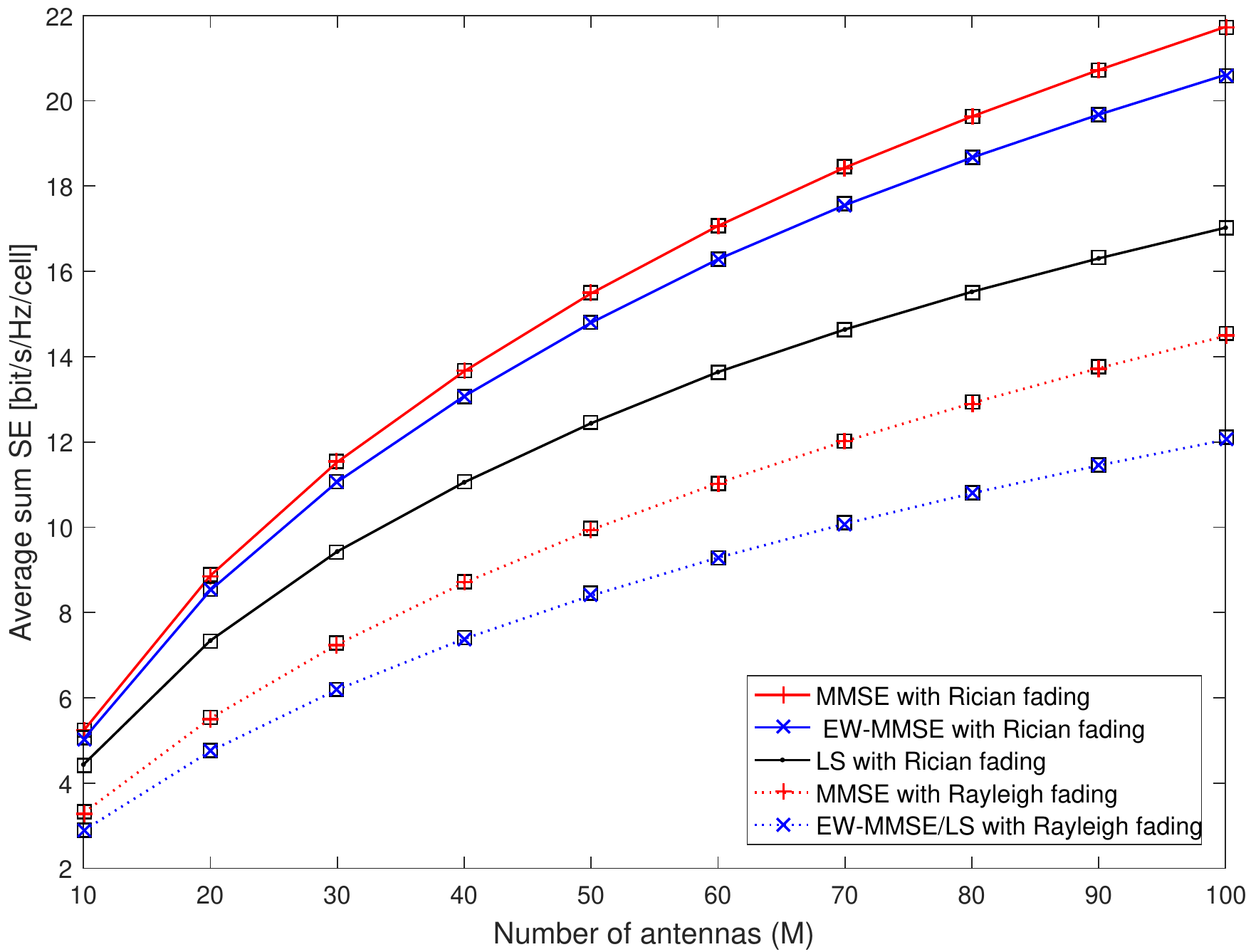}

    	\caption{Average DL sum SE for $K=10$ as a function of the number of BS antennas for different channel estimators.}

	\label{fig3}
\end{figure}

\begin{figure}[h!]
	\centering
	\includegraphics[width=\columnwidth]{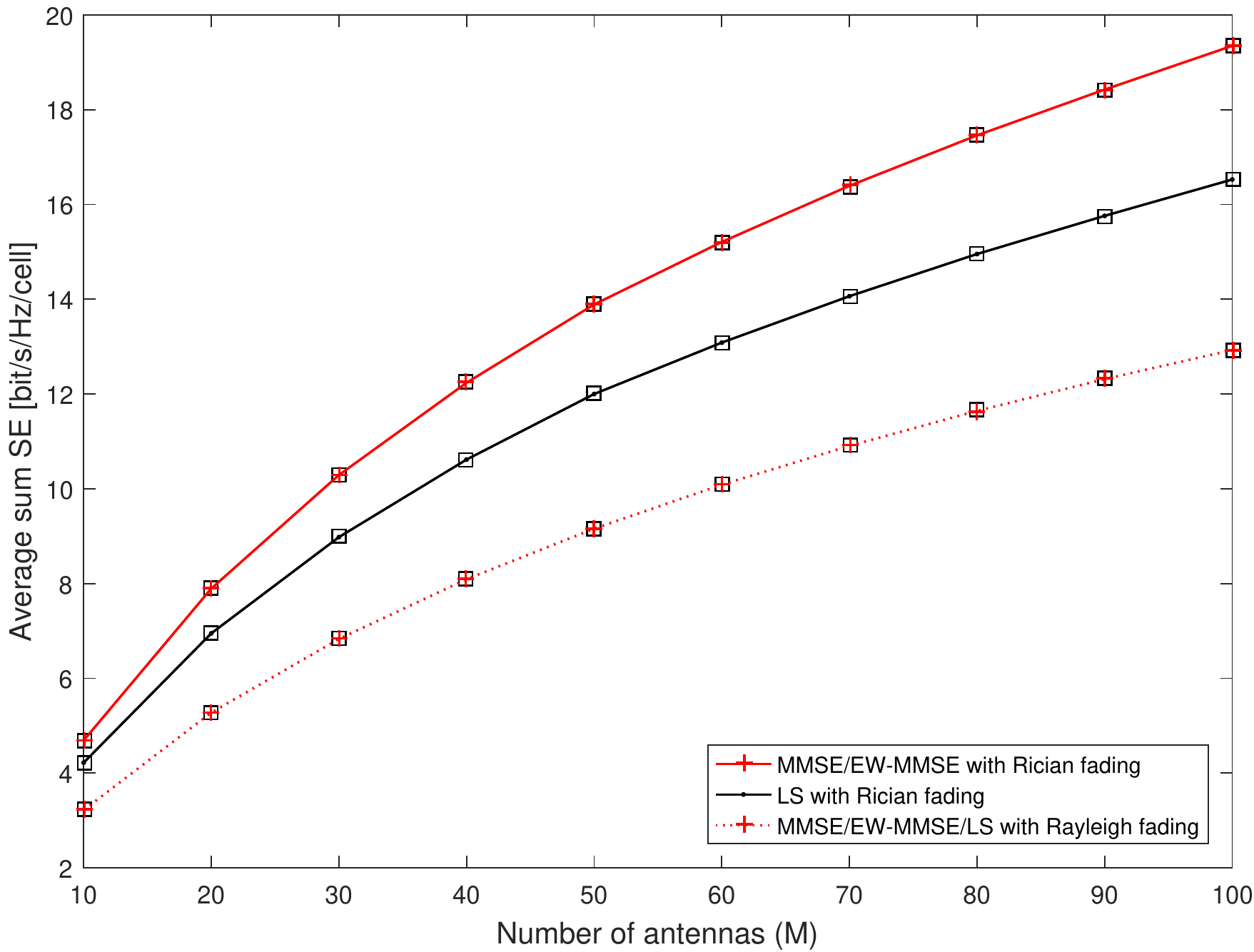}

	\caption{Average UL SE with spatially uncorrelated Rician and Rayleigh fading with $\mathbf{R}^j_{li}= \beta^{j,\mathrm{NLoS}}_{li}\mathbf{I}_M$.}
	\label{fig4}
\end{figure}

\begin{figure}[t!]
	\centering
	\includegraphics[width=\columnwidth]{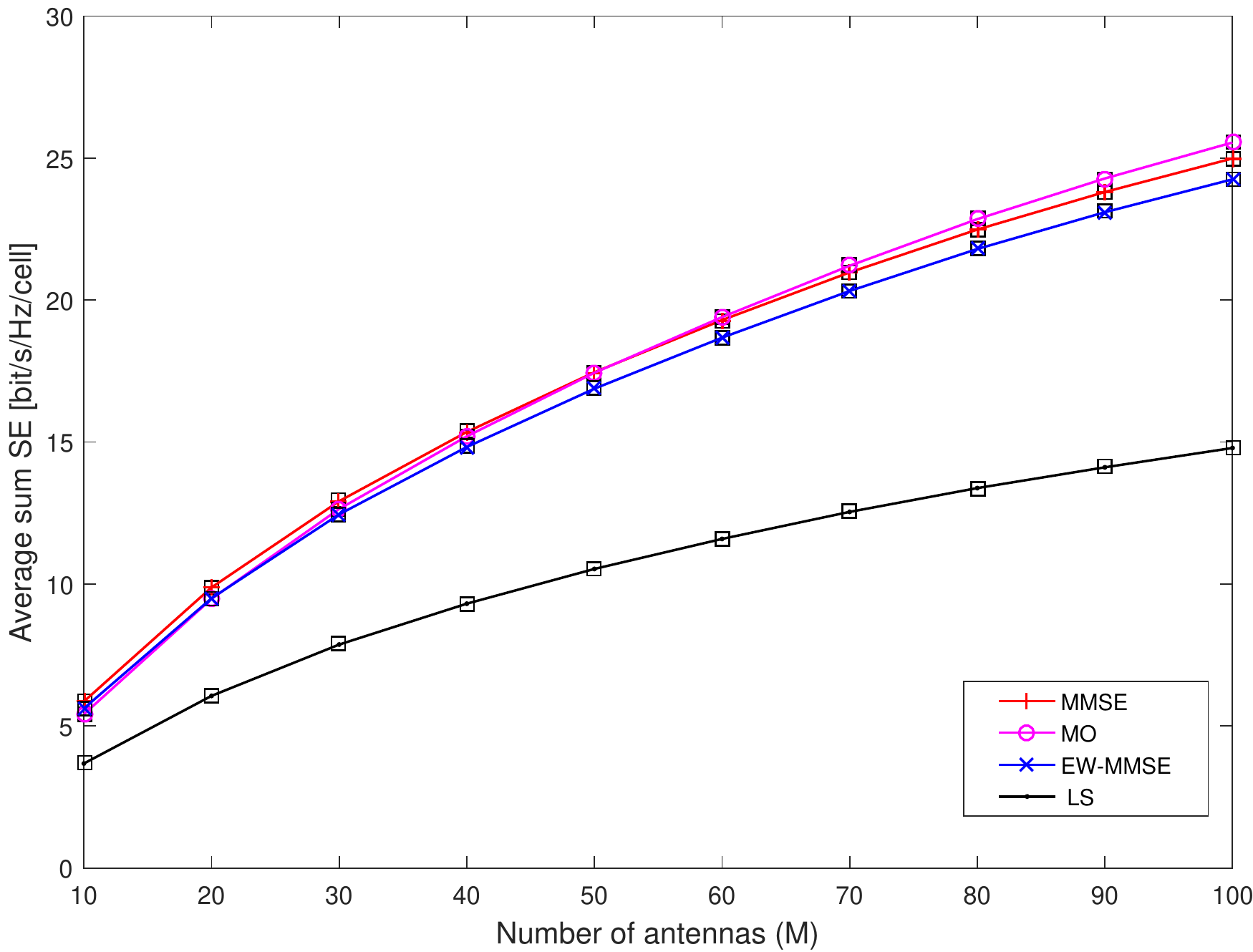}
	\caption{Average UL sum SE for K=10 as a function of BS antennas for different channel estimators where all UE-BS pairs have an LoS path.}
	\label{fig5}
\end{figure}

\begin{figure}[t!]
	\centering
	\includegraphics[width=\columnwidth]{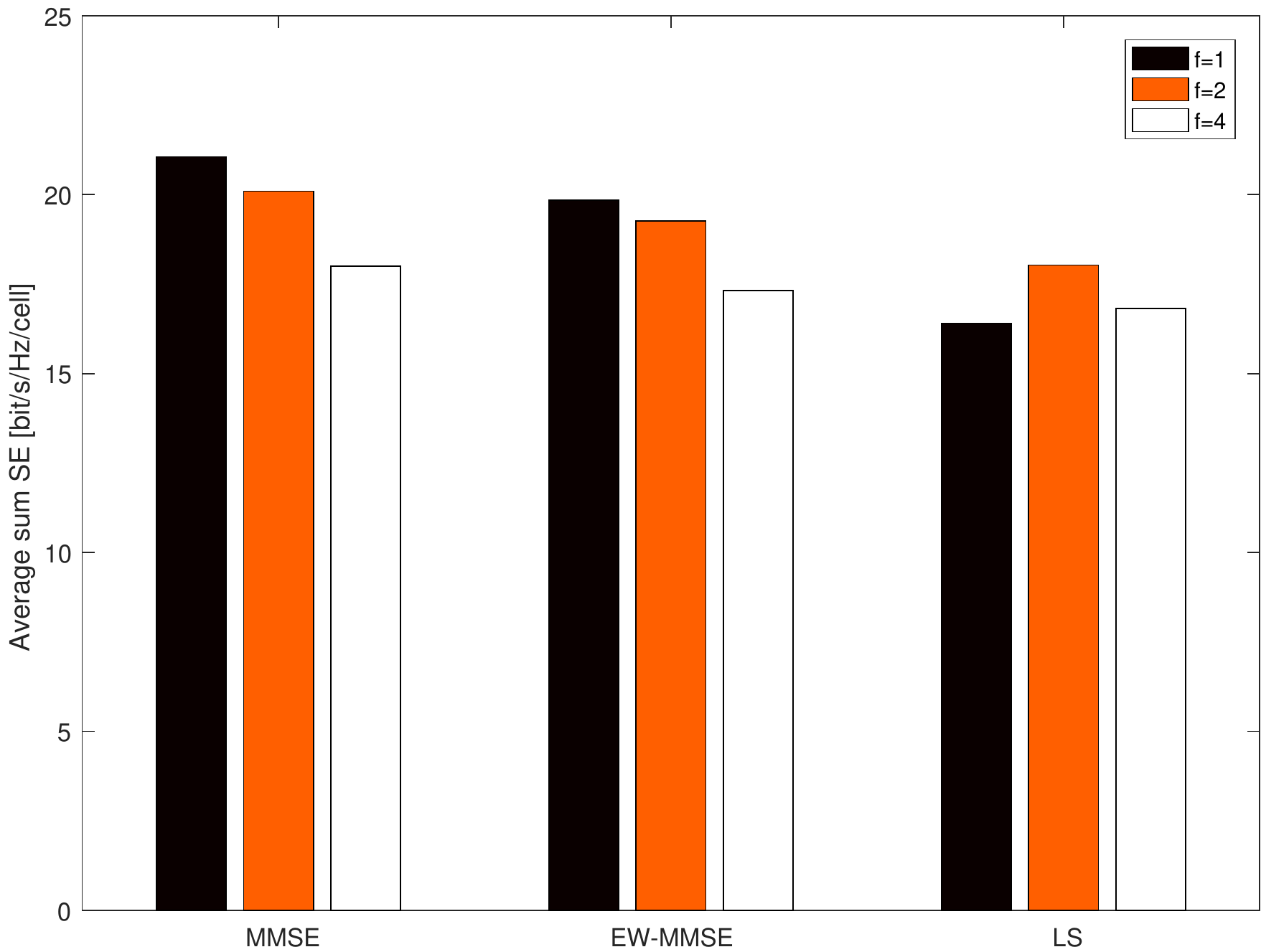}
	\caption{Average UL sum SE for different pilot reuse factors and different estimators.}
	\label{fig6}
\end{figure}

Fig.~\ref{fig2} shows cumulative distribution function (CDF) curves for the SE per UE. The randomness is due to random UE locations and shadow fading realizations. For UEs with good channels, the MMSE and EW-MMSE estimators give the same SE since the estimation errors are anyway small. On the other hand, there is a noticeable difference between MMSE and EW-MMSE for the UEs with the weakest channel conditions.
 
Fig.~\ref{fig3} shows the average sum DL SE over different UE locations and shadow fading realizations, with MR precoding based on the MMSE, LS or EW-MMSE estimators. The same behaviors are observed as in the UL. Fig.~\ref{fig4} shows the average UL SE with uncorrelated fading as a function the number of antennas. In this case, EW-MMSE and MMSE coincide since the spatial covariance matrices are diagonal. These estimators are better than LS when having Rician fading since the mean vectors are utilized to improve the estimates. In contrast, all the vectors give the same performance in the case of Rayleigh fading, since the estimates are equal up to a deterministic scaling factor, which cancel out in the SINR expressions.

In Fig.~\ref{fig5},  we consider a scenario with spatially correlated Rician fading channels between every pair of BSs and UEs which may not be possible in practice. Compared the case where only some UEs have LoS (e.g. Fig.~1), the average UL sum SEs are higher since the existence of an LoS component improves the SE. Note that if the MO estimator is used and some UEs in the cell have Rayleigh fading channels then their SEs would become zero with the MO estimator, which is why we do not consider this case in the other simulations.

Fig.~\ref{fig6} shows the effect of three different ways to reuse pilot sequences across the cells on the average sum UL SEs for different estimators. The integer $f$ denotes the pilot reuse factor where $\tau_p =f K$. This means that there are $f$ times more pilots than UEs per cell and the same subset of pilots is reused in a fraction $1/f$ of the cells. 	The increased number of pilots reduces the pre-log factor in \eqref{eq:SEexpression} since $\tau_u = \tau_c -\tau_p$, but it also increases the instantaneous SINR in \eqref{sec4eq1}. 	There is a trade-off, in that a larger reuse factor implies less pilot contamination but larger pre-log penalty. The LS estimator is more sensitive to pilot contamination since it cannot suppress pilot interference by using spatial processing, while the MMSE estimators can suppress this interference and therefore function even with a tighter reuse, and that this explains why the optimal reuse factor is different for the different estimators.

\section{Conclusion}

This paper studied the UL and DL SE of a multi-cell Massive MIMO system with spatially correlated Rician fading channels. We derived rigorous closed-form  achievable SE expressions when using either MMSE, EW-MMSE, or LS estimation. The expressions provide exact insights into the operation and interference behavior when having Rician fading channels.   These expressions can be utilized for more efficient  user scheduling, pilot allocation, power control, and other resource allocation in practical systems.  We observed that the spatial correlation plays an important role and  the existence of an LoS component improves the achievable SE in Massive MIMO. In addition, the MMSE estimator performs better than the other estimators for both spatially correlated Rayleigh and Rician fading, while the LS estimator gives the lowest SE. In practice, the covariance matrices and the mean vectors might not be known perfectly. Hence, the practical performance lies between the MMSE/EW-MMSE and LS estimators since it is highly probable that the mean is known up to a random phase-shift and covariance matrices are known with some error.


\appendices
\section{Useful results}
\begin{lemma}\label{lemmaMath1}
	Consider the vectors $ \mathbf{x} \sim \mathcal{N}_\mathbb{C}\left( \bar{\mathbf{x}},  \mathbf{R}_{x} \right)$, with mean vector $\bar{\mathbf{x}} \in \mathbb{C}^{N}$ and covariance matrix $\mathbf{R}_{x} \in \mathbb{C}^{N \times N}$, and $ \mathbf{y} \sim \mathcal{N}_\mathbb{C}\left( \bar{\mathbf{y}},  \mathbf{R}_{y} \right)$  with mean vector $\bar{\mathbf{y}} \in \mathbb{C}^{N}$ and covariance matrix $\mathbf{R}_{y} \in \mathbb{C}^{N \times N}$. Also, $\mathbf{B} \in \mathbb{C}^{N \times N}$ is a deterministic matrix and $ \mathbf{x} $ and $ \mathbf{y} $ are independent vectors. It holds that
	\begin{align}
	\mathbb{E}\left\lbrace |\mathbf{x}^H \mathbf{B}\mathbf{y}|^2 \right\rbrace =  \mathrm{tr}\left( \mathbf{B} \mathbf{R}_{y} \mathbf{B}^H \mathbf{R}_{x}\right)  + \bar{\mathbf{x}}^H \mathbf{B} \mathbf{R}_{y} \mathbf{B}^H \bar{\mathbf{x}}  \nonumber \\
	+ \bar{\mathbf{y}}^H \mathbf{B}^H \mathbf{R}_{x} \mathbf{B} \bar{\mathbf{y}}  + |\bar{\mathbf{x}}^H \mathbf{B} \bar{\mathbf{y}}|^2.
	\end{align}
\end{lemma}
\begin{IEEEproof}
	The proof is given in Appendix \ref{ProofLemmaMath1}.
\end{IEEEproof}

\begin{lemma}\label{lemmaMath2}
	Consider the vectors $ \mathbf{x} \sim \mathcal{N}_\mathbb{C}\left( \bar{\mathbf{x}},  \mathbf{R}_{x} \right)$, with mean vector $\bar{\mathbf{x}} \in \mathbb{C}^{N}$ and covariance matrix $\mathbf{R}_{x} \in \mathbb{C}^{N \times N}$, and $ \mathbf{y} \sim \mathcal{N}_\mathbb{C}\left( \bar{\mathbf{y}},  \mathbf{R}_{y} \right)$  with mean vector $\bar{\mathbf{y}} \in \mathbb{C}^{N}$ and covariance matrix $\mathbf{R}_{y} \in \mathbb{C}^{N \times N}$. Also, $\mathbf{B} \in \mathbb{C}^{N \times N}$ is a deterministic matrix. The vectors $ \mathbf{x} $ and $ \mathbf{y} $ are correlated and they are constructed as $\mathbf{x}=\mathbf{R}^{\frac{1}{2}}_{x} \mathbf{w} + \bar{\mathbf{x}}$ and $\mathbf{y}=\mathbf{R}^{\frac{1}{2}}_{y} \mathbf{w} + \bar{\mathbf{y}}$ where $\mathbf{w} \sim \mathcal{N}_\mathbb{C}\left( \mathbf{0},  \mathbf{I}_{N} \right)$. It holds that
	\begin{align}
\!\!	&\mathbb{E}\left\lbrace |\mathbf{x}^H \mathbf{B}\mathbf{y}|^2 \right\rbrace =  \left| \mathrm{tr}\left( (\mathbf{R}_{x}^H)^{\frac{1}{2}} \mathbf{B}  \mathbf{R}^{\frac{1}{2}}_{y}\right)\right| ^2  + \mathrm{tr}\left( \mathbf{B} \mathbf{R}_{y} \mathbf{B}^H \mathbf{R}_{x}\right) + |\bar{\mathbf{x}}^H \mathbf{B} \bar{\mathbf{y}}|^2 \nonumber \\
\!\!\!	&+2\mathrm{Re}\left\lbrace  \mathrm{tr}\left( (\mathbf{R}_{x}^H)^{\frac{1}{2}} \mathbf{B}  \mathbf{R}^{\frac{1}{2}}_{y} \right)  \bar{\mathbf{y}}^H \mathbf{B}^H \bar{\mathbf{x}} \right\rbrace +\bar{\mathbf{x}}^H \mathbf{B} \mathbf{R}_{y} \mathbf{B}^H \bar{\mathbf{x}} + \bar{\mathbf{y}}^H \mathbf{B}^H \mathbf{R}_{x} \mathbf{B} \bar{\mathbf{y}}.
	\end{align}
\end{lemma}
\begin{IEEEproof}
	The proof is given in Appendix \ref{ProofLemmaMath1}.
\end{IEEEproof}

\begin{lemma}\label{lemmaMath3}\cite[Lemmas B.7-8]{EmilsBook}
Consider the positive semi-definite matrices $\mathbf{A} \in \mathbb{C}^{N\times N} $ and $\mathbf{B} \in \mathbb{C}^{N\times N} $. It holds that 
\begin{equation}
	\mathrm{tr}\left( \mathbf{A}\mathbf{B}\right) \leq \|\mathbf{A}\|_2  	\mathrm{tr}\left(\mathbf{B}\right),
\end{equation}
\begin{equation}
\mathrm{tr}\left( \mathbf{A}^{-1}\mathbf{B}\right) \geq \frac{1}{\|\mathbf{A}\|_2  }	\mathrm{tr}\left(\mathbf{B}\right),
\end{equation}
where $\| .\|_2$ denotes the spectral norm which gives the largest eigenvalue of $\mathbf{A}$.
\end{lemma}

\section{Proof of Lemma \ref{lemmaMath1} and Lemma \ref{lemmaMath2}}\label{ProofLemmaMath1}

For the proof of Lemma \ref{lemmaMath1}, note that $\mathbf{x}=\mathbf{R}^{\frac{1}{2}}_{x} \mathbf{w}_x + \bar{\mathbf{x}}$ and $\mathbf{y}=\mathbf{R}^{\frac{1}{2}}_{y} \mathbf{w}_y + \bar{\mathbf{y}}$ where $\mathbf{w}_x \sim \mathcal{N}_\mathbb{C}\left( \mathbf{0},  \mathbf{I}_{N} \right)$ and $\mathbf{w}_y \sim \mathcal{N}_\mathbb{C}\left( \mathbf{0},  \mathbf{I}_{N} \right)$ are independent vectors. Hence,
\begin{align}
&\mathbb{E}\left\lbrace |\mathbf{x}^H \mathbf{B}\mathbf{y}|^2    \right\rbrace =\nonumber \\
\!\!\!&{\mathbb{E}\left\lbrace \left| \underbrace{\bar{\mathbf{x}}^H \mathbf{B} \mathbf{R}^{\frac{1}{2}}_{y}\mathbf{w}_y}_{\text{d}}  +  \underbrace{\mathbf{w}_x^H (\mathbf{R}_{x}^H)^{\frac{1}{2}} \mathbf{B}  \mathbf{R}^{\frac{1}{2}}_{y} \mathbf{w}_y}_{\text{b}} +  \underbrace{\bar{\mathbf{x}}^H \mathbf{B}\bar{\mathbf{y}}}_{\text{c}} +  \underbrace{\mathbf{w}_x^H (\mathbf{R}_{x}^H)^{\frac{1}{2}} \mathbf{B} \bar{\mathbf{y}}}_{\text{f}} \right|^2\right\rbrace} .
\end{align}
We compute each term as  $cc^* = |\bar{\mathbf{x}}^H \mathbf{B} \bar{\mathbf{y}}|^2 $ and
\begin{align}
&\!\!\! \mathbb{E}\left\lbrace dd^*\right\rbrace=\mathbb{E}\left\lbrace \bar{\mathbf{x}}^H \mathbf{B} \mathbf{R}^{\frac{1}{2}}_{y}\mathbf{w}_y   \mathbf{w}_y^H (\mathbf{R}_{y}^H)^{\frac{1}{2}} \mathbf{B}^H \bar{\mathbf{x}} \right\rbrace = \bar{\mathbf{x}}^H \mathbf{B} \mathbf{R}_{y} \mathbf{B}^H \bar{\mathbf{x}} \\
&\!\!\! \mathbb{E}\left\lbrace ff^*\right\rbrace=\mathbb{E}\left\lbrace \bar{\mathbf{y}}^H \mathbf{B}^H \mathbf{R}^{\frac{1}{2}}_{x}\mathbf{w}_x   \mathbf{w}_x^H (\mathbf{R}_{x}^H)^{\frac{1}{2}} \mathbf{B} \bar{\mathbf{y}} \right\rbrace =\bar{\mathbf{y}}^H \mathbf{B}^H \mathbf{R}_{x} \mathbf{B} \bar{\mathbf{y}}\\
&\mathbb{E}\left\lbrace bb^*\right\rbrace= \mathbb{E}\left\lbrace \left| \mathbf{w}_x^H (\mathbf{R}_{x}^H)^{\frac{1}{2}} \mathbf{B}  \mathbf{R}^{\frac{1}{2}}_{y} \mathbf{w}_y\right| ^2 \right\rbrace \nonumber\\
&=  \mathbb{E}\left\lbrace\mathbb{E}\left\lbrace  \mathbf{w}_x^H (\mathbf{R}_{x}^H)^{\frac{1}{2}} \mathbf{B}  \mathbf{R}^{\frac{1}{2}}_{y} \mathbf{w}_y \mathbf{w}^H_y  (\mathbf{R}^{\frac{1}{2}}_{y})^H \mathbf{B}^H  (\mathbf{R}_{x}^{\frac{1}{2}}) \mathbf{w}_x \bigg| \mathbf{w}_x\right\rbrace\right\rbrace \nonumber \\
&=\mathbb{E}\left\lbrace  \mathbf{w}_x^H (\mathbf{R}_{x}^H)^{\frac{1}{2}} \mathbf{B}  \mathbf{R}_{y}  \mathbf{B}^H  (\mathbf{R}_{x}^{\frac{1}{2}}) \mathbf{w}_x \right\rbrace=\mathrm{tr}\left( \mathbf{B} \mathbf{R}_{y} \mathbf{B}^H \mathbf{R}_{x}\right)
\end{align}
 The remaining terms are zero due to the circular symmetry properties and independence of $\mathbf{w}_x$ and $\mathbf{w}_y$. This completes the proof of Lemma \ref{lemmaMath1}.

For the proof of Lemma \ref{lemmaMath2}, note that $\mathbf{x}=\mathbf{R}^{\frac{1}{2}}_{x} \mathbf{w} + \bar{\mathbf{x}}$ and $\mathbf{y}=\mathbf{R}^{\frac{1}{2}}_{y} \mathbf{w} + \bar{\mathbf{y}}$, where $\mathbf{w} \sim \mathcal{N}_\mathbb{C}\left( \mathbf{0},  \mathbf{I}_{N} \right)$ same. Hence,
\begin{align}
&\mathbb{E}\left\lbrace |\mathbf{x}^H \mathbf{B}\mathbf{y}|^2    \right\rbrace   = \nonumber \\
&{\mathbb{E}\left\lbrace \left| \underbrace{\bar{\mathbf{x}}^H \mathbf{B} \mathbf{R}^{\frac{1}{2}}_{y}\mathbf{w}}_{\text{d}}  +  \underbrace{\mathbf{w}^H (\mathbf{R}_{x}^H)^{\frac{1}{2}} \mathbf{B}  \mathbf{R}^{\frac{1}{2}}_{y} \mathbf{w}}_{\text{b}} +  \underbrace{\bar{\mathbf{x}}^H \mathbf{B}\bar{\mathbf{y}}}_{\text{c}} +  \underbrace{\mathbf{w}^H (\mathbf{R}_{x}^H)^{\frac{1}{2}} \mathbf{B} \bar{\mathbf{y}}}_{\text{f}} \right|^2\right\rbrace}.
\end{align}
Similar to the proof of Lemma \ref{lemmaMath1}, we compute $
\mathbb{E}\left\lbrace dd^*\right\rbrace=\|\bar{\mathbf{x}}^H \mathbf{B} \mathbf{R}^{\frac{1}{2}}_{y}\|^2$, $
\mathbb{E}\left\lbrace ff^*\right\rbrace=\|\bar{\mathbf{y}}^H \mathbf{B}^H \mathbf{R}^{\frac{1}{2}}_{x}\|^2$, $
\mathbb{E}\left\lbrace bc^*\right\rbrace= \mathrm{tr}\left( (\mathbf{R}_{x}^H)^{\frac{1}{2}} \mathbf{B}  \mathbf{R}^{\frac{1}{2}}_{y} \right) \bar{\mathbf{y}}^H \mathbf{B}^H \bar{\mathbf{x}}$, $
\mathbb{E}\left\lbrace cb^*\right\rbrace=  \mathrm{tr}\left( (\mathbf{R}_{y}^H)^{\frac{1}{2}} \mathbf{B}^H \mathbf{R}^{\frac{1}{2}}_{x} \right) \bar{\mathbf{x}}^H \mathbf{B} \bar{\mathbf{y}} $, $
cc^*= |\bar{\mathbf{x}}^H \mathbf{B} \bar{\mathbf{y}}|^2$ and
\begin{align}
&\mathbb{E}\left\lbrace bb^*\right\rbrace=
\mathbb{E}\left\lbrace \left| \mathbf{w}^H (\mathbf{R}_{x}^H)^{\frac{1}{2}} \mathbf{B}  \mathbf{R}^{\frac{1}{2}}_{y} \mathbf{w}\right|^2 \right\rbrace \nonumber \\
& = \left| \mathrm{tr}\left( (\mathbf{R}_{x}^H)^{\frac{1}{2}} \mathbf{B}  \mathbf{R}^{\frac{1}{2}}_{y}\right)\right| ^2 +  \mathrm{tr}\left( \mathbf{B} \mathbf{R}_{y} \mathbf{B}^H \mathbf{R}_{x}\right).
\end{align}
 The other terms are zero due to the circular symmetry property of $\mathbf{w}$. This finishes the proof of Lemma \ref{lemmaMath2}.

\begin{figure*}[tp]
	\normalsize
		\vspace*{-2pt}
	\setcounter{equation}{88}
	\begin{align}\label{term3part1}
	&\mathbb{E}\left\lbrace \left| \left( \hat{\mathbf{h}}^j_{jk}\right)^H \hat{\mathbf{h}}^j_{li}  \right| ^2 \right\rbrace= p_{jk} p_{li}  \tau^2_p \left|  \mathrm{tr}\left( \mathbf{R}^j_{li} \boldsymbol{\Psi}^j_{jk} \mathbf{R}^j_{jk}\right) \right| ^2 + p_{jk} \tau_p \mathrm{tr}\left( \left( \mathbf{R}^j_{li} -\mathbf{C}_{li}\right)  \mathbf{R}^j_{jk} \boldsymbol{\Psi}^j_{jk} \mathbf{R}^j_{jk}\right) + \left|  (\bar{\mathbf{h}}^j_{jk})^H    \bar{\mathbf{h}}^j_{li}   \right| ^2 \nonumber  \\
	&\!\!\!\!\!\!+ 2 \sqrt{ p_{jk} p_{li}}\tau_p \mathrm{Re}\left\lbrace  \mathrm{tr}\left(    \mathbf{R}^j_{li} \boldsymbol{\Psi}^j_{jk} \mathbf{R}^j_{jk}  \right) \left( \bar{\mathbf{h}}^j_{li}\right)^H \bar{\mathbf{h}}^j_{jk}  \right\rbrace +  p_{jk} \tau_p \left( \bar{\mathbf{h}}^j_{li}\right)^H    \mathbf{R}^j_{jk} \boldsymbol{\Psi}^j_{jk} \mathbf{R}^j_{jk} \bar{\mathbf{h}}^j_{li} 
	+  \left( \bar{\mathbf{h}}^j_{jk}\right)^H \left( \mathbf{R}^j_{li} -\mathbf{C}_{li}\right) \bar{\mathbf{h}}^j_{jk} , 
	\end{align}
	
	\hrulefill
	\vspace*{-2pt}
\end{figure*}	
 \begin{figure*}[!h]
	\normalsize
	\setcounter{equation}{92}
	\vspace*{-2pt}
	\begin{align}\label{eq1EWMMSE}
	&\mathbb{E}\left\lbrace \left| \mathbf{v}^H_{jk} \mathbf{h}^j_{li} \right|^2  \right\rbrace= \mathbb{E}\left\lbrace (\mathbf{h}^j_{li})^H \bar{\mathbf{h}}^j_{jk} ( \bar{\mathbf{h}}^j_{jk})^H \mathbf{h}^j_{li}\right\rbrace +\sqrt{p_{jk}} \sqrt{p_{li}}\tau_p\mathbb{E}\left\lbrace (\mathbf{h}^j_{li})^H \bar{\mathbf{h}}^j_{jk} ( \mathbf{h}^j_{li} - \bar{\mathbf{h}}^j_{li})^H  \boldsymbol{\Lambda}^{j}_{jk} \mathbf{D}^j_{jk}  \mathbf{h}^j_{li} \right\rbrace \nonumber \\
	&+{p_{jk}} \mathbb{E}\left\lbrace (\mathbf{h}^j_{li})^H \mathbf{D}^j_{jk} \boldsymbol{\Lambda}^{j}_{jk} \left( \mathbf{x}_{jk}- \bar{\mathbf{x}}_{jk}\right) \left( \mathbf{x}_{jk} - \bar{\mathbf{x}}_{jk}\right)^H  \boldsymbol{\Lambda}^{j}_{jk} \mathbf{D}^j_{jk}  \mathbf{h}^j_{li}\right\rbrace  +\sqrt{p_{jk}} \sqrt{p_{li}}\tau_p\mathbb{E}\left\lbrace  (\mathbf{h}^j_{li})^H \mathbf{D}^j_{jk} \boldsymbol{\Lambda}^{j}_{jk}  ( \mathbf{h}^j_{li} - \bar{\mathbf{h}}^j_{li})  ( \bar{\mathbf{h}}^j_{jk})^H \mathbf{h}^j_{li}\right\rbrace \nonumber\\
	& +{p_{jk}} {p_{li}}\tau^2_p \mathbb{E}\left\lbrace  (\mathbf{h}^j_{li})^H \mathbf{D}^j_{jk} \boldsymbol{\Lambda}^{j}_{jk}  ( \mathbf{h}^j_{li} - \bar{\mathbf{h}}^j_{li}) \left( \mathbf{h}^j_{li} - \bar{\mathbf{h}}^j_{li}\right)^H  \boldsymbol{\Lambda}^{j}_{jk} \mathbf{D}^j_{jk}  \mathbf{h}^j_{li}\right\rbrace.  
	\end{align}
	
	\hrulefill
	\vspace*{-2pt}
\end{figure*}
	\setcounter{equation}{80}

\section{Proof of Lemma \ref{lemmaLS}}\label{ProoflemmaLS}
The LS estimate is clearly Gaussian distributed. Its mean value and covariance matrix can be calculated as $
\mathbb{E}\left\lbrace \hat{\mathbf{h}}^j_{li} \right\rbrace = \mathbb{E}\left\lbrace \frac{1}{\sqrt{p_{li}} \tau_p} \mathbf{y}^p_{jli} \right\rbrace =  \frac{1}{\sqrt{p_{li}} \tau_p} \bar{\mathbf{y}}^p_{jli}$ and $
\mathrm{Cov}\left\lbrace \hat{\mathbf{h}}^j_{li} \right\rbrace = \frac{1}{p_{li} \tau^2_p}	\mathbb{E}\left\lbrace \left( \mathbf{y}^p_{jli} - \bar{\mathbf{y}}^p_{jli} \right) \left( \mathbf{y}^p_{jli} - \bar{\mathbf{y}}^p_{jli} \right)^H \right\rbrace =  \frac{1}{p_{li} \tau_p} (\boldsymbol{\Psi}^j_{li})^{-1}$. Similarly, the mean and covariance of the estimation error can be computed as 
\begin{equation}
	\mathbb{E}\left\lbrace \tilde{\mathbf{h}}^j_{li} \right\rbrace = \mathbb{E}\left\lbrace {\mathbf{h}}^j_{li} -\hat{\mathbf{h}}^j_{li} \right\rbrace = \bar{\mathbf{h}}^j_{li} - \frac{1}{\sqrt{p_{li}} \tau_p} \bar{\mathbf{y}}^p_{jli} ,
\end{equation}
 \begin{align}\label{lscov}
\!\!\!\!\!\! \mathrm{Cov}\left\lbrace \tilde{\mathbf{h}}^j_{li} \right\rbrace = \mathbb{E}\left\lbrace {\mathbf{h}}^j_{li} ({\mathbf{h}}^j_{li})^H \right\rbrace - \mathbb{E}\left\lbrace {\mathbf{h}}^j_{li} (\hat{\mathbf{h}}^j_{li})^H \right\rbrace - \mathbb{E}\left\lbrace \hat{\mathbf{h}}^j_{li} ({\mathbf{h}}^j_{li})^H \right\rbrace \nonumber \\+\mathbb{E}\left\lbrace \hat{\mathbf{h}}^j_{li} (\hat{\mathbf{h}}^j_{li})^H \right\rbrace -  \mathbb{E}\left\lbrace \tilde{\mathbf{h}}^j_{li} \right\rbrace \mathbb{E}\left\lbrace (\tilde{\mathbf{h}}^j_{li} )^H\right\rbrace 
 \end{align}
where 
\begin{align}
&\mathbb{E}\left\lbrace \hat{\mathbf{h}}^j_{li} ({\mathbf{h}}^j_{li})^H \right\rbrace = \frac{1}{\sqrt{p_{li}} \tau_p}\mathbb{E}\left\lbrace  \mathbf{y}^p_{jli} ({\mathbf{h}}^j_{li})^H\right\rbrace \nonumber \\
&=\frac{1}{\sqrt{p_{li}} \tau_p}\mathbb{E}\left\lbrace \sum_{(l',i') \in \mathcal{P}_{li}} \sqrt{p_{l'i'}}\tau_p \mathbf{h}^j_{l'i'}({\mathbf{h}}^j_{li})^H  +\mathbf{N}^p_j \phi^*_{li}  ({\mathbf{h}}^j_{li})^H \right\rbrace \nonumber \\
&= \mathbf{R}^j_{li} + \sum_{(l',i') \in \mathcal{P}_{li} } \frac{\sqrt{p_{l'i'}}}{\sqrt{p_{li}}} \bar{\mathbf{h}}^j_{l'i'}  (\bar{\mathbf{h}}^j_{li} )^H
\end{align}
and $\mathbb{E}\left\lbrace {\mathbf{h}}^j_{li} ({\mathbf{h}}^j_{li})^H \right\rbrace=\mathbf{R}^j_{li} + \bar{\mathbf{h}}^j_{li}(\bar{\mathbf{h}}^j_{li})^H$. Inserting all terms into \eqref{lscov} gives the covariance matrix of the estimation error as shown in Lemma \ref{lemmaLS}.

\section{Proof of Theorem \ref{thULMMSE}}\label{ProofthULMMSE}

We need to characterize the terms $\mathbb{E}\left\lbrace \mathbf{v}^H_{jk} \mathbf{h}^j_{jk}  \right\rbrace$, $\mathbb{E}\left\lbrace  \| \mathbf{v}_{jk} \|^2 \right\rbrace$ and $\mathbb{E}\left\lbrace \left| \mathbf{v}^H_{jk} \mathbf{h}^j_{li} \right|^2  \right\rbrace$  in \eqref{sec4eq1}. We begin with calculating 

\begin{equation}
\!\!\!\mathbb{E}\left\lbrace \mathbf{v}^H_{jk} \mathbf{h}^j_{jk}  \right\rbrace = \mathbb{E}\left\lbrace ( \hat{\mathbf{h}}^j_{jk})^H  \mathbf{h}^j_{jk}  \right\rbrace = \mathbb{E}\left\lbrace  ( \hat{\mathbf{h}}^j_{jk})^H  \hat{\mathbf{h}}^j_{jk}   \right\rbrace+ \mathbb{E}\left\lbrace ( \hat{\mathbf{h}}^j_{jk})^H \tilde{\mathbf{h}}^j_{jk}   \right\rbrace.
\end{equation}
The second term $\mathbb{E}\left\lbrace ( \hat{\mathbf{h}}^j_{jk})^H \tilde{\mathbf{h}}^j_{jk}   \right\rbrace$ is zero due to the independence of estimate and estimation error when using the MMSE estimator. The first term is identical with $\mathbb{E}\left\lbrace  \| \mathbf{v}_{jk} \|^2 \right\rbrace = \mathbb{E}\left\lbrace  \left( \hat{\mathbf{h}}^j_{jk}\right)^H  \hat{\mathbf{h}}^j_{jk}   \right\rbrace$. Then,
\begin{align}
&\mathbb{E}\left\lbrace \mathbf{v}^H_{jk} \mathbf{h}^j_{jk}  \right\rbrace=\mathbb{E}\left\lbrace  \| \mathbf{v}_{jk} \|^2 \right\rbrace = \mathrm{tr}\left(  \mathbb{E}\left\lbrace  \hat{\mathbf{h}}^j_{jk}   \left(  \hat{\mathbf{h}}^j_{jk}\right)^H   \right\rbrace\right) \nonumber \\
&= p_{jk} \tau_p \mathrm{tr}\left( \mathbf{R}^j_{jk} \boldsymbol{\Psi}^j_{jk} \mathbf{R}^j_{jk}\right) + \| \bar{\mathbf{h}}^j_{jk}   \|^2.
\end{align}

The last term in \eqref{sec4eq1} can be written as
\begin{align}\label{term3}
&\mathbb{E}\left\lbrace \left| \mathbf{v}^H_{jk} \mathbf{h}^j_{li} \right|^2  \right\rbrace = \mathbb{E}\left\lbrace \left| \left( \hat{\mathbf{h}}^j_{jk}\right)^H \mathbf{h}^j_{li}  \right| ^2 \right\rbrace \nonumber \\
& = \mathbb{E}\left\lbrace \left|\left( \hat{\mathbf{h}}^j_{jk}\right)^H \hat{\mathbf{h}}^j_{li}  \right| ^2 \right\rbrace+ \mathbb{E}\left\lbrace \left| \left( \hat{\mathbf{h}}^j_{jk}\right)^H \tilde{\mathbf{h}}^j_{li}  \right| ^2 \right\rbrace ,
\end{align}
since  $\tilde{\mathbf{h}}^j_{li}$ and the pair $(\hat{\mathbf{h}}^j_{jk}, \hat{\mathbf{h}}^j_{li})$ of estimates are independent and estimation error has zero mean.

For $ (l,i) \in \mathcal{P}_{jk}$, $\hat{\mathbf{h}}^j_{jk}$ and $ \hat{\mathbf{h}}^j_{li} $ are not independent.  The channel estimates are $\hat{\mathbf{h}}^j_{jk} = \bar{\mathbf{h}}^j_{jk} + \sqrt{p_{jk}}  \mathbf{R}^j_{jk} \boldsymbol{\Psi}^j_{jk} (\mathbf{y}^p_{jjk} - \bar{\mathbf{y}}^p_{jjk})  $ and  $\hat{\mathbf{h}}^j_{li} = \bar{\mathbf{h}}^j_{li} + \sqrt{p_{li}}  \mathbf{R}^j_{li} \boldsymbol{\Psi}^j_{jk} (\mathbf{y}^p_{jjk} - \bar{\mathbf{y}}^p_{jjk})  $ where $\mathbf{y}^p_{jjk}  \sim \mathcal{N}_\mathbb{C}\left(\bar{\mathbf{y}}^p_{jjk}, \tau_p \left( \boldsymbol{\Psi}^j_{jk}\right)^{-1}  \right)$.  We reformulate these estimates using a matrix $\mathbf{Q}^{-1/2} = \frac{1}{\sqrt{\tau_p}} \left( \boldsymbol{\Psi}^j_{jk}\right)^{1/2} $ that gives $
\mathbf{Q}^{-1/2} (\mathbf{y}^p_{jjk} - \bar{\mathbf{y}}^p_{jjk})=\mathbf{w}   \sim \mathcal{N}_\mathbb{C}\left( \mathbf{0},  \mathbf{I}_{N} \right) $.
Then, the channel estimates are written as 
\begin{align}
&\hat{\mathbf{h}}^j_{jk} = \sqrt{p_{jk}}  \mathbf{R}^j_{jk} \boldsymbol{\Psi}^j_{jk}  \mathbf{Q}^{1/2} \mathbf{Q}^{-1/2}    (\mathbf{y}^p_{jjk} - \bar{\mathbf{y}}^p_{jjk}) + \bar{\mathbf{h}}^j_{jk} \nonumber \\
&= \sqrt{p_{jk}\tau_p}  \mathbf{R}^j_{jk} \left( \boldsymbol{\Psi}^j_{jk}\right)^{1/2} \mathbf{w} + \bar{\mathbf{h}}^j_{jk},
\end{align}
\begin{equation}
\hat{\mathbf{h}}^j_{li} = \sqrt{p_{li} \tau_p}  \mathbf{R}^j_{li} \left( \boldsymbol{\Psi}^j_{jk}\right)^{1/2} \mathbf{w} + \bar{\mathbf{h}}^j_{li}.
\end{equation}
	\setcounter{equation}{89}
Using  Lemma \ref{lemmaMath2} for  $\mathbf{R}^{\frac{1}{2}}_{x}=\sqrt{p_{jk} \tau_p}  \mathbf{R}^j_{jk} \left( \boldsymbol{\Psi}^j_{jk}\right)^{1/2} $, $\bar{\mathbf{x}}= \bar{\mathbf{h}}^j_{jk} $ and $\mathbf{R}^{\frac{1}{2}}_{y}= \sqrt{p_{li} \tau_p}  \mathbf{R}^j_{li} \left( \boldsymbol{\Psi}^j_{jk}\right)^{1/2}$, $\bar{\mathbf{y}}= \bar{\mathbf{h}}^j_{li} $ gives \eqref{term3part1} at the top of this page by using  the fact that $  p_{li}  \tau_p\mathbf{R}^j_{li} \boldsymbol{\Psi}^j_{jk} \mathbf{R}^j_{li}=\mathbf{R}^j_{li} -\mathbf{C}_{li}$. To  calculate $\mathbb{E}\left\lbrace \left| \left( \hat{\mathbf{h}}^j_{jk}\right)^H \tilde{\mathbf{h}}^j_{li}  \right| ^2 \right\rbrace$ in \eqref{term3},
we use Lemma \ref{lemmaMath1} since $\hat{\mathbf{h}}^j_{jk}$ and $\tilde{\mathbf{h}}^j_{li} $ are independent. Noting that the estimation error  $\tilde{\mathbf{h}}^j_{li}$ has distribution $  \mathcal{N}_\mathbb{C}\left( \mathbf{0},  \mathbf{C}_{li} \right) $, we obtain 
\begin{equation}\label{term3part2}
\mathbb{E}\left\lbrace \left| \left( \hat{\mathbf{h}}^j_{jk}\right)^H \tilde{\mathbf{h}}^j_{li}  \right| ^2 \right\rbrace =  p_{jk} \tau_p \mathrm{tr}\left(\mathbf{C}_{li} \mathbf{R}^j_{jk} \boldsymbol{\Psi}^j_{jk} \mathbf{R}^j_{jk}\right) +  \left( \bar{\mathbf{h}}^j_{jk}\right)^H \mathbf{C}_{li} \bar{\mathbf{h}}^j_{jk}.
\end{equation} 
Substituting \eqref{term3part1} and \eqref{term3part2}  back  into \eqref{term3}  gives the final result for  $ (l,i) \in \mathcal{P}_{jk}$  as in \eqref{sec4eq4}.

For the case of $ (l,i) \notin \mathcal{P}_{jk}$,  $\hat{\mathbf{h}}^j_{jk}$ and  $\mathbf{h}^j_{li}$ are independent. Using Lemma \ref{lemmaMath1}, we get
\begin{align}
&\mathbb{E}\left\lbrace \left| \left( \hat{\mathbf{h}}^j_{jk}\right)^H \mathbf{h}^j_{li}  \right| ^2 \right\rbrace =  p_{jk} \tau_p \left( \bar{\mathbf{h}}^j_{li}\right)^H    \mathbf{R}^j_{jk} \boldsymbol{\Psi}^j_{jk} \mathbf{R}^j_{jk} \bar{\mathbf{h}}^j_{li}  \\
\!\!&+ p_{jk} \tau_p \mathrm{tr}\left( \mathbf{R}^j_{li} \mathbf{R}^j_{jk} \boldsymbol{\Psi}^j_{jk} \mathbf{R}^j_{jk}\right) +  \left( \bar{\mathbf{h}}^j_{jk}\right)^H    \mathbf{R}^j_{li} \bar{\mathbf{h}}^j_{jk}  + \left|  (\bar{\mathbf{h}}^j_{jk})^H    \bar{\mathbf{h}}^j_{li}   \right| ^2. \nonumber
\end{align}
 This finishes the proof of Theorem \ref{thULMMSE}.

\begin{figure*}[tp]
	\normalsize
	\setcounter{equation}{100}
	\begin{equation}\label{asyeq1}
	\frac{{\xi}^{\mathrm{ul}}_{li}}{M_j} \leq  \frac{ p_{jk} \tau_p {\|\mathbf{R}^j_{li}\|_2\mathrm{tr}\left( \mathbf{R}^j_{jk} \boldsymbol{\Psi}^j_{jk} \mathbf{R}^j_{jk}\right)}+ p_{jk} \tau_p  \| \bar{\mathbf{h}}^j_{li} \|^2 \|\mathbf{R}^j_{jk} \boldsymbol{\Psi}^j_{jk} \mathbf{R}^j_{jk}\|_2 + \|\mathbf{R}^j_{li} \|_2 \|\bar{\mathbf{h}}^j_{jk} \|^2+\left|  (\bar{\mathbf{h}}^j_{jk})^H    \bar{\mathbf{h}}^j_{li}   \right| ^2}{ p_{jk} \tau_p M_j\mathrm{tr}\left( \mathbf{R}^j_{jk} \boldsymbol{\Psi}^j_{jk} \mathbf{R}^j_{jk}\right) + M_j\| \bar{\mathbf{h}}^j_{jk}   \|^2}. 
	\end{equation}
	\hrulefill
	\vspace*{-2pt}
\end{figure*}

\begin{figure*}[tp]
	\normalsize
	\begin{equation} \label{eq:coherent-inteference-derivation}
	\frac{{\Gamma}^{\mathrm{ul}}_{li}}{M_j}=    \frac{ \frac{p_{jk} p_{li}  \tau^2_p}{M_j}\left|  \mathrm{tr}\left( \mathbf{R}^j_{li} \boldsymbol{\Psi}^j_{jk} \mathbf{R}^j_{jk}\right) \right| ^2    }{p_{jk} \tau_p \mathrm{tr}\left( \mathbf{R}^j_{jk} \boldsymbol{\Psi}^j_{jk} \mathbf{R}^j_{jk}\right) +\| \bar{\mathbf{h}}^j_{jk}   \|^2} +  \frac{ 2 \sqrt{p_{li} p_{jk}} \mathrm{Re}\left\lbrace \mathrm{tr}\left( \mathbf{R}^j_{li} \boldsymbol{\Psi}^j_{jk} \mathbf{R}^j_{jk}\right) \frac{1}{M_j}  (\bar{\mathbf{h}}^j_{jk})^H    \bar{\mathbf{h}}^j_{li}  \right\rbrace   }{p_{jk} \tau_p \mathrm{tr}\left( \mathbf{R}^j_{jk} \boldsymbol{\Psi}^j_{jk} \mathbf{R}^j_{jk}\right) +\| \bar{\mathbf{h}}^j_{jk}   \|^2}.
	\end{equation}
	\hrulefill
	\vspace*{-2pt}
\end{figure*}	
	\setcounter{equation}{91}
\section{Proof of Theorem \ref{thEWMMSE}}\label{ProofthEWMMSE}

	We need to characterize the terms $\mathbb{E}\left\lbrace \mathbf{v}^H_{jk} \mathbf{h}^j_{jk}  \right\rbrace$, $\mathbb{E}\left\lbrace  \| \mathbf{v}_{jk} \|^2 \right\rbrace$ and $\mathbb{E}\left\lbrace \left| \mathbf{v}^H_{jk} \mathbf{h}^j_{li} \right|^2  \right\rbrace$  in \eqref{sec4eq1}. Calculations of $	\mathbb{E}\left\lbrace \mathbf{v}^H_{jk} \mathbf{h}^j_{jk}  \right\rbrace $ and $\mathbb{E}\left\lbrace  \| \mathbf{v}_{jk} \|^2 \right\rbrace$ are rather straightforward. For the third term $\mathbb{E}\left\lbrace \left| \mathbf{v}^H_{jk} \mathbf{h}^j_{li} \right|^2  \right\rbrace$, if $(l,i)\notin \mathcal{P}_{jk}$ then $\hat{\mathbf{h}}^j_{jk}$ and $\mathbf{h}^j_{li}$ are independent vectors. Directly applying Lemma \ref{lemmaMath1} gives 
	\begin{align}
		&\mathbb{E}\left\lbrace \left|  (\hat{\mathbf{h}}^j_{jk} )^H \mathbf{h}^j_{li} \right| ^2 \right\rbrace = \mathrm{tr}\left(\mathbf{R}^j_{li} \boldsymbol{\Sigma}^j_{jk}\right) + (\bar{\mathbf{h}}^j_{jk} )^H\mathbf{R}^j_{li} \bar{\mathbf{h}}^j_{jk} \nonumber \\
	&+ (\bar{\mathbf{h}}^j_{li} )^H \boldsymbol{\Sigma}^j_{jk} \bar{\mathbf{h}}^j_{li} + \left| (\bar{\mathbf{h}}^j_{jk} )^H \bar{\mathbf{h}}^j_{li}\right| ^2.
	\end{align}
 If $(l,i)\in \mathcal{P}_{jk}$ then $\hat{\mathbf{h}}^j_{jk} $ and $ {\mathbf{h}}^j_{li}$ are not independent vectors. Decomposing  ${\mathbf{y}}_{jjk}$ using ${\mathbf{x}}_{jk}={\mathbf{y}}_{jjk} - \sqrt{p_{li}}\tau_p \mathbf{h}^j_{li}$ as $ ( \mathbf{y}^p_{jjk} - \bar{\mathbf{y}}^p_{jjk} )  = \left( \mathbf{x}_{jk} - \bar{\mathbf{x}}_{jk} \right)^H  + \sqrt{p_{li}}\tau_p( \mathbf{h}^j_{li} - \bar{\mathbf{h}}^j_{li})$ gives \eqref{eq1EWMMSE} at the top of previous page. Computing each term in \eqref{eq1EWMMSE}  and noticing that $\mathrm{tr}\left(\mathbf{R}^j_{jk}  \boldsymbol{\Lambda}^{j}_{jk}\mathbf{D}^j_{jk} \right) =\mathrm{tr}\left( \mathbf{D}^j_{jk}  \boldsymbol{\Lambda}^{j}_{jk}\mathbf{D}^j_{jk}\right) $  leads to  \eqref{eq2EWMMSE}.

	\setcounter{equation}{93}
\section{Proof of Theorem \ref{thULLS}}\label{ProofthULLS}

 We need to characterize the terms $\mathbb{E}\left\lbrace \mathbf{v}^H_{jk} \mathbf{h}^j_{jk}  \right\rbrace$, $\mathbb{E}\left\lbrace  \| \mathbf{v}_{jk} \|^2 \right\rbrace$ and $\mathbb{E}\left\lbrace \left| \mathbf{v}^H_{jk} \mathbf{h}^j_{li} \right|^2  \right\rbrace$  in \eqref{sec4eq1}. The first two terms are given in Appendix \ref{ProoflemmaLS}. The last term that is needed to calculate in \eqref{sec4eq1}  is
	\begin{align}\label{LSproof1}
	&\mathbb{E}\left\lbrace \left| \mathbf{v}^H_{jk} \mathbf{h}^j_{li} \right| ^2 \right\rbrace = \frac{1}{{p_{jk}} \tau^2_p}\mathbb{E}\left\lbrace \left| (\mathbf{y}^p_{jjk})^H \mathbf{h}^j_{li}\right| ^2  \right\rbrace \\
	& = \frac{1}{{p_{jk}} \tau^2_p}\mathbb{E}\left\lbrace \left|   \sum_{(l',i') \in \mathcal{P}_{jk} } \sqrt{p_{l'i'}} \tau_p ({\mathbf{h}}^j_{l'i'})^H\mathbf{h}^j_{li}  +(\mathbf{N}^p_j \phi^*_{jk} )^H \mathbf{h}^j_{li}\right|^2  \right\rbrace,\nonumber 
	\end{align}
	where the noise term has distribution $\mathbf{N}^p_j \boldsymbol{\phi}^*_{jk} \sim \mathcal{N}_\mathbb{C}\left( \mathbf{0},  \sigma^2_{\mathrm{ul}} \tau_p\mathbf{I}_{M} \right)$ and $\|\boldsymbol{\phi}_{jk}\|^2 =\tau_p$.
	
	 If $(l,i) \notin \mathcal{P}_{jk}$, note that 
	$\mathbf{y}^p_{jjk}  \sim \mathcal{N}_\mathbb{C}\left(\bar{\mathbf{y}}^p_{jjk}, \tau_p \left( \boldsymbol{\Psi}^j_{jk}\right)^{-1}  \right)$ and $\mathbf{y}^p_{jjk}$ and $\mathbf{h}^j_{li}$ are independent random vectors. Directly utilizing Lemma \ref{lemmaMath1} gives
	\begin{align}
	&\mathbb{E}\left\lbrace \left| (\mathbf{y}^p_{jjk})^H \mathbf{h}^j_{li}\right| ^2\right\rbrace  =\tau_p \mathrm{tr}\left( \mathbf{R}^j_{li}\left( \boldsymbol{\Psi}^j_{jk}\right)^{-1}\right) + (\bar{\mathbf{y}}^p_{jjk})^H \mathbf{R}^j_{li}\bar{\mathbf{y}}^p_{jjk} \nonumber  \\
	&+ \tau_p (\bar{\mathbf{h}}^j_{li})^H \left( \boldsymbol{\Psi}^j_{jk}\right)^{-1}\bar{\mathbf{h}}^j_{li} +  \left|(\bar{\mathbf{y}}^p_{jjk})^H \bar{\mathbf{h}}^j_{li}\right|^2.
	\end{align}

	If $(l,i) \in \mathcal{P}_{jk} $, then $\mathbf{y}^p_{jjk}$ and $\mathbf{h}^j_{li}$ are not independent random vectors since $\mathbf{y}^p_{jjk}$ contains $\mathbf{h}^j_{li}$. We decompose the terms in \eqref{LSproof1}  by using ${\mathbf{x}}_{jk}={\mathbf{y}}_{jjk} - \sqrt{p_{li}}\tau_p \mathbf{h}^j_{li}$. Then we obtain
	\begin{align}
	&	\mathbb{E}\left\lbrace \left|   \mathbf{x}^H_{jk} \mathbf{h}^j_{li} + \sqrt{p_{li}} \tau_p ({\mathbf{h}}^j_{li})^H\mathbf{h}^j_{li} \right|^2  \right\rbrace =  \\
	& \mathbb{E}\left\lbrace \left|   \mathbf{x}^H_{jk} \mathbf{h}^j_{li} \right|^2  \right\rbrace        + p_{li} \tau^2_p \mathbb{E}\left\lbrace \left|  ({\mathbf{h}}^j_{li})^H\mathbf{h}^j_{li} \right|^2  \right\rbrace \nonumber \\
	&+ \sqrt{p_{li}} \tau_p    \mathbb{E}\left\lbrace   \mathbf{x}^H_{jk} \mathbf{h}^j_{li} ({\mathbf{h}}^j_{li})^H\mathbf{h}^j_{li}   \right\rbrace +\sqrt{p_{li}} \tau_p    \mathbb{E}\left\lbrace    ({\mathbf{h}}^j_{li})^H\mathbf{h}^j_{li} ({\mathbf{h}}^j_{li})^H \mathbf{x}_{jk}  \right\rbrace.\nonumber
	\end{align}
	 
	Further, we compute each term beginning with 
	\begin{align}
	&\mathbb{E}\left\lbrace \left|   \mathbf{x}^H_{jk} \mathbf{h}^j_{li} \right|^2  \right\rbrace= \tau_p \mathrm{tr}\left( \mathbf{R}^j_{li}(\boldsymbol{\Omega}^j_{jk})^{-1} \right) + \bar{\mathbf{x}}^H_{jk}\mathbf{R}^j_{li}\bar{\mathbf{x}}_{jk} \nonumber \\
   &+ \tau_p (\bar{\mathbf{h}}^j_{li})^H (\boldsymbol{\Omega}^j_{jk})^{-1}\bar{\mathbf{h}}^j_{li} +\left|\bar{\mathbf{x}}_{jk}^H \bar{\mathbf{h}}^j_{li}\right|^2
	\end{align}
 where  $(\boldsymbol{\Omega}^j_{jk})^{-1} = (\boldsymbol{\Psi}^j_{jk})^{-1} -p_{li}\tau_p \mathbf{R}^j_{li}  $.
The other terms are
	\begin{align}
	&\mathbb{E}\left\lbrace \left|  ({\mathbf{h}}^j_{li})^H\mathbf{h}^j_{li} \right|^2  \right\rbrace= \left| \mathrm{tr}\left(  \mathbf{R}^j_{li} \right)\right| ^2   +  \mathrm{tr}\left( \left(  \mathbf{R}^j_{li} \right)^2  \right) \nonumber \\ 
	&+ 2	 \|\bar{\mathbf{h}}^j_{li}\|^2 \mathrm{tr}\left(  \mathbf{R}^j_{li} \right)  + 2 \left( \bar{\mathbf{h}}^j_{li}\right)^H  {\mathbf{R}^j_{li} } \bar{\mathbf{h}}^j_{li} +   \|\bar{\mathbf{h}}^j_{li} \|^4,
	\end{align}
	\begin{align}
	&  \mathbb{E}\left\lbrace   \mathbf{x}^H_{jk} \mathbf{h}^j_{li} ({\mathbf{h}}^j_{li})^H\mathbf{h}^j_{li}   \right\rbrace = \mathbb{E}\left\lbrace   \mathbf{x}^H_{jk} \right\rbrace \mathbb{E}\left\lbrace\mathbf{h}^j_{li} ({\mathbf{h}}^j_{li})^H\mathbf{h}^j_{li}  \right\rbrace \nonumber \\
	& = \bar{\mathbf{x}}^H_{jk}  \bar{\mathbf{h}}^j_{li} \mathrm{tr}\left(  \mathbf{R}^j_{li} \right) + \bar{\mathbf{x}}^H_{jk} \mathbf{R}^j_{li} \bar{\mathbf{h}}^j_{li} +\bar{\mathbf{x}}^H_{jk}   \bar{\mathbf{h}}^j_{li} (\bar{\mathbf{h}}^j_{li})^H \bar{\mathbf{h}}^j_{li}, 
	\end{align}
	\begin{align}
	&  \mathbb{E}\left\lbrace    ({\mathbf{h}}^j_{li})^H\mathbf{h}^j_{li} ({\mathbf{h}}^j_{li})^H \mathbf{x}_{jk}  \right\rbrace =  (\bar{\mathbf{h}}^j_{li})^H \mathbf{R}^j_{li}\bar{\mathbf{x}}_{jk} \nonumber \\
	&+  \mathrm{tr}\left(  \mathbf{R}^j_{li} \right) (\bar{\mathbf{h}}^j_{li})^H \bar{\mathbf{x}}_{jk}  + (\bar{\mathbf{h}}^j_{li})^H \bar{\mathbf{h}}^j_{li}  (\bar{\mathbf{h}}^j_{li})^H  \bar{\mathbf{x}}_{jk}.  
	\end{align}
	
	 Combining the results for $(l,i) \in \mathcal{P}_{jk} $ and $(l,i) \notin \mathcal{P}_{jk} $	give the final result of $\mathbb{E}\left\lbrace \left| \mathbf{v}^H_{jk} \mathbf{h}^j_{li} \right| ^2 \right\rbrace$ in \eqref{lseq3}.

 \begin{figure*}[tp]
	\normalsize
	\setcounter{equation}{105}
	\begin{equation}\label{as:ls:derivation}
	\frac{\|\mathbf{R}^j_{li}\|_2 \mathrm{tr}\left( (\boldsymbol{\Psi}^j_{jk})^{-1} \right)}{M_j\left( \mathrm{tr}(\mathbf{R}^j_{jk} )+ \sum_{(l,i) \in \mathcal{P}_{jk} } \frac{\sqrt{p_{jk}}}{ \sqrt{p_{li}}} (\bar{\mathbf{h}}^j_{li} )^H \bar{\mathbf{h}}^j_{jk} \right)  }=\frac{\|\mathbf{R}^j_{li}\|_2 \left(  \sum_{(l,i) \in \mathcal{P}_{jk} } p_{li} \tau_p   \beta^j_{li} + \sigma_\mathrm{ul}\right) }{\mathrm{tr}(\mathbf{R}^j_{jk} )+ \sum_{(l,i) \in \mathcal{P}_{jk} } \frac{\sqrt{p_{jk}}}{ \sqrt{p_{li}}} (\bar{\mathbf{h}}^j_{li} )^H \bar{\mathbf{h}}^j_{jk} } ,
	\end{equation}
	\hrulefill
	\vspace*{2pt}
\end{figure*}	
\begin{figure*}[tp]
	\normalsize
	\begin{equation}\label{ls:derivation:2}
	\frac{2\sqrt{p_{li}} \tau_p  \mathrm{Re}\left\lbrace  (\bar{\mathbf{y}}^p_{jjk})^H \bar{\mathbf{h}}^j_{li} \mathrm{tr}\left(  \mathbf{R}^j_{li} \right) +  (\bar{\mathbf{y}}^p_{jjk})^H \mathbf{R}^j_{li} \bar{\mathbf{h}}^j_{li} \right\rbrace }{M_j\left( \mathrm{tr}(\mathbf{R}^j_{jk} )+ \sum_{(l,i) \in \mathcal{P}_{jk} } \frac{\sqrt{p_{jk}}}{ \sqrt{p_{li}}} (\bar{\mathbf{h}}^j_{li} )^H \bar{\mathbf{h}}^j_{jk} \right) } - \frac{2\sqrt{p_{li}} \tau_p   \|\bar{\mathbf{h}}^j_{li} \|^2 \mathrm{tr}\left(  \mathbf{R}^j_{li} \right) }{M_j\left( \mathrm{tr}(\mathbf{R}^j_{jk} )+ \sum_{(l,i) \in \mathcal{P}_{jk} } \frac{\sqrt{p_{jk}}}{ \sqrt{p_{li}}} (\bar{\mathbf{h}}^j_{li} )^H \bar{\mathbf{h}}^j_{jk} \right)  } \rightarrow 0 ,
	\end{equation}
	\hrulefill
	\vspace*{2pt}
\end{figure*}	

\section{Proof of Theorem \ref{thAsMMSE} and Theorem \ref{thAsDLMMSE}}\label{ProofthAsMMSE}
The proof begins with dividing the numerator and denominator of $\gamma^{\mathrm{ul,mmse}}_{jk}$ in \eqref{MMSEeq1} by $M_j$. The numerator becomes $ \frac{p^2_{jk} \tau_p}{M_j} \mathrm{tr}\left( \mathbf{R}^j_{jk} \boldsymbol{\Psi}^j_{jk} \mathbf{R}^j_{jk}\right) +\frac{p_{jk} }{M_j}\| \bar{\mathbf{h}}^j_{jk}   \|^2$. This term is strictly positive  and finite as $M_j \rightarrow \infty$  due to the Assumptions 1 and 2. The  non-coherent interference in \eqref{xi1} satisfies \eqref{asyeq1} at the top of next page.

	 If $(l,i)\neq (j,k)$ then the upper bound in \eqref{asyeq1} goes to zero as $M_j \rightarrow \infty$ due to Assumptions 1--3.  The LoS-related term $\frac{\nu_{jk}^{\mathrm{ul}} }{M_j}=\frac{\| \bar{\mathbf{h}}^j_{jk}   \|^4}{p_{jk} \tau_p M_j \mathrm{tr}\left( \mathbf{R}^j_{jk} \boldsymbol{\Psi}^j_{jk} \mathbf{R}^j_{jk}\right) + M_j \| \bar{\mathbf{h}}^j_{jk}   \|^2} $ is strictly finite as $M_j \rightarrow \infty $ due to Assumption 2. This term cancels out the non-coherent interference term ${\xi}^{\mathrm{ul}}_{li}/M_j$ for $(l,i)=(j,k)$  in \eqref{asyeq1}.  
 The noise term $\frac{\sigma^2_{\mathrm{ul}}}{M_j}$ also goes to zero as $M_j \rightarrow \infty$. 
 Considering the coherent interference in \eqref{coherentInterference}, we obtain \eqref{eq:coherent-inteference-derivation}  at the top of this page.

\setcounter{equation}{102}

The second term in \eqref{eq:coherent-inteference-derivation}  goes to zero due to Assumptions 1--3. The first term is bounded, since the expression ${ p_{jk} \tau_p\mathrm{tr}\left( \mathbf{R}^j_{jk} \boldsymbol{\Psi}^j_{jk} \mathbf{R}^j_{jk}\right) +\| \bar{\mathbf{h}}^j_{jk}   \|^2}$ in the denominator scales with $M_j$ and the traces  in the numerator can not grow faster than $M_j$ due to Assumption 1.  We note that 
\begin{equation}
	\frac{1}{M_j} \left| \mathrm{tr}\left( \mathbf{R}^j_{li} \boldsymbol{\Psi}^j_{jk} \mathbf{R}^j_{jk}\right)\right|  \leq \frac{1}{M_j } \left\| \boldsymbol{\Psi}^j_{jk} \right\| _2\mathrm{tr}\left( \mathbf{R}^j_{li} \mathbf{R}^j_{jk}\right).
\end{equation}

 It goes to zero  if $\frac{1}{M_j}\mathrm{tr}\left( \mathbf{R}^j_{li} \mathbf{R}^j_{jk}\right) \rightarrow 0$  for all  $(l,i) \in \mathcal{P}_{jk} \backslash (j,k) $, which happens under asymptotic spatial orthogonality. In this case, the SINR grows without bound as specified in the theorem. Otherwise, the SINR will asymptotically only contain these terms in the denominator.
This finishes the proof for the UL SE with MMSE estimator.
 In the DL, the expressions for $\gamma^{\mathrm{dl,mmse}}_{jk}$ contains the same matrix expressions as $\gamma^{\mathrm{ul,mmse}}_{jk}$, except that indices $(l,i)$ and $(j,k)$ are swapped in the interference terms. The proofs follows analogously.

\section{Proof of Theorem \ref{thAsEWMMSE} and Theorem \ref{thAsDLEWMMSE}}\label{ProofthAsEWMMSE}

The proof begins with dividing the numerator and denominator of $\gamma^{\mathrm{ul,ew}}_{jk}$ in \eqref{gammaew} by $M_j$  and $  p_{jk} \tau_p \mathrm{tr}\left(  \mathbf{D}^j_{jk}  \boldsymbol{\Lambda}^j_{jk} \mathbf{D}^j_{jk}\right) +\| \bar{\mathbf{h}}^j_{jk}   \|^2$. Then, the numerator is strictly positive and finite as $M_j \rightarrow \infty$ due to Assumptions 1 and 2. The interference term becomes $\chi^{\mathrm{ul}}_{li}\big/ M_j (  p_{jk} \tau_p \mathrm{tr}\left(  \mathbf{D}^j_{jk}  \boldsymbol{\Lambda}^j_{jk} \mathbf{D}^j_{jk}\right) + \| \bar{\mathbf{h}}^j_{jk}   \|^2) $ and the same procedure is applied as in the proof of Theorem \ref{thAsMMSE}. The non-coherent interference goes asymptotically zero due to Assumptions 1-3.
 For the noise term, note that $
\frac{  \mathrm{tr}\left(  \boldsymbol{\Sigma}^j_{jk} \right) + \| \bar{\mathbf{h}}^j_{jk}   \|^2 }{ p_{jk} \tau_p \mathrm{tr}\left(  \mathbf{D}^j_{jk}  \boldsymbol{\Lambda}^j_{jk} \mathbf{D}^j_{jk}\right) +\| \bar{\mathbf{h}}^j_{jk}   \|^2}	$ is bounded since the trace expressions can not grow faster than $M_j$ due to Assumption 1.  Then the noise term $
\frac{\sigma^2_{\mathrm{ul}} \left( \mathrm{tr}\left(  \boldsymbol{\Sigma}^j_{jk} \right) + \| \bar{\mathbf{h}}^j_{jk}   \|^2\right) }{M_j (  p_{jk} \tau_p \mathrm{tr}\left(  \mathbf{D}^j_{jk}  \boldsymbol{\Lambda}^j_{jk} \mathbf{D}^j_{jk}\right) +\| \bar{\mathbf{h}}^j_{jk}   \|^2)}$ goes to zero as $M_j \rightarrow \infty$ due to Assumptions 1 and 2.
 The coherent interference term
\begin{equation}
  \frac{ \frac{p_{jk} p_{li}  \tau^2_p}{M_j}\left(\mathrm{tr}\left( \mathbf{D}^j_{li} \boldsymbol{\Lambda}^j_{jk} \mathbf{D}^j_{jk}\right) \right)  ^2    }{p_{jk} \tau_p \mathrm{tr}\left( \mathbf{D}^j_{jk} \boldsymbol{\Lambda}^j_{jk} \mathbf{D}^j_{jk}\right) +\| \bar{\mathbf{h}}^j_{jk}   \|^2} 
\end{equation}
is bounded since the expression ${ p_{jk} \tau_p\mathrm{tr}\left( \mathbf{D}^j_{jk} \boldsymbol{\Lambda}^j_{jk} \mathbf{D}^j_{jk}\right) +\| \bar{\mathbf{h}}^j_{jk}   \|^2}$ in the denominator scales with $M_j$ and the traces  in the numerator can not grow faster than $M_j$ due to Assumption 1.  For this term, we note that
\begin{equation}
\frac{1}{M_j} \mathrm{tr}\left( \mathbf{D}^j_{li} \boldsymbol{\Lambda}^j_{jk} \mathbf{D}^j_{jk}\right) \leq \frac{1}{M_j } \left\| \boldsymbol{\Lambda}^j_{jk} \right\| _2\mathrm{tr}\left( \mathbf{D}^j_{li} \mathbf{D}^j_{jk}\right) .
\end{equation}
  
  It goes to zero  if $\frac{1}{M_j}\mathrm{tr}\left( \mathbf{D}^j_{li} \mathbf{D}^j_{jk}\right) \rightarrow 0$  for all  $(l,i) \in \mathcal{P}_{jk} \backslash (j,k) $, which happens under asymptotic spatial orthogonality. In this case, the whole denominator vanishes and the SINR goes asymptotically to infinity. Otherwise, it is only these terms that remain asymptotically in the denominator.
 This completes the proof of the UL part. In the DL, we divide the numerator and denominator of $\gamma^{\mathrm{dl,ew}}_{jk}$ in \eqref{dlgammaew} by $  p_{jk} \tau_p \mathrm{tr}\left(  \mathbf{D}^j_{jk}  \boldsymbol{\Lambda}^j_{jk} \mathbf{D}^j_{jk}\right) + \| \bar{\mathbf{h}}^j_{jk}   \|^2$. The numerator becomes strictly positive and finite due to Assumptions 1 and 2. The interference term is $
	\frac{\chi^{\mathrm{dl}}_{li} }{\left( \mathrm{tr}\left(  \boldsymbol{\Sigma}^l_{li} \right) + \| \bar{\mathbf{h}}^l_{li}   \|^2\right) \left(   p_{jk} \tau_p \mathrm{tr}\left(  \mathbf{D}^j_{jk}  \boldsymbol{\Lambda}^j_{jk} \mathbf{D}^j_{jk}\right) +\| \bar{\mathbf{h}}^j_{jk}   \|^2\right) }.$ By applying similar process as in the proof of Theorem \ref{thAsDLMMSE}, we can complete the proof of Theorem \ref{thAsDLEWMMSE}.

\setcounter{equation}{107}
\section{Proof of Theorem \ref{thAsLS}} \label{proofthAsLS}
The proof begins with dividing the numerator and denominator of $\gamma^{\mathrm{ul,ls}}_{jk}$ in \eqref{LSeq1} by $M_j$ and $ \left| \mathrm{tr}(\mathbf{R}^j_{jk} )+ \sum_{(l,i) \in \mathcal{P}_{jk} } \frac{\sqrt{p_{jk}}}{\sqrt{p_{li}} } (\bar{\mathbf{h}}^j_{li} )^H \bar{\mathbf{h}}^j_{jk} \right| $. The numerator becomes $ \left| \frac{p_{jk}}{M_j}\mathrm{tr}(\mathbf{R}^j_{jk} )+ \sum_{(l,i) \in \mathcal{P}_{jk} } \frac{\sqrt{p_{jk}}}{M_j\sqrt{p_{li}} } (\bar{\mathbf{h}}^j_{li} )^H \bar{\mathbf{h}}^j_{jk} \right| $. Note that $\frac{p_{jk}}{M_j}\mathrm{tr}(\mathbf{R}^j_{jk} )>0$ due to Assumption 1 and  the only non-zero term in the second part is $\frac{1}{M_j}\|\bar{\mathbf{h}}^j_{jk}\|^2$ due to  Assumption 3.
The noise term $
\frac{\sigma^2_{\mathrm{ul}}}{p_{jk} \tau^2_p  M_j}\frac{\tau_p\mathrm{tr}\left( (\boldsymbol{\Psi}^j_{jk})^{-1} \right) +  \|\bar{\mathbf{y}}^p_{jjk}\|^2 }{ \left|  \mathrm{tr}(\mathbf{R}^j_{jk} )+ \sum_{(l,i) \in \mathcal{P}_{jk} } \frac{\sqrt{p_{li}}}{\sqrt{p_{jk}} } (\bar{\mathbf{h}}^j_{li} )^H \bar{\mathbf{h}}^j_{jk}\right| } $ goes asymptotically  to zero since the first factor goes to zero and the second factor is bounded  due to Assumptions 1--3.
 For the interference term, by using $\mathrm{tr}\left( \mathbf{R}^j_{li}(\boldsymbol{\Psi}^j_{jk})^{-1} \right)\leq \|\mathbf{R}^j_{li}\|_2 \mathrm{tr}\left( (\boldsymbol{\Psi}^j_{jk})^{-1} \right)$, the upper bound of the interference term is \eqref{as:ls:derivation} at the top of previous page and it goes  asymptotically to zero due to Assumptions 1 and 2. Another term is \eqref{ls:derivation:2} at the top of previous page asymptotically due to Assumption 1--3. The inequalities
 \begin{align}
& \left( \bar{\mathbf{y}}^p_{jjk}\right)^H  {\mathbf{R}^j_{li} } \bar{\mathbf{y}}^p_{jjk} \leq ||\mathbf{R}^j_{li} ||_2 \|\bar{\mathbf{y}}^p_{jjk} \|^2  \\
&(\bar{\mathbf{h}}^j_{li})^H (\boldsymbol{\Omega}^j_{jk})^{-1}\bar{\mathbf{h}}^j_{li} \leq \|(\boldsymbol{\Omega}^j_{jk})^{-1} \|_2 \|\bar{\mathbf{h}}^j_{li} \|^2\\
& (\bar{\mathbf{h}}^j_{li})^H (\boldsymbol{\Psi}^j_{jk})^{-1}\bar{\mathbf{h}}^j_{li} \leq \|(\boldsymbol{\Psi}^j_{jk})^{-1} \|_2 \|\bar{\mathbf{h}}^j_{li} \|^2 \\
& \bar{\mathbf{x}}_{jk}^H   \bar{\mathbf{h}}^j_{li} (\bar{\mathbf{h}}^j_{li})^H \bar{\mathbf{h}}^j_{li} \leq  \|\bar{\mathbf{h}}^j_{li} \bar{\mathbf{x}}^H_{jk} \|_2 \|\bar{\mathbf{h}}^j_{li} \|^2
 \end{align}
 are also useful. Using these inequalities, Assumptions 1 and 2, and noting that 
 \begin{equation}
 	\frac{p_{li} \tau^2_p\left(  \mathrm{tr}(\mathbf{R}^j_{li} )\right) ^2 + p_{li}\tau^2_p \|\bar{\mathbf{h}}^j_{li}\|^4 }{M_jp_{jk} \tau^2_p\left( \mathrm{tr}(\mathbf{R}^j_{jk} )+ \sum_{(l,i) \in \mathcal{P}_{jk} } \frac{\sqrt{p_{jk}}}{ \sqrt{p_{li}}} (\bar{\mathbf{h}}^j_{li} )^H \bar{\mathbf{h}}^j_{jk}\right)  }
 \end{equation}
 has a positive finite limit as $M_j \rightarrow \infty$, we obtain \eqref{aslseq1}. This finishes the proof for the UL.

In the DL, we divide the numerator and denominator of $\gamma^{\mathrm{dl,ls}}_{jk}$ in \eqref{lsdlgamma} by $ M_j $. The numerator is  positive and finite as $M_j \rightarrow \infty$ due to Assumptions 1 and 2. The noise term $\frac{\sigma^2_\mathrm{dl}}{M_j}$ goes  to zero as $M_j$ goes to infinity. For interference term, we follow the approach from the UL part to obtain \eqref{asdleq1}.

\bibliographystyle{IEEEtran}
\bibliography{IEEEabrv,references}

%

\begin{IEEEbiography}[{\includegraphics[width=1in,height=1.25in,clip,keepaspectratio]{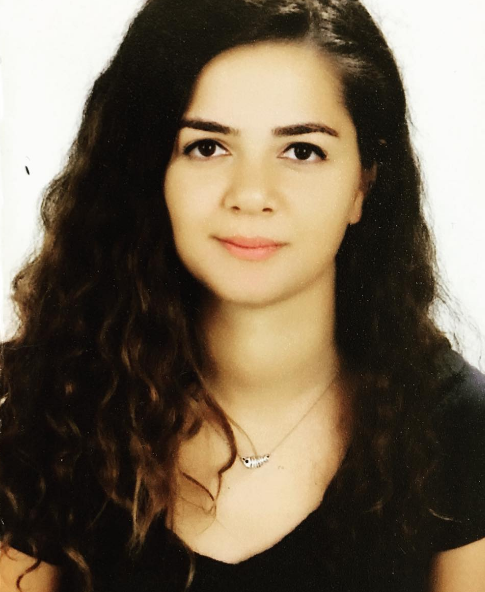}}]{\"Ozgecan \"Ozdogan}
(S'18) received her B.Sc and M.Sc. degrees in Electronics and Communication Engineering from İzmir Institute of Technology, Turkey in
2015 and 2017 respectively. She is currently pursuing the Ph.D. degree in communication systems at Link\"oping University, Sweden.
\end{IEEEbiography}
\vfill
\begin{IEEEbiography}[{\includegraphics[width=1in,height=1.25in,clip,keepaspectratio]{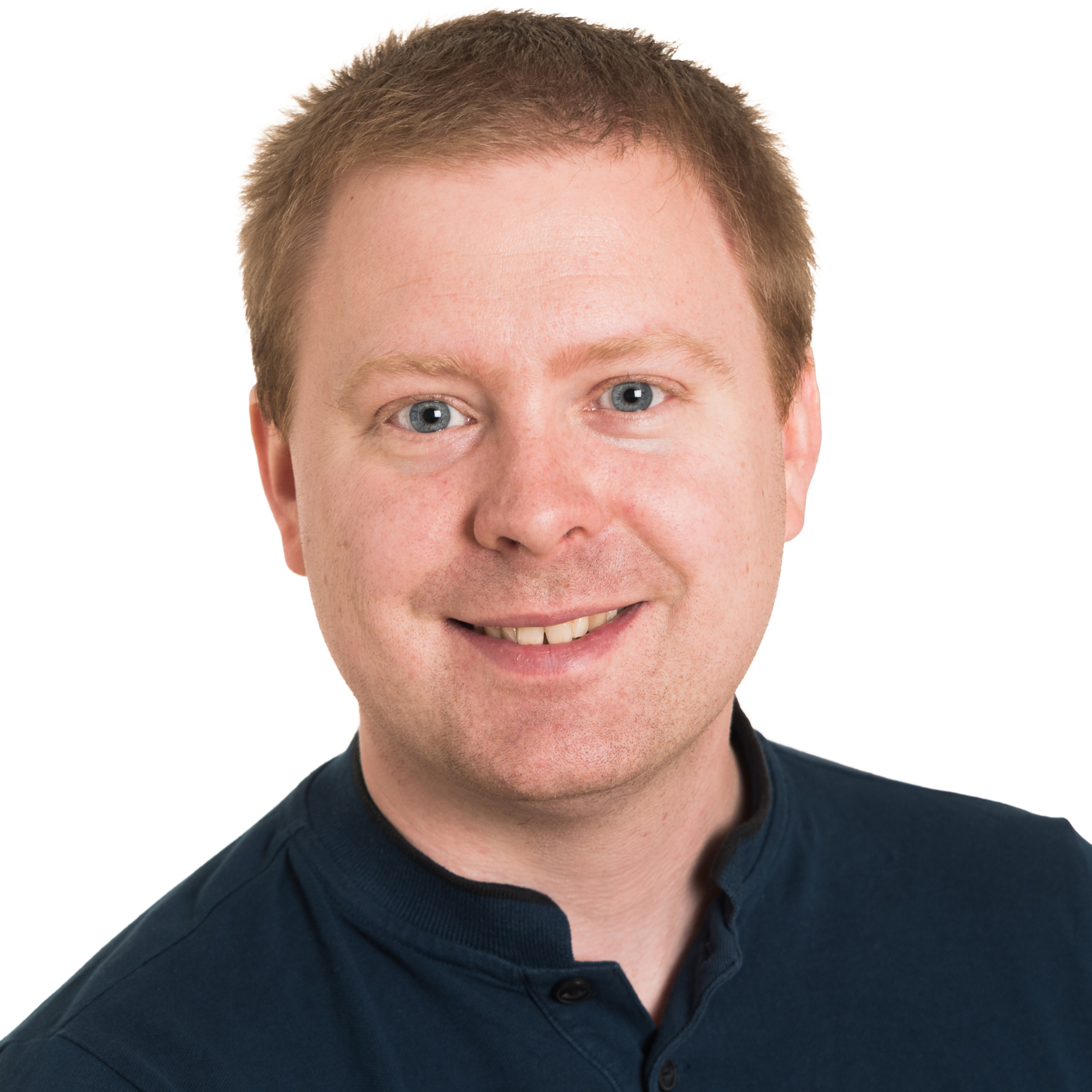}}]{Emil Bj\"ornson }(S'07-M'12-SM'17)
 received the M.S. degree in Engineering Mathematics from Lund University, Sweden, in 2007. He received the Ph.D. degree in Telecommunications from KTH Royal Institute of Technology, Sweden, in 2011. From 2012 to mid 2014, he was a joint postdoc at the Alcatel-Lucent Chair on Flexible Radio, SUPELEC, France, and at KTH. He joined Link\"oping University, Sweden, in 2014 and is currently Associate Professor and Docent at the Division of Communication Systems.

He performs research on multi-antenna communications, Massive MIMO, radio resource allocation, energy-efficient communications, and network design. He is on the editorial board of the IEEE Transactions on Communications (since 2017) and the IEEE Transactions on Green Communications and Networking (since 2016). He is the first author of the textbooks ``Massive MIMO Networks: Spectral, Energy, and Hardware Efficiency'' (2017)  and ``Optimal Resource Allocation in Coordinated Multi-Cell Systems'' from 2013. He is dedicated to reproducible research and has made a large amount of simulation code publicly available.

Dr. Bj\"ornson has performed MIMO research for more than ten years and has filed more than ten related patent applications. He received the 2018 Marconi Prize Paper Award in Wireless Communications, the 2016 Best PhD Award from EURASIP, the 2015 Ingvar Carlsson Award, and the 2014 Outstanding Young Researcher Award from IEEE ComSoc EMEA. He also co-authored papers that received best paper awards at the conferences WCSP 2017, IEEE ICC 2015, IEEE WCNC 2014, IEEE SAM 2014, IEEE CAMSAP 2011, and WCSP 2009.
\end{IEEEbiography}
\vfill

\begin{IEEEbiography}[{\includegraphics[width=1in,height=1.25in,clip,keepaspectratio]{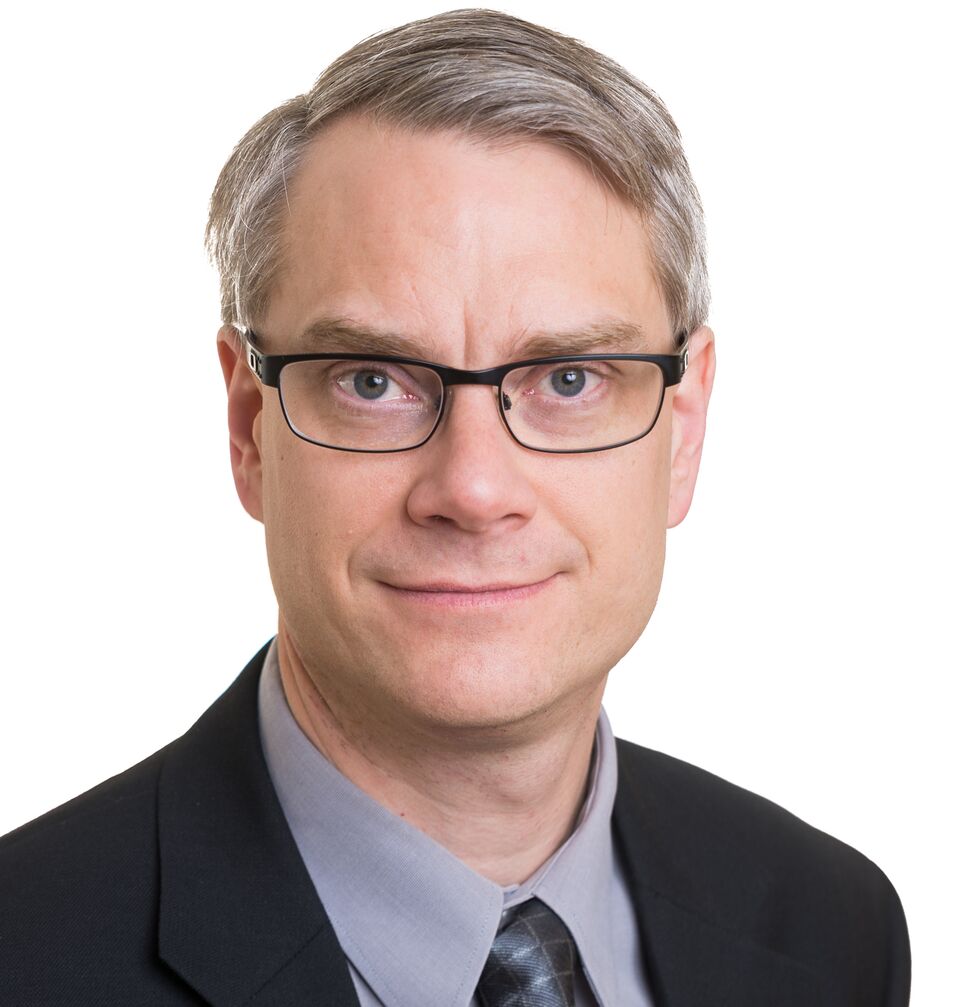}}]{Erik G. Larsson} (S'99-M'03-SM'10-F'16) received the Ph.D. degree from Uppsala University,
Uppsala, Sweden, in 2002.
He is currently Professor of Communication Systems at Link\"oping
University (LiU) in Link\"oping, Sweden. He was with the KTH Royal
Institute of Technology in Stockholm, Sweden, the George Washington
University, USA, the University of Florida, USA, and Ericsson
Research, Sweden.   His main professional interests are
within the areas of wireless communications and signal processing. He
has co-authored some 170 journal papers on these topics, and 
the two Cambridge University Press textbooks
\emph{Space-Time Block Coding for Wireless Communications} (2003) and
\emph{Fundamentals of Massive MIMO} (2016). He is co-inventor on 19
issued U.S. patents.

He is a member of the IEEE Signal Processing Society Awards Board
(2017--2019),  an editorial board member of the
\emph{IEEE Signal Processing Magazine} (2018--2020),
and a member of the steering committee for the 
\emph{IEEE Transactions on Wireless Communications}
(2019--2022). 
From 2015 to 2016 he was chair of the IEEE
Signal Processing Society SPCOM technical committee.  From 2014 to
2015 he was chair of the steering committee for the \emph{IEEE
	Wireless Communications Letters}.  He was   General Chair of the
Asilomar Conference on Signals, Systems and Computers in 2015, and its
Technical Chair in 2012.  He was Associate Editor for, among others,
the \emph{IEEE Transactions on Communications} (2010-2014) and the
\emph{IEEE Transactions on Signal Processing} (2006-2010).

He received the IEEE Signal Processing Magazine Best Column Award
twice, in 2012 and 2014, the IEEE ComSoc Stephen O. Rice Prize in
Communications Theory in 2015, the IEEE ComSoc Leonard G. Abraham
Prize in 2017, and the IEEE ComSoc Best Tutorial Paper Award in 2018. 
\end{IEEEbiography}





\end{document}